\documentclass[%
 reprint,
superscriptaddress,
 amsmath,amssymb,
 aps,
prb,
floatfix,
longbibliography
]{revtex4-2}

\usepackage{silence}%https://tex.stackexchange.com/questions/180762/revtex4-1-warning-repair-the-float-package
\usepackage{graphicx}% Include figure files
\usepackage{dcolumn}% Align table columns on decimal point
\usepackage{bm}% bold math
\usepackage{bbm} % \mathbbm{1}
\usepackage{mathtools}
\usepackage{leftindex} % \leftindex https://tex.stackexchange.com/questions/11542/left-and-right-subscript-superscript
\graphicspath{{./Figures01/},{./Figures01_OLD/}} 

\usepackage{float} % Paquete que permite la orden [H] = Pon esta figura aqu\'{i} bajo mi responsabilidad por muy feo que quede
\usepackage{fix-cm} % Caracteres de cualquier tamaño!!!
\usepackage[dvipsnames]{xcolor}
\usepackage{xcolor}
\usepackage[normalem]{ulem}
\usepackage[english]{babel}
\usepackage{graphicx}% Include figure files
\usepackage{dcolumn}% Align table columns on decimal point
\usepackage{verbatim}
\usepackage{mathrsfs}
\usepackage{cancel}
\usepackage{epstopdf}
\usepackage{color}
\usepackage{tensor}
\usepackage{relsize} % Escalar caracteres
\usepackage{ytableau}
\usepackage{slashed} % For Feynamnn slash notation in Dirac operators \slashed{p}
\usepackage{boldline} % For using V{3} and \hlineB{4}. https://tex.stackexchange.com/questions/425803/how-to-place-thick-vertical-line-in-table

\usepackage{pifont}% http://ctan.org/pkg/pifont
\usepackage[dvipsnames]{xcolor} 
\definecolor{Verde}{RGB}{0, 150, 0} %RGB de 0 a 255

\DeclareFontFamily{U}{mathc}{}
\DeclareFontShape{U}{mathc}{m}{it}%
{<->s*[1.03] mathc10}{}
\DeclareMathAlphabet{\mathscr}{U}{mathc}{m}{it}

\newcommand{\bra}[1]{\Big\langle {#1} \Big|}
\newcommand{\ket}[1]{\Big| {#1} \Big\rangle}

\newcommand{\eq}[1]{(\ref{#1})}
\newcommand{\sgn}{\text{sgn}}

\newcommand{\dd}{\mathrm{d}}
\newcommand{\ii}{\mathrm{i}}

\newcommand{\qq}{\mathrm{q}}
\newcommand{\kk}{\mathrm{k}}

\newcommand{\Tr}[1]{\text{Tr}\left(#1\right)}
\newcommand{\Abs}[1]{\left\vert #1\right\vert}

\newcommand{\surf}{\parallel}
\newcommand{\dlim}{\displaystyle\lim}
\newcommand{\dint}{\displaystyle\int}
\newcommand{\dsum}{\displaystyle\sum}
\newcommand{\mean}[1]{\big\langle #1\big\rangle}
\newcommand{\abs}[1]{\vert #1\vert}

\newcommand{\Imag}[1]{\mathbb{I}\text{m}\left[ #1 \right]}
\newcommand{\Real}[1]{\mathbb{R}\text{e}\left[ #1 \right]}
\newcommand{\Eq}[1]{Eq.~\eqref{#1}}
\newcommand{\Eqs}[1]{Eqs.~\eqref{#1}}
\newcommand{\Fig}[1]{Fig.~\ref{#1}}
\newcommand{\Tab}[1]{tab.~\ref{#1}}
\newcommand{\Sect}[1]{Sec.~\ref{#1}}

\newcommand{\NEW}[1]{\textcolor{red}{#1}}

\newcommand{\eV}{\text{ eV}}
\newcommand{\KK}{\text{ K}}

\usepackage[usenames,dvipsnames]{pstricks}
\usepackage{epsfig}
\usepackage{pst-grad} % For gradients
\usepackage{pst-plot} % For axes

\usepackage{hyperref}

\usepackage{verbatim}

\begin{document}

\preprint{APS/123-QED}

%\title{Graphene electric conductivity: Kubo model versus QFT model}
\title{Electric conductivity in graphene: Kubo model versus a nonlocal quantum field theory model}
%\title{Role of non-locality and dispersion in Casimir effect of graphene}

\author{Pablo Rodriguez-Lopez}
\email{pablo.ropez@urjc.es}
\affiliation{{\'A}rea de Electromagnetismo and Grupo Interdisciplinar de Sistemas Complejos (GISC), Universidad Rey Juan Carlos, 28933, M{\'o}stoles, Madrid, Spain}
\affiliation{Laboratoire Charles Coulomb (L2C), UMR 5221 CNRS-University of Montpellier, F-34095 Montpellier, France}

\author{Jian-Sheng Wang}
\affiliation{Department of Physics, National University of Singapore, Singapore 117551, Republic of Singapore}

\author{Mauro Antezza}
\email{mauro.antezza@umontpellier.fr}
\affiliation{Laboratoire Charles Coulomb (L2C), UMR 5221 CNRS-University of Montpellier, F-34095 Montpellier, France}
\affiliation{Institut Universitaire de France, Ministère de  l’Enseignement Supérieur et de la Recherche, 1 rue Descartes, F-75231, Paris, France}

\date{\today}% It is always \today, today,

\begin{abstract}
We compare three models of graphene electric conductivity: a non-local Kubo model, a local model derived by Falkovsky, and finally, a non-local quantum field theory (QFT) polarization-based model. These models are supposed to provide consistent results since they are derived from the same Hamiltonian. While we confirm that the local model is a proper $\bm{q}\to\bm{0}$ limit of both the non-local Kubo and the non-local QFT model (once losses are added to this last model), we find hard inconsistencies in the non-local QFT model as derived and currently used in literature. In particular, in the genuine non-local region ($\bm{q}\neq\bm{0}$), the available QFT model shows an intrinsic non-physical plasma-like behavior for the interband transversal electric conductivity at low frequencies (even after introducing the unavoidable losses). The Kubo model, instead, shows the expected behavior, i.e., an almost constant electric conductivity as a function of frequency $\omega$ with a gap for frequencies $\hbar\omega < \sqrt{(\hbar v_{F}q)^{2} + 4m^{2}}$. We show that the Kubo and QFT models can be expressed using an identical Polarization operator $\Pi_{\mu\nu}(\omega,\bm{q})$, but they employ different expressions for the electric conductivity $\sigma_{\mu\nu}(\omega,\bm{q})$. In particular, the Kubo model uses a standard regularized expression, a direct consequence of Ohm's Law and causality, as we rigorously re-derive. We show that, once the standard regularized expression for $\sigma_{\mu\nu}(\omega,\bm{q})$ is used in the QFT model, and losses are included, the Kubo and QFT model coincide, and all its anomalies naturally disappear. Our findings show the necessity to appropriately define and regularize the electric conductivity to connect it with the available QFT model. This can be relevant for theory, predictions, and experimental tests in the nanophotonics and Casimir effect communities.%\cite{PhysRevB.111.115428}
\end{abstract}

\maketitle

\section{Introduction}
Since graphene was isolated in 2004 \cite{doi:10.1126/science.1102896}, the electric conductivity of graphene has been of great interest due to the wide range of potential applications of this emergent 2D material \cite{RevModPhys.81.109}\cite{RevModPhys.88.045003}\cite{DasSarmaRMP2011}.

The general expression linking the total induced electric current to the most general electromagnetic field is \cite{Fialkovsky2011}\cite{Fialkovsky2016}\cite{Fialkovsky2012}\cite{Bordag2015b}:
\begin{eqnarray}\label{Def_J_PIA}
\mean{ J^{{\rm tot}}_{\mu} } = - \Pi_{\mu\nu}A^{\nu},
\end{eqnarray}
where $\Pi_{\mu\nu}$ is the polarization tensor defined by the current-current correlation, and $A^{\nu}$ is the electromagnetic vector potential (we use here the $\phi = c A^{0} =0$ gauge). In this paper, we address the problem of the derivation of the electric conductivity $\sigma_{\mu\nu}$ of graphene, which is defined as the transport coefficient that relates only the electric field $E^{\nu}$ with the electric current it induces, according to Ohm's Law \cite{Bordag2015b}\cite{Klimchitskaya2018}:
\begin{eqnarray}\label{Def_J_SE}
\mean{ J_{\mu} } = \sigma_{\mu\nu}E^{\nu}.
\end{eqnarray}
It is worth stressing that the current in \Eq{Def_J_SE} is not the total one since, differently from the one in \Eq{Def_J_PIA}, it does not consider the system's response to magnetic fields.
There are several different models for the electric conductivity of graphene depending on the level of detail and formalism applied. Here, we are interested in two particular ones: the first is based on the Kubo formula \cite{Kubo1957}\cite{4463896}\cite{Luttinger1968} and has been first derived in \cite{non-local_Graphene_Lilia_Pablo}, providing, for the electric conductivity, the following regularized expression:
\begin{eqnarray}%\label{True_Relation_Polarization_conductivity_maintext}
\sigma_{\mu\nu}^{\rm{K}}(\omega,\bm{q}) = \dfrac{\Pi_{\mu\nu}(\omega,\bm{q}) - \dlim_{\omega\to0^{+}}\Pi_{\mu\nu}(\omega,\bm{q})}{-\ii\omega}.
\end{eqnarray}
The second model, based on a Quantum Field Theory (QFT) formulation, provides a non-regularized (NR) expression for the electric conductivity:
\cite{Fialkovsky2016}\cite{Fialkovsky2012}\cite{Bordag2015b}\cite{Klimchitskaya2018}\cite{universe6090150}\cite{Klimchitskaya2016}\cite{Klimchitskaya2016b}\cite{Klimchitskaya2017}\cite{Gusynin2007b}
\begin{eqnarray}%\label{Relation_Polarization_conductivity}
\sigma_{\mu\nu}^{\rm{NR}}(\omega,\bm{q}) = \dfrac{\Pi_{\mu\nu}(\omega,\bm{q})}{-\ii\omega}.
\end{eqnarray}

We will focus on comparing the Kubo and QFT models. These two models have never been directly compared.  Hence, it is interesting to make a comparative analysis and discuss the origin of their differences and which one provides the correct expression for electric conductivity. In this article, we show that by construction, the Kubo formula provides regularized results that guarantee the fulfillment of the condition $\dlim_{\omega\to0} \langle J_\mu \rangle = 0$ for zero electric field $E_{\nu}(\omega)$ \cite{Falkovsky2007} and includes the effect of dissipation of electronic quasiparticles in the electric conductivity. On the other hand, the QFT model, in the form it was developed and used in literature \cite{PRL_Mohideen}\cite{Bordag2015}\cite{Bimonte2017}, not only does not consider unavoidable effects of losses on the electric conductivity, but also predicts an additional divergent dissipation-less plasma behavior that cannot be cured by adding losses. This dissipation-less plasma behavior that appears in the non-local transverse interband electric conductivity \cite{PRL_Mohideen}\cite{Bordag2015}\cite{Bimonte2017} even when the chemical potential $\mu$ is inside the mass gap of the band spectrum, is not an acceptable result in normal materials, since it would lead to unobserved dissipation-less permanent currents in graphene, independently from the value of the mass gap, of the chemical potential, of the temperature and dissipation, in close analogy to superconductors \cite{Annett:730995}\cite{London}, but without a proper microscopic theory \cite{Bardeen1951}\cite{Coleman2015}. 
We show that this non-physical dissipation-less plasma behavior is naturally removed using a standard regularization (which we rigorously derive for this case). This makes the regularized QFT model \emph{identical} to the Kubo model. The Kubo and the QFT models are not different; none have a particular ``first-principle'' advantage.  They start from precisely the same Hamiltonian and they provide exactly the same final electric conductivity/Polarization result once the regularization is employed. 

It is worth stressing that the Kubo and the QFT models are identical (once losses are included in the QFT model) in the local limit ($q=0$). Their longitudinal components are also identical in the non-local limit.

Several different models of the electric response have been used to study graphene electric conductivity \cite{RevModPhys.81.109}\cite{RevModPhys.88.045003}\cite{RevModPhys.82.2673}. For low frequencies, as the Dirac point is close to the chemical potential $\mu$, the tight-binding model of graphene can be approximated by two $(2+1)D$ massless four-spinors or to a sum of four two-spinors. Due to its simplicity and adequacy to experimental results, the local limit of the Kubo formula, derived by Falkovsky et al. \cite{Falkovsky2007} (see also \cite{Gusynin2006} and \cite{Gusynin2009}) has been widely used \cite{Fialkovsky2011}\cite{Fialkovsky2012}\cite{Fialkovsky2008}. This model takes into account the (real or imaginary) frequency $\omega$, the chemical potential $\mu$, the temperature $T$, and the dissipation rate $\Gamma = \tau^{-1}$ of the electronic quasiparticles for the Drude electric conductivity. However, the dissipation for interband transitions and the non-zero mass gap cases are not considered in the model.

In ref.~\cite{non-local_Graphene_Lilia_Pablo}, by using the Kubo formula \cite{ALLEN2006165} and the two-spinor representation, the generalization to non-local electric conductivities of \cite{Falkovsky2007} for finite mass gaps and non-zero dissipation rate of the interband electric conductivity was performed. The authors presented closed analytical results for all complex frequencies of the imaginary positive complex plane in the zero temperature limit. From these results, the electric conductivity for finite temperature is easily obtained.

Another different approach based on Quantum Field Theory (QFT) of the Dirac four spinor in $(2+1)D$ and on the random phase approximation (RPA), like in  \cite{PhysRevB.34.979}\cite{doi:10.1142/S0217984993001612}\cite{Wunsch_2006}\cite{GONZALEZ1994595}\cite{Gonzalez1999}\cite{Pyatkovskiy_2009}, gives the electric conductivity derived from the Polarization operator \cite{Fialkovsky2016}\cite{Fialkovsky2012}\cite{Klimchitskaya2018}\cite{universe6090150}\cite{PRL_Mohideen}\cite{Bordag2015}\cite{Bimonte2017}\cite{Klimchitskaya2020}\cite{Klimchitskaya2014}\cite{Sernelius2012}\cite{Bordag2009}\cite{MostepanenkoReflConfPol}, just to cite a small set references. In this case, the dissipation rate $\Gamma = \tau^{-1}$ of the electronic quasiparticles is not considered (being equivalent to set to zero), but the results are valid for finite chemical potential $\mu$, temperature $T$ and non-topological Dirac masses $m$ (note that, in \cite{Fialkovsky2016}, the effect of topological Dirac masses was described). According to the authors of these papers, the results of those QFT models are ``obtained on the solid foundation of quantum field theory and do not use any phenomenology'' \cite{MostepanenkoReflConfPol}. They lead to double poles at zero frequency for the dielectric susceptibility ``of doubtless physical significance'' \cite{MostepanenkoReflConfPol}. This double pole $1/\omega^{2}$ in $\epsilon$ implies a single pole $1/\omega$ for $\sigma$. We will show in this paper that this ``plasma-like'' single pole $1/\omega$ is not a physical one. Indeed, it is the sum of two different $1/\omega$ divergences. One of them can be cured by adding losses, while the second one is cured by the standard regularization that must be applied to $\sigma$ (see \Sect{KubosigmaderivationAppendix}). 

% What we do
In this article, we compare the three different derivations of the electric conductivity of graphene in the small $\bm{k}\cdot\bm{p}$ limit, and show that the local result of Falkovsky et al.\ can be derived from the non-local Kubo result. We show how the QFT results are related to the non-local Kubo ones, explain in detail the origin of their differences, and how the disagreement can be fixed. We hope this study will clarify the approximations used in each model and their similarities and differences. 

The article is organized as follows: In \Sect{Section_Graphene}, we introduce the tight-binding model of graphene and the approximations used in the rest of the article. 
In \Sect{Section_EM_response}, we derive and present the formulas used to obtain the Polarization and electric conductivity of graphene in the three different models. In particular, in \Sect{KubosigmaderivationAppendix}, we rigorously derive the regularization for electric conductivity.
In \Sect{Section_Tensor_decomposition}, we show how to relate the different quantities obtained in the non-local Kubo model (the longitudinal and transversal electric conductivities) with the quantities obtained in the QFT model (the pure temporal term and trace of the Polarization operator).
In \Sect{Section_non_local_Kubo}, the non-local model of electric conductivity derived from the Kubo formula is shown. 
In \Sect{Section_yes_local_Kubo}, the Falkovsky local model of electric conductivity is presented, and its convergence with the non-local Kubo model is shown. 
In \Sect{Section_QFT_Model}, the QFT model for the Polarization (and therefore the electric conductivity) of graphene is presented. We re-derive the results shown in other articles and explicitly show the relation of this model to the non-local Kubo model when the results coincide; also, when and why they do not; we compare numerically the three different models, highlighting their similarities and differences.
We finish in \Sect{Section_Conclusions} with the conclusions.
In the appendices, we provide several computational details.

\section{Tight-Binding model of Graphene}\label{Section_Graphene}
In this section, we will derive the tight-binding model of graphene. The goal is to show which approximations are needed to obtain the $(2+1)$D Dirac Hamiltonian and the sum of four $(2+1)$D 2-spinor Hamiltonian. The relation between the two formulas for the electric conductivity we are discussing pivots around those two different representations and their Green functions.

%\begin{table}[H]
%\centering
%\begin{tabular}{V{3} c V{3} c|c|c V{3}}
%\hlineB{4}
% & $(\lambda, G)$ & $(\lambda, \nu)$ & $(E, G)$\\
%\hlineB{4}
%$\lambda = $ &  &  &  \\
%\hline
%$G = $ &  & $\lambda\frac{1 - 2\nu}{\nu}$ & $3\frac{K - \lambda}{2}$ \\
%\hlineB{4}
%\end{tabular}
%\caption{ Table with the notation used along this paper. }
%\label{fig:Comparison_dissipation}
%\end{table}
\begin{table}[ht]
\centering
\begin{tabular}{V{3} c V{3} }
\hlineB{4}
$\begin{array}{rcl|rcl}
k_{0} & = & \omega    &  q_{0} & = & \Omega = \omega + \ii\Gamma\\
\tilde{k}_{i}   & = & \hbar v_{F}k_{i}   &  \tilde{q}_{0} & = & \hbar\Omega = \hbar\omega + \ii\hbar\Gamma\\
\tilde{k}_{\mu} & = & \tilde{k}_{0} + \mu   &  \tilde{k}_{0} & = & \hbar\omega = \ii\hbar\xi = \ii\Xi\\
\tilde{\bm{k}}_{\surf} & = & \hbar v_{F}\bm{k}_{\surf}  & \bm{k}_{\surf} & = & (k_{1}, k_{2})\\
\tilde{k}_{\surf} & = & \hbar v_{F} k_{\surf} = \sqrt{ \tilde{k}_{1}^{2} + \tilde{k}_{2}^{2}}  & k_{\surf} & = & \sqrt{ k_{1}^{2} + k_{2}^{2}}\\
\tilde{\bm{q}}_{\surf} & = & \hbar v_{F}\bm{q}_{\surf}  & \bm{q}_{\surf} & = & (q_{1}, q_{2})\\
\tilde{q}_{\surf} & = & \hbar v_{F} q_{\surf} = \sqrt{ \tilde{q}_{1}^{2} + \tilde{q}_{2}^{2}}  & q_{\surf} & = & \sqrt{ q_{1}^{2} + q_{2}^{2}}\\
\tilde{k}_{z} & = & \hbar v_{F} k_{z} = \sqrt{ \tilde{k}_{0}^{2} - \tilde{k}_{\surf}^{2} }  & k_{z} & = & \sqrt{ \left(\frac{\omega}{v_{F}}\right)^{2} - k_{\surf}^{2} }\\
\tilde{q}_{z} & = & \hbar v_{F} q_{z} = \sqrt{ \tilde{q}_{0}^{2} - \tilde{q}_{\surf}^{2} }  & q_{z} & = & \sqrt{ \left(\frac{\Omega}{v_{F}}\right)^{2} - q_{\surf}^{2} }\\
\tilde{\kappa}_{z} & = & \ii\tilde{k}_{z} = \sqrt{ \Xi^{2} + \tilde{k}_{\surf}^{2} } & \kk & = & k_{\mu} = (k_{0}, \tilde{\bm{k}}_{\surf})\\
\tilde{\theta}_{z} & = & \ii\tilde{q}_{z} = \sqrt{ \Xi^{2} + \tilde{q}_{\surf}^{2} } & \qq & = & q_{\mu} = (q_{0}, \tilde{\bm{q}}_{\surf})\\
\tilde{\gamma}_{\mu} & = & (\tilde{\gamma}_{0}, \tilde{\bm{\gamma}}) = (\hbar\gamma_{0}, \hbar v_{F}\bm{\gamma}) &  \tilde{\gamma}_{i} & = & \hbar v_{F}\gamma_{i}\\
\dint_{k} & = & \dint_{-\infty}^{\infty}\dfrac{\dd k_{0}}{2\pi}\dint_{\bm{k}}  & \dint_{\bm{k}} & = & \dint_{BZ}\dfrac{\dd^{2}\bm{k}_{\surf}}{(2\pi)^{2}}\\
\delta & = & \dfrac{2\abs{m}}{\tilde{\theta}_{z}}  &  \gamma & = & \dfrac{\Xi}{\tilde{\theta}_{z}}
\end{array}$\\
\hlineB{4}
$\begin{array}{rcccccl}
K_{\nu} & = & \kk + (\mu,\bm{0}) & = & (\tilde{k}_{\mu}, \tilde{\bm{k}}_{\surf}) & = & (k_{0} + \mu, \hbar v_{F}\bm{k}_{\surf})\\
S_{\mu} & = & K_{\mu} + q_{\mu} & = & (\tilde{s}_{0}, \tilde{\bm{s}}_{\surf}) & = & (k_{0} + q_{0} + \mu, \tilde{\bm{k}}_{\surf} + \tilde{\bm{q}}_{\surf})
\end{array}$\\
\hlineB{4}
\end{tabular}
\caption{Notation used in this article. }
\label{Notation}
\end{table}

Graphene is a 2D material with a honeycomb lattice, whose unit cell consists of two nonequivalent carbon atoms in $sp^{2}$ electronic configuration. We model the electronic excitations of graphene in the grand-canonical ensemble with a tight-binding model of a bidimensional honeycomb lattice \cite{RevModPhys.81.109}\cite{CastroNeto2018}\cite{Gusynin2007}\cite{Ezawa2015}\cite{Ezawa2013PRB}. This lattice is made by two non-equivalent triangular lattices (denoted as $A$ and $B$ here). The position of the atoms in each sublattice (or of the unit cell) can be specified by a vector $\bm{R}_{n_{1},n_{2}} = n_{1}\bm{a}_{1} + n_{2}\bm{a}_{2}$ ($n_{i}\in\mathbb{Z}$), with lattice vectors
\begin{eqnarray}
\bm{a}_{1} = \frac{\sqrt{3}a}{2}\left(\begin{array}{c}
1\\
\sqrt{3}
\end{array}\right),
\hspace{3pt}
\bm{a}_{2} = \frac{\sqrt{3}a}{2}\left(\begin{array}{c}
1\\
-\sqrt{3}
\end{array}\right),
\end{eqnarray}
being $a=1.42\textup{~\AA}$ the carbon-carbon interatomic distance in graphene. The nearest neighbors of an atom of the sublattice $A$ are given by the vectors
\begin{eqnarray}
\bm{\delta}_{1} = a\left(\begin{array}{c}
0\\
1
\end{array}\right),
\hspace{3pt}
\bm{\delta}_{2} = \frac{a}{2}\left(\begin{array}{c}
\sqrt{3}\\
-1
\end{array}\right),
\hspace{3pt}
\bm{\delta}_{3} = \frac{a}{2}\left(\begin{array}{c}
-\sqrt{3}\\
-1
\end{array}\right).
\end{eqnarray}
The reciprocal lattice is also a honeycomb lattice, whose fundamental translation vectors $\bm{b}_{j}$ are defined by the relation $\bm{a}_{i}\cdot\bm{b}_{j} = 2\pi\delta_{ij}$, resulting in
\begin{eqnarray}
\bm{b}_{1} = \frac{2\pi}{3a}\left(\begin{array}{c}
\sqrt{3}\\
1
\end{array}\right),
\hspace{3pt}
\bm{b}_{2} = \frac{2\pi}{3a}\left(\begin{array}{c}
\sqrt{3}\\
-1
\end{array}\right).
\end{eqnarray}
The tight-binding Hamiltonian of graphene in real space is
\begin{eqnarray}
H = \sum_{\bm{n}\in\mathbb{Z}^{d}}\sum_{\ell,\ell'}\sum_{j=1}^{\abs{\bm{\delta}_{j}}<\delta_{M}}
\hat{c}^{\dagger}_{\ell,\bm{R}_{\bm{n}}}t_{\ell,\ell'}(\bm{\delta}_{j})\hat{c}_{\ell',\bm{R}_{\bm{n}} + \bm{\delta}_{j}},
\end{eqnarray}
where $\hat{c}^{\dagger}_{\ell,\bm{R}_{\bm{n}}}$ is the creation operator of an electron placed at $\bm{R}_{\bm{n}} = \bm{a}_{1}n_{1} + \bm{a}_{2}n_{2}$, with the three quantities spin $s \in \{\uparrow, \downarrow\}$, triangular sub-lattice $\mathcal{L}\in\{A,B\}$ and orbital $\mathcal{O}$ (note that $\mathcal{O}= 2p_{z}$ in our case for graphene), labeled by the single symbol $\ell=\{\mathcal{L},\mathcal{O},s\}$ in the unit cell.
$\hat{c}_{\ell,\bm{R}_{\bm{n}}}$ is the annihilation operator of the electron. $t_{\ell,\ell'}(\bm{\delta}_{j})$ is the tight-binding coupling between an electron placed at $\bm{R}_{\bm{n}}$, with spin, sub-lattice and orbital fixed by $\ell$ and another electron placed at $\bm{R}_{\bm{n}}+\bm{\delta}_{j}$, with spin, sub-lattice and orbital fixed by $\ell'$. Those coefficients can be calculated for each particular case. $\delta_{M}$ is the maximum hopping distance between atoms we consider in the model.
As the chemical potential $\mu$ is close to the Dirac points, and we are interested in relatively small frequencies, we only consider the $\pi$ and $\pi^{*}$ bands in our model and first neighbors coupling only. Therefore, the tight-binding Hamiltonian operator of graphene in real space is reduced to
\begin{eqnarray}
H & = & - t\sum_{s = \pm 1}\sum_{\left\langle i\vert j\right\rangle}\hat{c}_{i,s}^{\dagger}\hat{c}_{j,s}.
% - \mu\sum_{s = \pm 1}\sum_{i}\hat{c}_{i,s}^{\dagger}\hat{c}_{j,s}
\end{eqnarray}
For clarity, we will add the contribution of the chemical potential to the Hamiltonian later. $t = V_{pp\pi}\approx 2.8\eV$ is the nearest-neighbor hopping energy \cite{RevModPhys.81.109}\cite{Ezawa2015}, $\hat{c}_{i,s}^{\dagger}$ and $\hat{c}_{j,s}$ are the creation and annihilation operators of electronic excitations with spin $s\in\{\uparrow,\downarrow\}$ at site $i$, and in $\left\langle i\vert j\right\rangle$, $i$ run to all the atoms of the lattice while $j$ run all over the nearest neighbors hopping sites of $i$.

In the momentum space, the Hamiltonian becomes
\begin{eqnarray}
H = \sum_{s = \pm 1}\dint_{\bm{p}}
\hat{\chi}_{s}^{\dagger}(\bm{p}_{\surf})
\hat{H}_{s}(\bm{p}_{\surf})
\hat{\chi}_{s}(\bm{p}_{\surf}),
\end{eqnarray}
where we have defined $\dint_{\bm{p}} = \displaystyle\int_{\rm BZ}\dfrac{\dd^{2}\bm{p}_{\surf}}{(2\pi)^{2}}$ as the momentum integral over the Brillouin Zone $\rm BZ$ (see \Tab{Notation}),
\begin{eqnarray}
\hat{H}_{s}(\bm{p}_{\surf}) = - t\left(\begin{array}{cc}
0 & f(\bm{p}_{\surf})\\
f^{*}(\bm{p}_{\surf}) & 0
\end{array}\right),
\end{eqnarray}
and the bi-spinor in the sublattice space as
\begin{eqnarray}
\hat{\chi}_{s}(\bm{p}_{\surf}) & = & \left(\begin{array}{c}
\hat{c}_{A,s}(\bm{p}_{\surf})\\
\hat{c}_{B,s}(\bm{p}_{\surf})
\end{array}\right),
\end{eqnarray}
where $\hat{c}_{A,s}(\bm{p}_{\surf})$ is the annihilation operator of electronic excitations in the sublattice $A$ with spin $s$ and momentum $\bm{p}_{\surf}$, and
\begin{eqnarray}
f(\bm{p}_{\surf})
& = & \sum_{j=1}^{3}e^{\ii\bm{p}_{\surf}\cdot\bm{\delta}_{j}}\nonumber\\
& = & e^{-\ii a p_{y}} + 2 e^{\ii\frac{a p_{y}}{2}}\cos\left(\frac{\sqrt{3}}{2}a p_{x}\right).
\end{eqnarray}
Diagonalizing $\hat{H}$ in momentum space gives the energy spectrum as
\begin{eqnarray}
E_{\lambda}(\bm{p}_{\surf}) = \lambda t\abs{f(\bm{p}_{\surf})},% - \mu,
\end{eqnarray}
with $\lambda = \pm 1$ representing the conduction ($\lambda=+1$) and valence ($\lambda=-1$) bands respectively, and,
\begin{eqnarray}
\abs{f(\bm{p}_{\surf})}^{2} & = & 1 + 4\cos^{2}\left(\frac{\sqrt{3}ap_{x}}{2}\right)\nonumber\\
& & +  4\cos\left(\frac{3ap_{x}}{2}\right)\cos\left(\frac{\sqrt{3}ap_{y}}{2}\right).
\end{eqnarray}
%This function is zero at the 6 vertices of the hexagon in the reciprocal space $K_{n}$, placed at
%\begin{eqnarray}
%\bm{K}_{n} = \frac{4\pi}{3\sqrt{3}a}\left(\begin{array}{c}
%\cos\left(n\frac{\pi}{3}\right)\\
%\sin\left(n\frac{\pi}{3}\right)
%\end{array}\right).
%\end{eqnarray}
Inside the Brillouin Zone defined by the parallelogram $\bm{b}_{1}\otimes\bm{b}_{2}$, this function is zero at the $K_{\pm}$ points defined as
\begin{eqnarray}
{\bm K}_{\eta} & = & \frac{2\pi}{\sqrt{3}a}\left(\begin{array}{c}
1 - \frac{\eta}{3}\\
0
\end{array}\right),
\end{eqnarray}
with $\eta=\pm 1$ the valley index. As the chemical potential $\mu$ is close to the crossing points between $\pi$ and $\pi^{*}$ bands at the $\bm{K}_{\eta}$ points, the dispersion of the bands can be approximated as \cite{Wallace1947}
\begin{eqnarray}\label{Hamiltonian_valley_points}
\hat{H}_{s}(\bm{K}_{\eta} + \bm{k}_{\surf}) & = &\frac{3at}{2}
\left( \eta k_{1}\tau_{1} + k_{2}\tau_{2} \right),
%- \mu\tau_{0},
\end{eqnarray}
where $\tau_{i}$ is the $i^{\underline{th}}$ Pauli matrix of the sublattice pseudo-spin for the A and B sites for electronic excitations of spin $s$. From this expression the Fermi velocity is \cite{Ezawa2015}\cite{Wallace1947}
\begin{eqnarray}
v_{F} = \frac{3at}{2\hbar} \approx \frac{c}{300}.
\end{eqnarray}
Then, the dispersion band for each valley is approximated by
\begin{eqnarray}
\hat{H}_{s}(\bm{K}_{\eta} + \bm{k}_{\surf}) = \hat{H}_{s}^{\eta}(\bm{k}_{\surf}) = \hbar v_{F}\left( \eta k_{1}\tau_{1} + k_{2}\tau_{2} \right).
%- \mu\tau_{0},
\end{eqnarray}
After applying this small $(\bm{k}_{\surf}\cdot\bm{p}_{\surf})$ expansion, the electronic Hamiltonian can be approximated by a family of four ($4=g_{s}g_{v}$, where $g_{s} = 2$ because of spin degeneration and $g_{v} = 2$ because of the two different valleys) 2D Dirac Hamiltonians placed in the continuum limit as \cite{Ezawa2015}
\begin{eqnarray}\label{bispinor_Graphene}
H = \sum_{s,\eta = \pm}\dint_{\bm{k}}
\chi_{s}^{\eta,\dagger}(\bm{k}_{\surf})
\hat{H}_{s}^{\eta}(\bm{k}_{\surf})
\chi_{s}^{\eta}(\bm{k}_{\surf}).
\end{eqnarray}
This Hamiltonian represents a set of four equal Dirac cones, labeled by their valley $\eta$ and spin $s$. Consequently, in addition to the discrete $CPT$ symmetry, the Hamiltonian possesses a global continuous $U(4)$ symmetry that operates in the valley, sublattice, and spin spaces \cite{Gusynin2007}.

Finally, we combine the bi-spinors of the same spin of the two valleys to form a Dirac four-spinor (note the exchange of sublattices of the $\eta=-1$ valley terms) \cite{Gusynin2007}
\begin{eqnarray}
\Psi_{s}(\bm{k}_{\surf}) & \hspace{-2pt} = \hspace{-2pt} \left(\hspace{-2pt}\begin{array}{c}
\phantom{\tau_{1}}\chi_{s}(\bm{K}_{+} + \bm{k}_{\surf})\\
\tau_{1}\chi_{s}(\bm{K}_{-} + \bm{k}_{\surf})
\end{array}\hspace{-2pt}\right)
 & \hspace{-2pt} = \hspace{-2pt} \left(\hspace{-2pt}\begin{array}{c}
c_{A,s}(\bm{K}_{+} + \bm{k}_{\surf})\\
c_{B,s}(\bm{K}_{+} + \bm{k}_{\surf})\\
c_{B,s}(\bm{K}_{-} + \bm{k}_{\surf})\\
c_{A,s}(\bm{K}_{-} + \bm{k}_{\surf})
\end{array}\hspace{-2pt}\right),
\end{eqnarray}
resulting into
\begin{eqnarray}\label{Dirac_Graphene}
H & = & \sum_{s = \pm}\dint_{\bm{k}}
\Psi_{s}^{\dagger}(\bm{k}_{\surf})
\hat{H}_{s}^{D}(\bm{k}_{\surf})
\Psi_{s}(\bm{k}_{\surf}),
\end{eqnarray}
\begin{eqnarray}
\hat{H}_{s}^{D}(\bm{k}_{\surf})\hspace{-5pt} & = & \hspace{-6pt}\hbar v_{F}\hspace{-2pt}\left(\hspace{-2pt}{\displaystyle\begin{array}{cc|cc}
0 & k_{1} - \ii k_{2} & 0 & 0\\
k_{1} + \ii k_{2} & 0 & 0 & 0\\
\hline
0 & 0 & 0 & - k_{1} + \ii k_{2}\\
0 & 0 & - k_{1} - \ii k_{2} & 0
\end{array}}\hspace{-2pt}\right)\nonumber\\
& = & \hspace{-4pt}\hbar v_{F}\hspace{-2pt}\left( \alpha^{1}k_{1} + \alpha^{2}k_{2} \right).
\end{eqnarray}
Here the $\alpha^{\mu}$ matrices are the Dirac matrices. For $i \in \{1, 2, 3\}$, we have
\begin{eqnarray}
\alpha^{i} = \tilde{\tau}_{3}\otimes\tau_{i} = \left(\begin{array}{c|c}
\tau_{i} & 0\\
\hline
0 & - \tau_{i}
\end{array}\right),
\end{eqnarray}
where $\tilde{\tau}_{i}$ is the $i^{\underline{th}}$ Pauli matrix of the valley pseudo-spin $\eta$ \cite{Gusynin2007}. It will be useful to define $\alpha^{0}$ as
\begin{eqnarray}
\alpha^{0} = \left(\begin{array}{c|c}
\tau_{0} & 0\\
\hline
0 & \tau_{0}
\end{array}\right) = \tilde{\tau}_{0}\otimes\tau_{0},
\end{eqnarray}
and we define the $\beta$ matrix as $\alpha^{4}$ in what follows
\begin{eqnarray}
\alpha^{4} = \beta = \tilde{\tau}_{1}\otimes\tau_{0} = \left(\begin{array}{c|c}
0 & \tau_{0}\\
\hline
\tau_{0} & 0
\end{array}\right).
\end{eqnarray}
From those definitions, we have that $\alpha^{0}$ is a $4\times 4$ identity matrix, and the anticommutation relations
\begin{eqnarray}
\{\alpha^{i}, \alpha^{j}\} = 2\delta^{ij}\alpha^{0}
\hspace{10pt}
\forall\, i, j\in\{1,2,3,4\}.
\end{eqnarray}
In conclusion, we obtain two equivalent descriptions of the Hamiltonian of graphene, one in \Eq{bispinor_Graphene}, as the sum of $g_{s}g_{v} = 4$ bi-spinors in the sub-lattice space, and another one in \Eq{Dirac_Graphene} the sum of $g_{s} = 2$ four-spinors in the sub-lattice-valley space.

\subsection{Action of graphene}
To make a connection with the covariant QFT description of graphene \cite{PRL_Mohideen}\cite{Bordag2015}\cite{Bimonte2017}\cite{Gusynin2007}, we write the full space-temporal second quantized action of graphene given in \Eq{Dirac_Graphene} as
\begin{eqnarray}%\label{}
S_{0} & = & \sum_{s = \pm}\dint_{\kk}
\Psi_{s}^{\dagger}(\bm{k}_{\surf})
\left[ \hbar\omega\alpha^{0} - \hat{H}_{s}^{D}(\bm{k}_{\surf}) \right]
\Psi_{s}(\bm{k}_{\surf})\nonumber\\
& = & \sum_{s = \pm}\dint_{\kk}
\Psi_{s}^{\dagger}(\bm{k}_{\surf})
\hat{\mathcal{H}}_{s}^{D}(\bm{k}_{\surf})
\Psi_{s}(\bm{k}_{\surf}),
\end{eqnarray}
\begin{eqnarray}
\hat{\mathcal{H}}_{s}^{D}(\bm{k}_{\surf}) = \tilde{k}_{0}\alpha^{0} - \hbar v_{F}\left( \alpha^{1}k_{1} + \alpha^{2}k_{2} \right),
\end{eqnarray}
with $\tilde{k}_{0} = \hbar\omega$ and $\dint_{\kk} = \dint_{-\infty}^{\infty}\dfrac{\dd \tilde{k}_{0}}{2\pi}\dint_{\bm{k}}$ (see \Tab{Notation}). This is the Dirac representation of the action of the $(2+1)D$ Dirac field. To write this action in a full covariant way by using the Weyl representation, we define the $\gamma$ matrices as $\gamma^{\mu} = \alpha^{4}\alpha^{\mu}$. With this prescription, we have $\gamma^{0} = \alpha^{4}\alpha^{0} = \alpha^{4}$,  $\gamma^{4} = \alpha^{4}\alpha^{4} = \alpha^{0}$ and, for $i \in \{1, 2, 3\}$
\begin{eqnarray}
\gamma^{i} = \alpha^{4}\alpha^{i} = (-\ii\tilde{\tau}_{2})\otimes\tau_{i} = \left(\begin{array}{c|c}
0 & - \tau_{i}\\
\hline
\tau_{i} & 0
\end{array}\right).
\end{eqnarray}
The usual anti-commutation relations are fulfilled
\begin{eqnarray}
\{\gamma^{\mu}, \gamma^{\nu}\} = 2g^{\mu\nu}\gamma^{4}
\hspace{10pt}
\forall \{\mu,\nu\}\in\{0,1,2,3\},
\end{eqnarray}
where $g^{\mu\nu} = \text{diag}(+1,-1,-1,-1)$ is the metric tensor. The Dirac conjugated spinor is defined as $\bar{\Psi}_{s}(\bm{k}_{\surf}) = \Psi_{s}^{\dagger}(\bm{k}_{\surf})\alpha^{4} = \Psi_{s}^{\dagger}(\bm{k}_{\surf})\gamma^{0}$. Then, the action is now represented as
\begin{eqnarray}%\label{}
S_{0} & = & \sum_{s = \pm}\dint_{\kk}
\bar{\Psi}_{s}(\bm{k}_{\surf})
\hat{\mathcal{H}}_{s}^{W}(\bm{k}_{\surf})
\Psi_{s}(\bm{k}_{\surf}),
\end{eqnarray}
with
\begin{eqnarray}
\hat{\mathcal{H}}_{s}^{W}(\bm{k}_{\surf}) = \tilde{k}_{0}\gamma^{0} - \hbar v_{F}\left( \gamma^{1}k_{1} + \gamma^{2}k_{2} \right).
\end{eqnarray}
When this Hamiltonian is perturbed, depending on the breaking of the $CPT$ discrete symmetries and on the generators of the $U(4)$ symmetry used, different kinds of gaps in the Dirac bands can be induced \cite{Gusynin2007}\cite{Ezawa2013PRB}. Here, we will focus on two kinds of non-topological mass gaps, which we will denote as $m_{z}$ and $m$ in what follows. The full Hamiltonian becomes
\begin{eqnarray}\label{Dirac_H}
\hat{\mathcal{H}}_{s}^{D}(\bm{k}_{\surf}) = \tilde{k}_{0}\alpha^{0} - \left( \alpha^{1}\tilde{k}_{1} + \alpha^{2}\tilde{k}_{2} \right) - \alpha^{3}m_{z} - \alpha^{4}m,
\end{eqnarray}
\begin{eqnarray}\label{Weyl_H}
\hat{\mathcal{H}}_{s}^{W}(\bm{k}_{\surf}) = \tilde{k}_{0}\gamma^{0} - \left( \gamma^{1}\tilde{k}_{1} + \gamma^{2}\tilde{k}_{2} \right) - \gamma^{3}m_{z} - \gamma^{4}m,
\end{eqnarray}
where $m_{z}$ is a hopping term between fermions of the same valley \cite{Ezawa2015}\cite{Ezawa2013PRB}, while $m$ couples quasi-particles of different valleys. Being it a non-local interaction, it is a possible result of the symmetry-breaking interaction over graphene \cite{Gusynin2007}.

\subsection{Grand-canonical ensemble}
As we are working in the grand canonical ensemble, we add a term to the action of our field
\begin{eqnarray}
S_{\mu} = \sum_{s = \pm}\dint_{\kk}\mu\hat{\mathcal{N}},
\end{eqnarray}
where $\mu$ is the chemical potential and $\hat{\mathcal{N}}$ is the number of particles operator. This term can be written in each representation as
\begin{eqnarray}
S_{\mu} & = & \mu\sum_{s = \pm}\dint_{\kk}
\Psi_{s}^{\dagger}(\bm{k}_{\surf})
\alpha^{0}
\Psi_{s}(\bm{k}_{\surf})\nonumber\\
& = & \mu\sum_{s = \pm}\dint_{\kk}
\bar{\Psi}_{s}(\bm{k}_{\surf})
\gamma^{0}
\Psi_{s}(\bm{k}_{\surf}).
\end{eqnarray}
To compare the different models of the Polarization operator, we are going to use three different grand-canonical Hamiltonians for graphene, the bi-spinor expression from \Eq{bispinor_Graphene}, where
\begin{eqnarray}%\label{}
S_{1} \hspace{-4pt}& = & \hspace{-6pt}\sum_{s,\eta = \pm}\dint_{\kk}
\chi_{s}^{\eta,\dagger}(\bm{k}_{\surf})
\left[ \tau_{0}(\hbar\omega + \mu) - \hat{H}_{s}^{\eta}(\bm{k}_{\surf})\right]
\chi_{s}^{\eta}(\bm{k}_{\surf}),\nonumber\\
\end{eqnarray}
\begin{eqnarray}\label{Eq_Hamiltonian_spinor}
\hat{H}_{s}^{\eta}(\bm{k}_{\surf}) = \hbar v_{F}\left[\eta k_{1}\tau_{1} + k_{2}\tau_{2} \right] + \tau_{3}\Delta_{s}^{\eta},
\end{eqnarray}
with $m_{z} = \Delta_{s}^{\eta}$, the Dirac form of the Dirac Hamiltonian (using \Eq{Dirac_H}) as a bridge between the 2 formalisms
\begin{eqnarray}\label{Dirac_Action}
S_{D} & = & \sum_{s = \pm}\dint_{\kk}
\Psi_{s}^{\dagger}(\bm{k}_{\surf})
\left[ \hat{\mathcal{H}}_{s}^{D}(\bm{k}_{\surf}) + \alpha^{0}\mu\right]
\Psi_{s}(\bm{k}_{\surf}),
\end{eqnarray}
and the covariant expression of the Dirac Hamiltonian (using \Eq{Weyl_H}) as
\begin{eqnarray}\label{Weyl_covariant_Dirac_Action}
S_{W} & = & \sum_{s = \pm}\dint_{\kk}
\bar{\Psi}_{s}^{\dagger}(\bm{k}_{\surf})
\left[ \hat{\mathcal{H}}_{s}^{W}(\bm{k}_{\surf}) + \gamma^{0}\mu\right]
\Psi_{s}(\bm{k}_{\surf}).
\end{eqnarray}
We have introduced those three equivalent representations here since these will be used in the derivation of the different results we will compare in the following.

\subsection{Effect of interactions}\label{Effect_interactions}
Electronic quasiparticles are subject to different possible interactions: phonons, scattering centers, the unavoidable Coulomb interaction, external fields, illumination, decoration (impurities) and so on \cite{RevModPhys.81.109}\cite{DasSarmaRMP2011}\cite{Gusynin2006}\cite{Marinko2009}\cite{RMPdissipation2022}\cite{BOOKAbrikosov}. When the effects of interactions is taken into account into the dynamics of the electronic quasiparticles, the Hamiltonian is modified by the causal self-energy $\Sigma(\omega,\bm{k}) = \Sigma_{R}(\omega,\bm{k}) + \ii\Sigma_{I}(\omega,\bm{k})$ \cite{Gusynin2006}\cite{BOOKAbrikosov}\cite{Ostrovsky2006}\cite{yanagisawa2004}. The electronic spectrum is modified by the presence of a real self-energy $\Sigma_{R}(\omega,\bm{k})$ (whose effect is considered here small and absorbed into the phenomenological constants of the Hamiltonian) and a non-positive imaginary part $\Sigma_{I}(\omega,\bm{k})$, which leads to a frequency-dependent finite dissipation of the electronic quasiparticles \cite{RevModPhys.81.109}\cite{DasSarmaRMP2011}\cite{RMPdissipation2022}.
Here we argue that the electronic dissipation is a small non-zero quantity. We assume that its effect on the electric conductivity can be safely treated using the finite lifetime approximation $\Sigma_{I}(\omega,\bm{k})\approx - \Gamma$, consisting in using a constant dissipation rate $\Gamma = \tau^{-1} \geq 0$ \cite{DasSarmaRMP2011}\cite{BOOKAbrikosov}\cite{yanagisawa2004}\cite{Adam2007}\cite{Szunyogh1999}.
Taking into account that the measured electrical conductivity of graphene is a high but finite quantity ($\sigma = 0.96\times10^{6}\Omega^{-1}\text{ cm}^{-1}$ in \cite{Cao2019}\cite{NobelPrize:2010-Physics}), and that the dissipation time has been estimated to be on the order of $\tau\backsim 6\times 10^{-13}\text{ s}$ \cite{DasSarmaRMP2011}\cite{Marinko2009}, we will take this quantity in our study.
Of course, in situations where the effect of interactions in graphene is of paramount relevance (beyond its non-zero nature), like in the study of the electron-phonon interaction \cite{Gusynin2006}, or the universal DC electric conductivity of graphene when $\mu=0$ \cite{Ostrovsky2006}\cite{Kashuba2008}\cite{Fritz2008}\cite{Link2016}\cite{TeberPhDThesis2018}, our interactions-naive phenomenological approach would not be enough and a more detailed study of the effect of interactions will be necessary.

\subsection{Green function}
For a general linear Hamiltonian $\hat{\mathcal{H}}(\bm{r})$, we have two different expressions of the same Green function. Defining $\tilde{k}_{0} = \hbar\omega$, the Green function fulfills
\begin{eqnarray}
\hat{\mathcal{H}}(\bm{r})\mathcal{G}_{0}(\bm{r},\bm{s}) = \delta(\bm{r} - \bm{s}),
\end{eqnarray}
then, in momentum space we have $\hat{\mathcal{H}}(\bm{k})\mathcal{G}_{0}(\bm{k}) = \mathbbm{1}$ and, therefore $\mathcal{G}_{0}(\bm{k}) = \hat{\mathcal{H}}^{-1}(\bm{k})$. For each of the four-spinor Hamiltonians, its inverse operator is
\begin{eqnarray}
\mathcal{G}_{0}^{D}(\bm{k}) = \hat{\mathcal{H}}_{D}^{-1}(\bm{k})
 = \frac{1}{\alpha^{\rho}K_{\rho} - \alpha^{4}m}
 = \frac{\alpha^{\rho}K_{\rho} + \alpha^{4}m}{K^{\rho}K_{\rho} - m^{2}},
\end{eqnarray}
\begin{eqnarray}\label{Weyl_covariant_Dirac_Green_function}
\mathcal{G}_{0}^{W}(\bm{k}) = \hat{\mathcal{H}}_{W}^{-1}(\bm{k})
 = \frac{1}{\gamma^{\rho}K_{\rho} - \gamma^{4}m}
 = \frac{\gamma^{\rho}K_{\rho} + \gamma^{4}m}{K^{\rho}K_{\rho} - m^{2}},
\end{eqnarray}
where we define $K_{\rho} = (\hbar\omega + \mu,\hbar v_{F}k_{1},\hbar v_{F}k_{2},m_{z})$ (see \Tab{Notation}), the Einstein summation convention is assumed and
\begin{eqnarray}
K^{\rho}K_{\rho} - m^{2}
= \prod_{\lambda=\pm}\left[\hbar\omega + \mu - \epsilon^{\lambda}_{\bm{k}}\right]
= \prod_{\lambda=\pm}\left[\hbar\omega - \xi^{\lambda}_{\bm{k}}\right],
\end{eqnarray}
where $\xi^{\lambda}_{\bm{k}} = \epsilon^{\lambda}_{\bm{k}} - \mu$,
\begin{eqnarray}\label{Eq_Energy_BiSpinor}
\epsilon^{\lambda}_{\bm{k}} = \lambda\sqrt{ \tilde{k}_{\surf}^{2} + m_{z}^{2} + m^{2} },
\end{eqnarray}
with $\tilde{k}_{\surf} = \hbar v_{F}k_{\surf}$ and $k_{\surf} = \sqrt{ k_{1}^{2} + k_{2}^{2} }$ (see \Tab{Notation}). There is another equivalent expression for a general linear Hamiltonian of the form $\hbar\omega\psi = \tilde{k}_{0}\psi = \hat{H}\psi$ in terms of the eigenvalues and eigenfunctions of the Hamiltonian, starting again from the equation for the Green function $\left( \tilde{k}_{0} - \hat{H}\right)G_{0}(\bm{r},\bm{s}) = \delta(\bm{r} - \bm{s})$, and from the eigenproblem
\begin{eqnarray}
\hat{H}\ket{u_{\bm{k}}^{\lambda}} = \left(\epsilon_{\bm{k}}^{\lambda} - \mu\right)\ket{u_{\bm{k}}^{\lambda}} = \xi_{\bm{k}}^{\lambda}\ket{u_{\bm{k}}^{\lambda}},
\end{eqnarray}
where the chemical potential $\mu$ has been included. The grand-canonical Green function in momentum space is
\begin{eqnarray}\label{Electric_Green0_function}
G_{0}(k) = \sum_{\lambda}\dfrac{\ket{u_{\bm{k}}^{\lambda}}\bra{u_{\bm{k}}^{\lambda}}}{ \tilde{k}_{0} - \xi_{\bm{k}}^{\lambda} },
\end{eqnarray}
which is an eigenvalue expansion of the Green function. In our case, the Hamiltonian is diagonal in spin. Therefore, the Green functions are multiplied by $\delta_{ss'}$.

\subsection{Presence of an electromagnetic field}
We introduce the coupling of the electronic quasiparticles of the lattice to the electromagnetic field via the Peierls substitution \cite{Gusynin2007}\cite{Millis2004}
\begin{eqnarray}
H = - \sum_{\bm{n}\in\mathbb{Z}^{d}}\sum_{\ell,\ell'}\sum_{j=1}^{\abs{\bm{\delta}_{j}}<\delta_{M}}
\hat{c}^{\dagger}_{\ell,\bm{R}_{\bm{n}}}t_{\ell,\ell'}(\bm{\delta}_{j})e^{-\ii Q\bm{A}_{\bm{R}_{\bm{n}}}\cdot\bm{\delta}_{j}}\hat{c}_{\ell',\bm{R}_{\bm{n}} + \bm{\delta}_{j}},
\end{eqnarray}
where we have approximated
\begin{eqnarray}
\exp\left(-\ii Q\int_{\bm{R}_{\bm{n}}}^{\bm{R}_{\bm{n}} + \bm{\delta}_{j}}\bm{A}(\bm{r})\cdot\dd\bm{r}\right) \approx e^{-\ii Q\bm{A}_{\bm{R}_{\bm{n}}}\cdot\bm{\delta}_{j}},
\end{eqnarray}
being $Q$ the electric charge of the quasi-particle described by the Hamiltonian, and where we have $Q = - e$ for electronic excitations. At the linear order in $\bm{A}_{\bm{q}}$, the Hamiltonian in reciprocal space is
\begin{eqnarray}\label{H_with_A}
H
& = & - \int_{\bm{q}}\int_{\bm{p}}
\sum_{\ell,\ell'}
\hat{c}_{\bm{p} + \bm{q}}^{\ell,\dagger}
\left[\sum_{j=1}^{\abs{\delta_{j}}<\delta_{M}}t_{\ell,\ell'}(\bm{\delta}_{j})
e^{\ii\left( \bm{p} - Q\bm{A}_{\bm{q}}\right)\cdot\bm{\delta}_{j}}\right]
\hat{c}_{\bm{p}}^{\ell'}
\nonumber\\
& = & \int_{\bm{q}}\int_{\bm{p}}
\sum_{\ell,\ell'}
\hat{c}_{\bm{p} + \bm{q}}^{\ell,\dagger}
\hat{H}^{\ell,\ell'}\left( \bm{p} - Q\bm{A}_{\bm{q}}\right)
\hat{c}_{\bm{p}}^{\ell'}.
\end{eqnarray}
It is clear that, at first order in $\bm{A}_{\bm{q}}$, the inclusion of the Peierls substitution leads to a minimal coupling of the momentum \cite{Gusynin2007}. To study electric conductivity, we need an expression for the current, understood as the potential vector's conjugated force. Then, expanding the Hamiltonian at linear order in $\bm{A}$, we obtain 
\begin{eqnarray}\label{H_perturbed_in_A}
H
& = & H_{0} - \int_{\bm{q}}J_{\mu,-\bm{q}}A^{\mu}_{\bm{q}} = H_{0} - \int_{\bm{q}}J^{*}_{\mu,\bm{q}}A^{\mu}_{\bm{q}}\nonumber\\
& \approx & H_{0} + \int_{\bm{q}}\dfrac{\delta\hat{H}}{\delta A^{\mu}_{\bm{q}}}A^{\mu}_{\bm{q}}.
\end{eqnarray}
Therefore, the second quantized current is defined as
\begin{eqnarray}
J_{\mu,\bm{q}}^{*} & = & - \dfrac{\delta\hat{H}}{\delta A^{\mu}_{\bm{q}}} = \int_{\bm{p}}
\sum_{\ell,\ell'}
\hat{c}_{\bm{p} + \bm{q}}^{\ell,\dagger}\hat{J}_{\mu}^{\ell,\ell'}\left(\bm{p}\right)\hat{c}_{\bm{p}}^{\ell'},
\end{eqnarray}
with the current operator given as
\begin{eqnarray}\label{Current_Operator_definition}
\hat{J}_{\mu}^{\ell,\ell'}\left(\bm{p}\right)
& = & - \left.\dfrac{\partial\hat{H}^{\ell,\ell'}\left( \bm{p} - Q\bm{A}_{\bm{q}}\right)}{\partial A_{\bm{q}}^{\mu}}\right\vert_{\bm{A}_{\bm{q}}\to\bm{0}} = Q\dfrac{\partial\hat{H}^{\ell,\ell'}\left(\bm{p}\right)}{\partial p^{\mu}}\nonumber\\
& = & \frac{Q}{\hbar}\dfrac{\partial\hat{H}^{\ell,\ell'}\left(\bm{k}\right)}{\partial k^{\mu}} = Q\hat{v}_{\mu},
\end{eqnarray}
where $\bm{p} = \hbar\bm{k}$, and $\hat{v}_{\mu}$ is the velocity operator of the electronic quasiparticles. It is worth stressing that while in \Eq{H_perturbed_in_A} we have a perturbation term associated with a generic electromagnetic field, for the definition of the electric conductivity, we have to consider a perturbation generated exclusively by an electric field, as we have done in \Sect{KubosigmaderivationAppendix}.

\section{Electric response of graphene: the regularized QFT model is identical to the Kubo model}\label{Section_EM_response}
In this section, we derive Ohm's Law starting from the electric field perturbation (the effect of the magnetic field is neglected in this derivation since we are interested only in the pure electric response part), we hence derive the definition of the electric conductivity including the natural regularization present in the Kubo formalism. We compare this Kubo electric conductivity with the one given usually in QFT. We show that the induced electric current in the case of zero electric field goes to zero when using the Kubo formula for the electric conductivity, while the same is not necessarily true when the QFT non-regularized expression is employed.

\subsection{Derivation of Ohm's Law and the Kubo formula for the electric conductivity} \label{KubosigmaderivationAppendix}
In this section, we are going to review the Kubo formula for the electric conductivity $J_{\mu} = \sigma_{\mu\nu}^{\rm{K}}E^{\nu}$ (compare with \Eq{OhmLaw}). We start from the interaction Hamiltonian
\begin{eqnarray}\label{electric_Ohm_eq0}
H = H_{0} + H_{I}(t),
\end{eqnarray}
with the perturbation due to only an electric field (Eqs. (2.1) and (5.7) of \cite{Kubo1957})
\begin{eqnarray}\label{E_perturbation}
H_{I}(t) = - \int_{\Omega}\dd\bm{x}\hat{d}_{\nu}(\bm{x})E^{\nu}(\bm{x},t),
\end{eqnarray}
where we are carrying out the integral over the volume $\Omega$, we have defined the dipolar moment operator as $\hat{d}_{\nu}(\bm{x}) = Q\hat{x}_{\nu}$ and the electric current as $\hat{J}_{\nu}(\bm{x}) = Q\hat{v}_{\nu} = Q\dot{\hat{x}}_{\nu}$, therefore, we have
\begin{eqnarray}\label{Relation_J_d}
\hat{J}_{\nu}(\bm{x}) = Q\dot{\hat{x}}_{\nu} = \dot{\hat{d}}_{\nu}(\bm{x}).
\end{eqnarray}
It is worth stressing that there is always a magnetic field associated with an oscillating electric field, and this magnetic field can induce current as well. Here, we are not considering these currents because they are not part of electric conductivity, defined as the transport coefficient that relates the induced electric current with the electric field (see \Eq{Def_J_SE}), the transport coefficient that relates the induced electric current with the magnetic field will be studied in a future work \cite{PabloJianShengMauroGauge2025}.
\begin{widetext}
Then, a direct application of the Kubo formula for linear transport theory (Eqs. (2.10), (2.17) and (2.19) of \cite{Kubo1957}),
\begin{eqnarray}\label{electric_Ohm_from_Kubo}
\mean{\hat{J}_{\mu}(\bm{x},t)}
& = & \mean{\hat{J}_{\mu}(\bm{x},t_{0})} - \dfrac{\ii}{\hbar}\int_{t_{0}}^{t}\dd\tau \Tr{\hat{\rho}_{\beta}\big[\hat{J}_{\mu}(\bm{x},t), \,H_{I}(\tau)\big]},
\end{eqnarray}
where $[\hat{J}_{\mu}(\bm{x}, t), H_{I}(\tau)]$ is the commutator between $\hat{J}_{\mu}(\bm{x}, t)$ and $H_{I}(\tau)$. To avoid any transitory to the new (non-equilibrium) steady state \cite{yanagisawa2004}, we make the switching starts at $t_{0}\to-\infty$ and impose $\dlim_{t_{0}\to-\infty}\mean{\hat{J}_{\mu}(\bm{x},t_{0})} = 0$. Using $H_{I}(t)$ as defined in \Eq{H_with_A}, we obtain the Kubo expression of \Eq{Def_J_PIA}
\begin{eqnarray}\label{Kubo_Pi_A}
\mean{\hat{J}_{\mu}(\bm{x},t)}
& = & \int_{-\infty}^{t}\dd\tau\int_{\Omega}\dd\bm{y} \dfrac{\ii}{\hbar}\Tr{\hat{\rho}_{\beta}\big[\hat{J}_{\mu}(\bm{x},t), \,\hat{J}_{\nu}(\bm{y},\tau)\big]}A^{\nu}(\bm{y},\tau) = - \int_{-\infty}^{t}\dd\tau\int_{\Omega}\dd\bm{y} \Pi_{\mu\nu}(\bm{x},t;\bm{y},\tau)A^{\nu}(\bm{y},\tau),
\end{eqnarray}
Using $H_{I}(t)$ as defined in \Eq{E_perturbation}, we obtain the microscopic Ohm's Law in position's space (Eq. (5.10) of \cite{Kubo1957})
\begin{eqnarray}\label{electric_Ohm_eq6}
\mean{\hat{J}_{\mu}(\bm{x},t)}
& = & \int_{-\infty}^{t}\dd\tau\int_{\Omega}\dd\bm{y} \dfrac{\ii}{\hbar}\Tr{\hat{\rho}_{\beta}\big[\hat{J}_{\mu}(\bm{x},t), \,\hat{d}_{\nu}(\bm{y},\tau)\big]}E^{\nu}(\bm{y},\tau) =  \int_{-\infty}^{t}\dd\tau\int_{\Omega}\dd\bm{y} \sigma_{\mu\nu}(\bm{x},t;\bm{y},\tau)E^{\nu}(\bm{y},\tau).
\end{eqnarray}
Now, using \Eq{Relation_J_d}, we obtain
\begin{eqnarray}\label{electric_Ohm_eq12}
\dfrac{\dd\phantom{\tau}}{\dd \tau}\dfrac{\ii}{\hbar}\Tr{\hat{\rho}_{\beta}\big[\hat{J}_{\mu}(\bm{x}, t), \,\hat{d}_{\nu}(\bm{y},\tau)\big]} = \dfrac{\ii}{\hbar}\Tr{\hat{\rho}_{\beta}\big[\hat{J}_{\mu}(\bm{x}, t), \,\hat{J}_{\nu}(\bm{y},\tau)\big]}.
\end{eqnarray}
\end{widetext}
Using that the polarization operator in the position's space is defined (equivalently to the definition of \Eq{PI_Definition}) as
\begin{eqnarray}\label{electric_Ohm_eq13}
\Pi_{\mu\nu}(\bm{x},t;\bm{y},\tau) = \dfrac{-\ii}{\hbar}\Tr{\hat{\rho}_{\beta}\big[\hat{J}_{\mu}(\bm{x},t), \,\hat{J}_{\nu}(\bm{y},\tau)\big]},
\end{eqnarray}
we obtain that the electric conductivity is related to the polarization operator by
\begin{eqnarray}\label{electric_Ohm_eq14}
\dfrac{\dd\phantom{\tau}}{\dd\tau}\sigma_{\mu\nu}(\bm{x},t;\bm{y},\tau) = - \Pi_{\mu\nu}(\bm{x},t;\bm{y},\tau).
\end{eqnarray}
Imposing $\dlim_{\tau\to-\infty}\sigma_{\mu\nu}(\bm{x},t;\bm{y},\tau) = 0$, we obtain
\begin{eqnarray}
\sigma_{\mu\nu}(\bm{x},t;\bm{y},T)  = - \dint_{-\infty}^{T}\dd \tau\Pi_{\mu\nu}(\bm{x},t;\bm{y},\tau).
\end{eqnarray}
where the electric conductivity tensor is (Eq. (5.14) of \cite{Kubo1957})
\begin{eqnarray}\label{Kubo_sigma}
\sigma_{\mu\nu}(\bm{x},t;\bm{y},T) = \dfrac{\ii}{\hbar}\dint_{-\infty}^{T}\dd \tau\Tr{\hat{\rho}_{\beta}\big[\hat{J}_{\mu}(\bm{x},t), \,\hat{J}_{\nu}(\bm{y},\tau)\big]}
\end{eqnarray}
Note that, for the electric conductivity, we only are considering electric currents generated by a non-zero electric field; it is to say, if $\bm{E}=\bm{0}$, we have $\mean{ \bm{J} } = \bm{0}$. In addition to that, we are not considering the effect of the diamagnetic term for the electric conductivity of graphene because the Dirac Hamiltonian is linear in momentum (see \Eq{Hamiltonian_valley_points}). Substituting $E^{\nu}(\tau, \bm{q}) = - \partial_{\tau}A^{\nu}(\tau, \bm{q})$ (we are using the Temporal Gauge in these calculations, as usual) in \Eq{electric_Ohm_eq6}, we get
\begin{widetext}
\begin{eqnarray}\label{j_eq2}
\mean{ J_{\mu}(\bm{x},t) } = - \dint_{-\infty}^{t}\dd\tau\int_{\Omega}\dd\bm{y}\sigma_{\mu\nu}(\bm{x},t;\bm{y},\tau)\partial_{\tau}A^{\nu}(\bm{y},\tau).
\end{eqnarray}
Here, we apply an integration by parts in time
\begin{eqnarray}\label{j_eq3}
\mean{ J_{\mu}(\bm{x},t) }
& = & - \int_{\Omega}\dd\bm{y}\Big[ \sigma_{\mu\nu}(\bm{x},t;\bm{y},\tau)A^{\nu}(\bm{y},\tau))\Big]_{-\infty}^{t} + (-1)^{2} \dint_{-\infty}^{t}\dd\tau\int_{\Omega}\dd\bm{y} A^{\nu}(\bm{y},\tau)\partial_{\tau}\sigma_{\mu\nu}(\bm{x},t;\bm{y},\tau).
\end{eqnarray}
Using that the integral of the derivative is the initial function, we obtain
\begin{eqnarray}\label{j_eq4}
\mean{ J_{\mu}(\bm{x},t) }
& = & - \Big[ \int_{\Omega}\dd\bm{y}A^{\nu}(\bm{y},\tau)\int_{-\infty}^{\tau}\dd\tau_{1}\partial_{\tau_{1}}\sigma_{\mu\nu}(\bm{x},t;\bm{y},\tau_{1})\Big]_{-\infty}^{t} + \int_{\Omega}\dd\bm{y}\dint_{-\infty}^{t}\dd\tau A^{\nu}(\bm{y},\tau)\partial_{\tau}\sigma_{\mu\nu}(\bm{x},t;\bm{y},\tau).
\end{eqnarray}
We can cancel out the boundary $t\to-\infty$ integral term
\begin{eqnarray}\label{j_eq5}
\mean{ J_{\mu}(\bm{x},t) }
& = & - \left[ \int_{\Omega}\dd\bm{y}A^{\nu}(\bm{y},t)\int_{-\infty}^{t}\dd\tau\partial_{\tau}\sigma_{\mu\nu}(\bm{x},t;\bm{y},\tau) - 0\right] + \int_{\Omega}\dd\bm{y}\dint_{-\infty}^{t}\dd\tau A^{\nu}(\bm{y},\tau)\partial_{\tau}\sigma_{\mu\nu}(\bm{x},t;\bm{y},\tau).
\end{eqnarray}
and this result, using \Eq{electric_Ohm_eq14}, can be simplified to
\begin{eqnarray}\label{j_eq6}
\mean{ J_{\mu}(\bm{x},t) }
& = & \int_{\Omega}\dd\bm{y}A^{\nu}(\bm{y},t)\int_{-\infty}^{t}\dd\tau\Pi_{\mu\nu}(\bm{x},t;\bm{y},\tau) - \int_{\Omega}\dd\bm{y}\dint_{-\infty}^{t}\dd\tau\Pi_{\mu\nu}(\bm{x},t;\bm{y},\tau)A^{\nu}(\bm{y},\tau).
\end{eqnarray}
\begin{eqnarray}\label{j_eq7}
\mean{ J_{\mu}(\bm{x},t) }
& = & - \dint_{-\infty}^{t}\dd\tau\int_{\Omega}\dd\bm{y}\Pi_{\mu\nu}(\bm{x},t;\bm{y},\tau)\Big[ A^{\nu}(\bm{y},\tau) - A^{\nu}(\bm{y},t) \Big].
\end{eqnarray}
Note that this result is already different from the linear relation between the electric conductivity and the potential vector proposed in \Eq{Kubo_Pi_A} and \Eq{JdependsonA}, which can be written as a function of time as
\begin{eqnarray}%\label{JdependsonAc}
\mean{ J_{\mu}(\bm{x},t) }
= - \dint_{-\infty}^{t}\dd\tau\int_{\Omega}\dd\bm{y}\Pi_{\mu\nu}(\bm{x},t;\bm{y},\tau)A^{\nu}(\bm{y},\tau).
\end{eqnarray}
We can write the integral part proportional to $A^{\nu}(t, \bm{q})$ of \Eq{j_eq7} as a kernel proportional to the retarded $A^{\nu}(\tau,\bm{q})$ by using
\begin{eqnarray}\label{transformation_Pi_kernel}
\dint_{-\infty}^{t}\dd\tau\Pi_{\mu\nu}(\bm{x},t;\bm{y},\tau)A^{\nu}(\bm{y},t)
& = & \dint_{-\infty}^{t}\dd\tau_{1}\Pi_{\mu\nu}(\bm{x},t;\bm{y},\tau_{1})A^{\nu}(\bm{y},t)\nonumber\\
& = & \dint_{-\infty}^{t}\dd\tau_{1}\Pi_{\mu\nu}(\bm{x},t;\bm{y},\tau_{1})\dint_{-\infty}^{t}\dd\tau\delta(\tau - t)A^{\nu}(\bm{y},t)\nonumber\\
& = & \dint_{-\infty}^{t}\dd\tau\dint_{-\infty}^{t}\dd\tau_{1}\Pi_{\mu\nu}(\bm{x},t;\bm{y},\tau_{1})\delta(\tau - t)A^{\nu}(\bm{y},t)\nonumber\\
& = & \dint_{-\infty}^{t}\dd\tau\delta(\tau - t)\left( \dint_{-\infty}^{\tau}\dd\tau_{1}\Pi_{\mu\nu}(\bm{x},\tau;\bm{y},\tau_{1}) \right)A^{\nu}(\bm{y},\tau).
\end{eqnarray}
where we have used the prescription $\dint_{-\infty}^{t} d\tau \delta(\tau-t) = 1$ for the Dirac delta function. Introducing \Eq{transformation_Pi_kernel} in \Eq{j_eq7}, we have
\begin{eqnarray}\label{j_eq8b}
\mean{ J_{\mu}(\bm{x},t) }
& = & - \int_{\Omega}\dd\bm{y}\dint_{-\infty}^{t}\dd\tau\left[ \Pi_{\mu\nu}(\bm{x},t;\bm{y},\tau) -\delta(\tau - t)\left( \dint_{-\infty}^{\tau}\dd\tau_{1}\Pi_{\mu\nu}(\bm{x},\tau;\bm{y},\tau_{1}) \right) \right]A^{\nu}(\bm{y},\tau).
\end{eqnarray}
Assuming homogeneity in space and using the homogeneity in time, we have
\begin{eqnarray}\label{j_eq8c}
\mean{ J_{\mu}(\bm{x},t) }
& = & - \int_{\Omega}\dd\bm{y}\dint_{-\infty}^{t}\dd\tau\left[ \Pi_{\mu\nu}(\bm{x} - \bm{y},t-\tau) - \delta(\tau - t)\left( \dint_{-\infty}^{\tau}\dd\tau_{1}\Pi_{\mu\nu}(\bm{x} - \bm{y},\tau - \tau_{1}) \right) \right]A^{\nu}(\bm{y},\tau).
\end{eqnarray}
Applying a Fourier transform in the position's coordinates, using the convolution theorem, we obtain
\begin{eqnarray}\label{j_eq8}
\mean{ J_{\mu}(t,\bm{q}) }
& = & - \dint_{-\infty}^{t}\dd\tau\left[ \Pi_{\mu\nu}(t-\tau,\bm{q}) - \delta(\tau - t)\left( \dint_{-\infty}^{\tau}\dd\tau_{1}\Pi_{\mu\nu}(\tau - \tau_{1},\bm{q}) \right) \right]A^{\nu}(\tau,\bm{q}).
\end{eqnarray}
In temporal frequency space, using the Fourier transform of the retarded operator ($A^{R}(T) = \Theta(T)A(T)$)
\begin{eqnarray}\label{def_Pi_R(w)}
\Pi_{\mu\nu}^{R}(\omega,\bm{q}) = \int_{-\infty}^{\infty}\dd\tau_{1}e^{\ii\omega\tau_{1}}\Theta(\tau - \tau_{1})\Pi_{\mu\nu}(\tau - \tau_{1},\bm{q}) = \int_{-\infty}^{\tau}\dd\tau_{1}e^{\ii\omega\tau_{1}}\Pi_{\mu\nu}(\tau - \tau_{1},\bm{q}),
\end{eqnarray}
and we obtain the relation
\begin{eqnarray}\label{j_eq9}
\mean{ J_{\mu}(\omega, \bm{q}) } = - \Bigg[ \Pi_{\mu\nu}^{R}(\omega,\bm{q}) - \left( \int_{-\infty}^{\tau}\dd\tau_{1}\Pi_{\mu\nu}(\tau - \tau_{1},\bm{q}) \right) \Bigg]A^{\nu}(\omega, \bm{q}),
\end{eqnarray}
\begin{eqnarray}\label{j_eq10}
\mean{ J_{\mu}(\omega, \bm{q}) } = - \Big[ \Pi_{\mu\nu}^{R}(\omega,\bm{q}) - \dlim_{\omega\to0}\Pi_{\mu\nu}^{R}(\omega,\bm{q}) \Big]A^{\nu}(\omega, \bm{q})
\neq - \Pi_{\mu\nu}^{R}(\qq)A^{\nu}(\omega,\bm{q}).
\end{eqnarray}
\end{widetext}
%\Pablo{Remember that our current is different from 0 only when E is different from zero; the total j that contains magnetic currents generated by magnetostatic terms is $sigma = Pi/-iw$. This j does not contain electric currents generated by static magnetic fields. As the complete J also contains the contribution of static magnetic fields, it results in $sigma = Pi/-iw$}

%This is a result in direct contradiction with the usual assumed relation provided in \Eq{JdependsonA}.
Finally, using the microscopic Ohm's Law in temporal frequency space and assuming spatial homogeneity, we have from \Eq{j_eq2}
\begin{eqnarray}\label{j_eq11}
\mean{ J_{\mu}(\omega, \bm{q}) } = \sigma_{\mu\nu}^{R}(\omega,\bm{q})(\ii\omega)A^{\nu}(\omega, \bm{q}).
\end{eqnarray}
Then we can compare \Eq{j_eq10} and \Eq{j_eq11} to obtain the so-called Luttinger formula \cite{Luttinger1968}\cite{yanagisawa2004}\cite{Szunyogh2002}
\begin{eqnarray}\label{True_Relation_Polarization_conductivity}
\sigma_{\mu\nu}^{R}(\omega,\bm{q}) = \dfrac{\Pi_{\mu\nu}^{R}(\omega,\bm{q}) - \dlim_{\omega\to0}\Pi_{\mu\nu}^{R}(\omega,\bm{q})}{ - \ii\omega}.
\end{eqnarray}
So we conclude that, assuming the microscopic Ohm's Law given in \Eq{j_eq1}, and the relation between the electric conductivity $\sigma_{\mu\nu}$ and the Polarization operator $\Pi_{\mu\nu}$ derived from the Kubo formula given in \Eq{Kubo_sigma}, the Fourier transform of the electric conductivity tensor is given by \Eq{True_Relation_Polarization_conductivity}. Therefore, we conclude that the electric conductivity should be obtained from \Eq{True_Relation_Polarization_conductivity}. Note that this subtraction, which we derived here simply from Ohm's Law and time causality, naturally implies that electric field-induced currents cannot exist when $E^{\nu}(\tau,\bm{y}) = 0$ or, equivalently, when $A_{\mu}(\omega=0,\bm{q}) \neq 0$. This is a strong physical requirement \cite{Falkovsky2007}.

Note that the same calculation can be performed with retarded operators from the very beginning, and we obtain the same result. Assuming spatial homogeneity and applying a Fourier transform in the position's coordinates, using the convolution theorem to the (causal) microscopic Ohm's Law in the position space \Eq{electric_Ohm_eq6}, we obtain Ohm's Law in momentum space for homogeneous systems as
\begin{eqnarray}\label{j_eq1}
\mean{J_{\mu}(t, \bm{q})} = \dint_{-\infty}^{t}\dd\tau\sigma_{\mu\nu}(t - \tau,\bm{q})E^{\nu}(\tau, \bm{q}),
\end{eqnarray}
where the conductivity tensor is (from \Eq{electric_Ohm_eq6} and \Eq{Kubo_sigma})
\begin{eqnarray}%\label{Kubo_sigma}
\sigma_{\mu\nu}(t,\bm{q}) = \dfrac{\ii}{\hbar}\Tr{\hat{\rho}_{\beta}\big[ \hat{J}_{\mu}(t - \tau,\bm{q}), \,\hat{d}_{\mu}^{*}(0,\bm{q}) \big]}\nonumber\\
= \dfrac{\ii}{\hbar}\dint_{-\infty}^{t}\dd\tau\Tr{\hat{\rho}_{\beta}\big[ \hat{J}_{\mu}(t - \tau,\bm{q}), \,\hat{J}_{\mu}^{*}(0,\bm{q}) \big]}\nonumber\\
 = - \dint_{-\infty}^{t}\dd T\Pi_{\mu\nu}(T,\bm{q}) = - \dint_{-\infty}^{\infty}\dd T\Pi^{R}_{\mu\nu}(T,\bm{q}),
\end{eqnarray}
with $T = t - \tau$ and the Polarization operator is
\begin{eqnarray}
\Pi_{\mu\nu}(T,\bm{q}) = \dfrac{- \ii}{\hbar}\Tr{\hat{\rho}_{\beta}\big[ J_{\mu}(T,\bm{q}),\,J_{\mu}^{*}(0,\bm{q}) \big]}.
\end{eqnarray}
$\Pi^{R}_{\mu\nu}(T,\bm{q}) = \Theta(T)\Pi_{\mu\nu}(T,\bm{q})$ is the retarded polarization operator and $\Theta(T)$ is the Heaviside Theta function. Assuming homogeneity, the (causal) microscopic Ohm's Law in the position space given in \Eq{electric_Ohm_eq6} can be written as
\begin{widetext}
\begin{eqnarray}%\label{j_eq1b}
\mean{ J_{\mu}(t, \bm{x}) }
& = & \dint_{-\infty}^{t}\dd\tau\dint_{\Omega}\dd^{d}\bm{y}\sigma_{\mu\nu}(t - \tau,\bm{x} - \bm{y})E^{\nu}(\tau, \bm{y}) = \dint_{-\infty}^{\infty}\dd\tau\dint_{\Omega}\dd^{d}\bm{y}\sigma_{\mu\nu}^{R}(t - \tau,\bm{x} - \bm{y})E^{\nu}(\tau, \bm{y}),
\end{eqnarray}
where the retarded electric conductivity operator is defined as $\sigma_{\mu\nu}^{R}(t - \tau,\bm{x} - \bm{y}) = \Theta(t - \tau)\sigma_{\mu\nu}(t - \tau,\bm{x} - \bm{y})$. Applying a Fourier transform in the position's coordinates, using the convolution theorem, and substituting $E^{\nu}(\tau, \bm{q}) = - \partial_{\tau}A^{\nu}(\tau, \bm{q})$ (we are using the Temporal Gauge in these calculations, as usual) we get (compare with \Eq{j_eq2})
\begin{eqnarray}%\label{j_eq2}
\mean{ J_{\mu}(t, \bm{q}) } = - \dint_{-\infty}^{\infty}\dd\tau\sigma_{\mu\nu}^{R}(t - \tau,\bm{q})\partial_{\tau}A^{\nu}(\tau, \bm{q}).
\end{eqnarray}
As before, we apply an integration by parts in time (compare with \Eq{j_eq3})
\begin{eqnarray}%\label{j_eq3}
\mean{ J_{\mu}(t, \bm{q}) }
& = & - \Big[ \sigma_{\mu\nu}^{R}(t - \tau,\bm{q})A^{\nu}(\tau, \bm{q})\Big]_{-\infty}^{\infty} + (-1)^{2} \dint_{-\infty}^{\infty}\dd\tau\big[\partial_{\tau}\sigma_{\mu\nu}^{R}(t - \tau,\bm{q})\big]A^{\nu}(\tau, \bm{q}).
\end{eqnarray}
Now, using that $\sigma_{\mu\nu}^{R}(t - \tau,\bm{q}) = \Theta(t - \tau)\sigma_{\mu\nu}(t - \tau,\bm{q})$ should be understood as a tempered distribution, that, because of its definition, $\dlim_{t\to\infty}\sigma_{\mu\nu}^{R}(t - \tau,\bm{q}) = 0$ and $\dlim_{t\to-\infty}A^{\nu}(\tau, \bm{q}) = 0$, we obtain (compare with \Eq{j_eq4}, \Eq{j_eq5} and \Eq{j_eq8})
\begin{eqnarray}%\label{j_eq4}
\mean{ J_{\mu}(t, \bm{q}) }
& = & - \left[ 0 - 0 \right] + \dint_{-\infty}^{\infty}\dd\tau\partial_{\tau}\big[\Theta(t - \tau)\sigma_{\mu\nu}(t - \tau,\bm{q})\big]A^{\nu}(\tau, \bm{q})\\
& = & \dint_{-\infty}^{\infty}\dd\tau\big[\sigma_{\mu\nu}(t - \tau,\bm{q})\partial_{\tau}\Theta(t - \tau) + \Theta(t - \tau)\partial_{\tau}\sigma_{\mu\nu}(t - \tau,\bm{q})\big]A^{\nu}(\tau, \bm{q})\\
& = & \dint_{-\infty}^{\infty}\dd\tau\big[ - \delta(t - \tau)\sigma_{\mu\nu}(t - \tau,\bm{q}) - \Theta(t - \tau)\Pi_{\mu\nu}(t - \tau,\bm{q})\big]A^{\nu}(\tau, \bm{q})\\
& = & \dint_{-\infty}^{\infty}\dd\tau\Bigg[ (-1)^{2} \delta(t - \tau)\left( \int_{-\infty}^{\tau}\dd\tau_{1}\Pi_{\mu\nu}(\tau - \tau_{1},\bm{q}) \right) - \Pi_{\mu\nu}^{R}(t - \tau,\bm{q})\Bigg]A^{\nu}(\tau, \bm{q}),
\end{eqnarray}
\end{widetext}
where we have used the definition provided in \Eq{Kubo_sigma}. In temporal frequency space, using \Eq{def_Pi_R(w)} and $\mathcal{F}[\delta(t)](\omega) = 1$, we obtain \Eq{j_eq10}, and from there we obtain the final result given in \Eq{True_Relation_Polarization_conductivity}.

\subsection{Constitutive relation of the electric conductivity}
Starting from the microscopic Ohm's Law, the electric current generated by an electric field is obtained as
\begin{eqnarray}\label{OhmLaw}
\mean{ J_{\mu}(\omega,\bm{q}) } = \sigma_{\mu\nu}(\omega,\bm{q})E^{\nu}(\omega,\bm{q}),
\end{eqnarray}
where $\sigma_{\mu\nu}(\qq)$ is the electric conductivity operator, $\qq = (\tilde{q}_{0},\tilde{\bm{q}})$ (see \Tab{Notation}) and $E^{\mu}(\omega,\bm{q}) = \ii\omega A^{\mu}(\omega,\bm{q})$ is the electric field in Temporal Gauge ($\phi = cA^{0} = 0$). Note that even if we are working in the specific Temporal Gauge here, we can restore Gauge invariance at the end of the calculations by writing all final results in explicitly Gauge-invariant quantities.
In \Sect{KubosigmaderivationAppendix}, by using the Kubo formula \cite{Kubo1957} and the microscopic Ohm's Law, and assuming causality being respected, we re-derive the standard Luttinger formula \cite{Luttinger1968}\cite{yanagisawa2004}\cite{Szunyogh2002}\cite{Rammer_2007}\cite{Ando-PhysRevB.74.155411} 
\begin{eqnarray}\label{True_Relation_Polarization_conductivity_maintext}
\sigma_{\mu\nu}(\qq) = \dfrac{\Pi_{\mu\nu}(\omega,\bm{q}) - \dlim_{\omega\to0^{+}}\Pi_{\mu\nu}(\omega,\bm{q})}{-\ii\omega} = \dfrac{\tilde{\Pi}_{\mu\nu}(\omega,\bm{q})}{-\ii\omega},
\end{eqnarray}
where the Polarization operator is defined (in the Matsubara formalism) as
\begin{eqnarray}\label{PI_Definition}
\Pi_{\mu\nu}(\qq) = \dfrac{-\ii}{\hbar}\int_{\kk}\Tr{\hat{J}_{\mu}(\kk)\hat{J}_{\nu}(\kk+\qq)},
\end{eqnarray}
Here we use the definition $\dint_{\kk} = \dint_{-\infty}^{\infty}\dfrac{\dd k_{0}}{2\pi}\dint_{\bm{k}}$ (see \Tab{Notation}), $\Tr{\hat{A}}$ is the trace of $\hat{A}$, and
\begin{eqnarray}\label{tPI_Definition}
\tilde{\Pi}_{\mu\nu}(\qq)
= \Pi_{\mu\nu}(\qq) - \dlim_{\omega\to0}\Pi_{\mu\nu}(\qq).
\end{eqnarray}
In \Sect{KubosigmaderivationAppendix} (see \Eq{j_eq10}) we also derive that
\begin{eqnarray}\label{j_eq10_maintext}
\mean{ J_{\mu}(\omega, \bm{q}) } = - \Big[ \Pi_{\mu\nu}(\omega,\bm{q}) - \dlim_{\omega\to0}\Pi_{\mu\nu}(\omega,\bm{q}) \Big]A^{\nu}(\omega, \bm{q}).
\end{eqnarray}
Note that this subtraction, which we derived here simply from Ohm's Law and time causality, naturally implies that there is no electric current $\mean{ J_{\mu}(t) }\neq0$ when the electric field is zero $E^{\nu}(t) = 0$ due to Ohm's Law. Indeed, since $E^{\nu}(t) = - \partial_{t} A^{\nu}(t)$, $E^{\nu}(t) = 0$ implies $A^{\nu}(t) = A_{0}^{\nu}$ to be a constant in time. Using the Fourier transform definition $f(\omega) = \int f(t) e^{\ii\omega t}\dd t$, this means that $A^{\nu}(\omega) = A_{0}^{\nu}\delta(\omega)$, and it is easy to see that \Eq{j_eq10_maintext} implies $\mean{ J_{\mu}(t) } = - A_{0}^{\nu}\dlim_{\omega\to0}\Big[ \Pi_{\mu\nu}(\omega,\bm{q}) - \dlim_{\omega\to0}\Pi_{\mu\nu}(\omega,\bm{q}) \Big] = 0$, independently from of the functional and tensorial form of $\Pi_{\mu\nu}(\omega,\bm{q})$. This strong physical requirement is already discussed in \cite{Falkovsky2007}.\\

However, another expression for the electric conductivity is derived from the general EM transport relation \cite{Fialkovsky2016}\cite{Fialkovsky2012}\cite{Bordag2015b}\cite{Klimchitskaya2018}\cite{universe6090150}\cite{Bimonte2017}
\begin{eqnarray}\label{JdependsonA}
\mean{ J_{\mu}(\omega, \bm{q}) } = - \Pi_{\mu\nu}(\omega,\bm{q})A^{\nu}(\omega, \bm{q}),
\end{eqnarray}
Indeed, this is a solution of the differential equation for the transport coefficient provided in \Eq{H_perturbed_in_A}. Since $E^{\nu}(\omega, \bm{q})=\ii\omega A^{\nu}(\omega, \bm{q})$, \Eq{JdependsonA} can be written as 
\begin{eqnarray}\label{JdependsonAss}
\mean{ J_{\mu}(\omega, \bm{q}) } = \frac{\Pi_{\mu\nu}(\omega,\bm{q})}{-\ii\omega}E^{\nu}(\omega, \bm{q})= \sigma^{\rm{NR}}_{\mu\nu}(\omega,\bm{q})E^{\nu}(\omega,\bm{q}),
\end{eqnarray}
introducing the definition of the polarization-based electric conductivity tensor as
\begin{eqnarray}\label{QFTconductivitydef}
\sigma^{\rm{NR}}_{\mu\nu}(\omega,\bm{q})=\frac{\Pi_{\mu\nu}(\omega,\bm{q})}{-\ii\omega}.
\end{eqnarray}
However, this definition of electric conductivity as the transport coefficient that relates the induced electric current with the electric field in the system is problematic because, from \Eq{JdependsonAss}, we can generate an electric current even without an electric field.
Indeed, by moving back from frequency to time representation, \Eq{JdependsonA} can be written as
\begin{eqnarray}\label{JdependsonAtime}
\mean{ J_{\mu}(t, \bm{q}) } = - \int_{-\infty}^{t}\dd\tau \Pi_{\mu\nu}(t-\tau,\bm{q})A^{\nu}(\tau,\bm{q}),
\end{eqnarray}
assuming now an electric field equal to zero at any time $E^{\nu}(t,\bm{q}) = 0$, since $E^{\nu}(t,\bm{q})= - \partial_t A^{\nu}(t,\bm{q})$, this correspond to a constant in time $A^{\nu}(t,\bm{q})=A^{\nu}_{0}(\bm{q})$. Hence, the induced electric current \eqref{JdependsonAtime} becomes 
\begin{eqnarray}\label{JdependsonAtime33b}
\mean{ J_{\mu}(t, \bm{q}) } & = & - A_{0}^{\nu}(\bm{q}) \int_{-\infty}^{t}\dd\tau\Pi_{\mu\nu}(t-\tau,\bm{q}) \nonumber\\
& = & - A_{0}^{\nu}(\bm{q}) \lim_{\omega\to0}\Pi_{\mu\nu}(\omega,\bm{q}),
\end{eqnarray}
%where we used that $\Pi_{\mu\nu}^{R}(t-\tau,\bm{q}) = \Theta(t-\tau)\Pi_{\mu\nu}(t - \tau,\bm{q})$ is a retarded function.
In this case, the electric current is induced by the static non-homogeneous magnetic field, not by an electric field. Therefore, this current has a pure magnetic origin and cannot be related to an electric conductivity transport coefficient. Hence, the polarization-based expression \eqref{JdependsonA} predicts the existence of permanent electric currents in the absence of an electric field; therefore, by definition provided in \Eq{QFTconductivitydef}, $\sigma^{\rm{NR}}_{\mu\nu}(\omega,\bm{q})$ can induce electric current even in the absence of an electric field. This pathological prediction of the QFT model as is developed and used in literature, explicitly using \eqref{JdependsonA}, will be discussed in detail in section \Sect{Section_QFT_Model}.

\subsection{Polarization Operator}\label{Sect_def_Polarization}
The Kubo formula for the Polarization operator given in \Eq{PI_Definition} can be reduced to the bubble Feynman diagram as \cite{WangKong2010}\cite{watson2023mathematical}\cite{Khalilov2015}
%\begin{widetext}
\begin{eqnarray}\label{def:Polarization_Operator}
\Pi_{\mu\nu}(\qq) = \dfrac{-\ii}{\hbar}\dint_{\kk}\Tr{ G_{0}^{\ell,\ell'}(\kk)\hat{J}_{\mu}^{\ell',\mathscr{m}}(\kk)G_{0}^{\mathscr{m},\tilde{\ell}}(\kk +\qq)\hat{J}_{\nu}^{\tilde{\ell},\tilde{\ell}'}(\kk+\qq) }.
\end{eqnarray}
%\end{widetext}
Here we use the definition $\dint_{\kk} = \dint_{-\infty}^{\infty}\dfrac{\dd k_{0}}{2\pi}\dint_{\bm{k}}$ (see \Tab{Notation}), $\hat{J}_{\mu}^{\ell,\tilde{\ell}}\left(\qq\right)$ is the current operator of electronic quasiparticles, defined in \Eq{Current_Operator_definition} and $G_{0}^{\ell,\tilde{\ell}}(\kk) = \mean{\mathcal{T}\hat{c}_{\kk}^{\ell}\hat{c}_{\kk}^{\tilde{\ell},\dagger}}_{0}$ is the Matsubara time-ordered Green function of the unperturbed Hamiltonian of the system. In the rest of the paper, the time ordering is implicitly assumed. For a general linear Hamiltonian of the form $\hbar\omega\psi = \tilde{k}_{0}\psi = \hat{H}\psi$, the Green function fulfills $\left( \tilde{k}_{0}\delta^{\ell}_{\phantom{\ell}\ell'} - \hat{H}^{\ell}_{\phantom{\ell}\ell'}\right)G_{0}^{\ell',\tilde{\ell}}(\bm{r},\bm{s}) = \delta^{\ell,\tilde{\ell}}\delta(\bm{r} - \bm{s})$, and we have the eigenproblem  
\begin{eqnarray}\label{Eigensystem}
\hat{H}^{\ell}_{\phantom{\ell}\ell'}\ket{u^{\lambda}}^{\ell'} = \left(\epsilon^{\lambda} - \mu\right)\ket{u^{\lambda}}^{\ell} = \xi^{\lambda}\ket{u^{\lambda}}^{\ell}.
\end{eqnarray}
The grand-canonical Green function in momentum space is
\begin{eqnarray}\label{Electric_Green0_functionb}
G_{0}^{\ell,\tilde{\ell}}(k) = \dsum_{\lambda}\dfrac{\ket{u_{\bm{k}}^{\lambda}}^{\ell}\bra{u_{\bm{k}}^{\lambda}}^{\tilde{\ell}}}{ \tilde{k}_{0} - \xi_{\bm{k}}^{\lambda} },
\end{eqnarray}
where $\ell=\{\mathcal{L},\mathcal{O},s\}$, and $\lambda$ is the band in the reciprocal space. Using a different (space-time covariant) form of the Green function for the Dirac Hamiltonian leads to different representations of the same result discussed here \cite{Bordag2015}. After introducing the Green function into the Polarization operator, using the cyclic property of trace, $\Tr{ABC} = \Tr{BCA}$, and applying the Matsubara formalism to carry out the $k_{0}$ integral for the fermionic case by using $\tilde{k}_{0} = \hbar\omega_{n} = \frac{2\pi}{\beta}\left(n + \frac{1}{2}\right)$ $\forall n\in\mathbb{Z}$, $\beta = (k_{B}T)^{-1}$ and $\tilde{q}_{0} = \ii\hbar\omega_{m} = \ii\frac{2\pi}{\beta}\left(m + \frac{1}{2}\right)$, we obtain the following auxiliary sum $M$ that will be used later.
\begin{eqnarray}\label{Matsubara_Sum}
M
& = & \dfrac{-\ii}{\hbar}\int_{-\infty}^{\infty}\dfrac{\dd k_{0}}{2\pi}\dfrac{1}{ \tilde{k}_{0} - \xi_{\bm{k}}^{\lambda} }\dfrac{1}{ \tilde{k}_{0} + \tilde{q}_{0} - \xi_{\bm{k}+\bm{q}}^{\lambda'} }\nonumber\\
& = & - \frac{1}{\beta}\sum_{n\in\mathbb{Z}}^{\text{Fermi}}
\dfrac{1}{ \ii\hbar\omega_{n} - \xi_{\bm{k}}^{\lambda} }
\dfrac{1}{ \ii\hbar\omega_{n} + \ii\hbar\omega_{m} - \xi_{\bm{k}+\bm{q}}^{\lambda'} }\nonumber\\
& = & \dfrac{n_{F}(\xi_{\bm{k}}^{\lambda}) - n_{F}(\xi_{\bm{k}+\bm{q}}^{\lambda'}) }{ \ii\hbar\omega_{m} + \xi_{\bm{k}}^{\lambda} - \xi_{\bm{k}+\bm{q}}^{\lambda'} },
\end{eqnarray}
where the Fermi-Dirac distribution is
\begin{eqnarray}\label{Fermi_Dirac_Distribution}
n_{F}(\xi_{\bm{k}}^{\lambda}) = \frac{1}{e^{\beta\xi_{\bm{k}}^{\lambda}} + 1} = \frac{1}{e^{\beta(\epsilon_{\bm{k}}^{\lambda} - \mu)} + 1}.
\end{eqnarray}
Note that we have used $n_{F}(\ii\hbar\omega_{m} + \xi_{\bm{k}+\bm{q}}^{\lambda'}) = n_{F}(\xi_{\bm{k}+\bm{q}}^{\lambda'})$. Using the definition of the electric current operator given in \Eq{Current_Operator_definition}, the Green function obtained in \Eq{Electric_Green0_function} and the Matsubara sum given in \Eq{Matsubara_Sum} analytically expanded to the whole upper complex plane by applying the formal change $\ii\omega_{m} = \omega\in\mathbb{C}^{+}$ into the definition of the Polarization operator \Eq{def:Polarization_Operator}, we finally obtain
\begin{widetext}
\begin{eqnarray}\label{Def_Polarization}
\Pi_{\mu\nu}(\qq, \mu, T)
& = & e^{2}\sum_{\lambda,\lambda'}\int\dfrac{\dd^{d}\bm{k}}{(2\pi)^{d}}
\dfrac{ \bra{u_{\bm{k}}^{\lambda}}\hat{v}_{\mu}
\ket{u_{\bm{k}+\bm{q}}^{\lambda'}}\bra{u_{\bm{k}+\bm{q}}^{\lambda'}}\hat{v}_{\nu}\ket{u_{\bm{k}}^{\lambda}} }{ \hbar\omega + \epsilon_{\bm{k}}^{\lambda} - \epsilon_{\bm{k}+\bm{q}}^{\lambda'} }
\left[n_{F}(\xi_{\bm{k}}^{\lambda}) - n_{F}(\xi_{\bm{k}+\bm{q}}^{\lambda'})\right],
\end{eqnarray}
where we have made explicit the trace operator because it is only applied in the $\bm{k}$-space and the $(\lambda,\lambda')$-bands space.

In order to connect this expression with the electric conductivity obtained from the Kubo formula  \Eq{True_Relation_Polarization_conductivity_maintext} we need to calculate $\tilde{\Pi}_{\mu\nu}(\qq)
= \Pi_{\mu\nu}(\qq) - \dlim_{\omega\to0}\Pi_{\mu\nu}(\qq)$ of \Eq{tPI_Definition}. We then use
\begin{eqnarray}
\frac{1}{ \hbar\omega + A} - \lim_{\omega\to0}\frac{1}{ \hbar\omega + A} = \frac{1}{ \hbar\omega + A} - \frac{1}{A} = \frac{- 1}{ \hbar\omega + A}\frac{\hbar\omega}{A},
%\end{eqnarray}
%\begin{eqnarray}
\hspace{0.75cm}
\Rightarrow
\hspace{0.75cm}
\frac{- 1}{ \hbar\omega + A}\frac{1}{A}
 = \frac{1}{\hbar\omega}\left[ \frac{1}{ \hbar\omega + A} - \lim_{\omega\to0}\frac{1}{ \hbar\omega + A}\right],
\end{eqnarray}
which directly provides
\begin{eqnarray}\label{General_Polarization_Operator}
\tilde{\Pi}_{\mu\nu}(\qq, \mu, T)
& = & - e^{2}\hbar\omega\sum_{\lambda,\lambda'}\int_{BZ}\dfrac{\dd^{d}\bm{k}}{(2\pi)^{d}}
\dfrac{ \bra{u_{\bm{k}}^{\lambda}}\hat{v}_{\mu}
\ket{u_{\bm{k}+\bm{q}}^{\lambda'}}\bra{u_{\bm{k}+\bm{q}}^{\lambda'}}\hat{v}_{\nu}\ket{u_{\bm{k}}^{\lambda}} }{ \hbar\omega + \epsilon_{\bm{k}}^{\lambda} - \epsilon_{\bm{k}+\bm{q}}^{\lambda'} }
\frac{n_{F}(\xi_{\bm{k}}^{\lambda}) - n_{F}(\xi_{\bm{k}+\bm{q}}^{\lambda'})}{\epsilon_{\bm{k}}^{\lambda} - \epsilon_{\bm{k}+\bm{q}}^{\lambda'}},
\end{eqnarray}
and finally, using \Eq{True_Relation_Polarization_conductivity_maintext} ($\sigma_{\mu\nu}(\qq) = \ii\tilde{\Pi}_{\mu\nu}(\qq)/\omega$), the electric conductivity \cite{Falkovsky2007}\cite{Fialkovsky2008}\cite{ALLEN2006165}\cite{Ludwig1994}\cite{universe7070237}:
\begin{eqnarray}\label{General_Kubo_Formula_dissipationless}
\sigma_{\mu\nu}(\qq, \mu, T) & = & - \ii e^{2}\hbar\sum_{\lambda,\lambda'}\int_{BZ}\frac{\dd^{d}\bm{k}}{(2\pi)^{d}}
\frac{\bra{u_{\bm{k}}^{\lambda}} \hat{v}_{\mu} \ket{u_{\bm{k}+\bm{q}}^{\lambda'}} \bra{u_{\bm{k}+\bm{q}}^{\lambda'}} \hat{v}_{\nu} \ket{u_{\bm{k}}^{\lambda}} }{ \hbar\omega + \epsilon_{\bm{k}}^{\lambda} - \epsilon_{\bm{k}+\bm{q}}^{\lambda'} }
\frac{n_{F}(\xi_{\bm{k}}^{\lambda}) - n_{F}(\xi_{\bm{k}+\bm{q}}^{\lambda'})}{\epsilon_{\bm{k}}^{\lambda} - \epsilon_{\bm{k}+\bm{q}}^{\lambda'}}.
\end{eqnarray}
As discussed in subsect.~\ref{Effect_interactions} the interactions are included by introducing a constant dissipation rate corresponding to a finite lifetime  $\Gamma^{\lambda} = \tau_{\lambda}^{-1}$. Then, the electronic quasienergies are modified as $\epsilon_{\bm{k}}^{\lambda} \to \epsilon_{\bm{k}}^{\lambda} + \ii\hbar\Gamma^{\lambda}$ and $\epsilon_{\bm{k}+\bm{q}}^{\lambda'}\to \epsilon_{\bm{k}+\bm{q}}^{\lambda',*} = \epsilon_{\bm{k}+\bm{q}}^{\lambda'} - \ii\hbar\Gamma^{\lambda'}$, therefore
\begin{eqnarray}\label{General_Kubo_Formula_dissipation}
\sigma_{\mu\nu}(\qq, \mu, T) & = & - \ii e^{2}\hbar\sum_{\lambda,\lambda'}\int_{BZ}\frac{\dd^{d}\bm{k}}{(2\pi)^{d}}
\frac{\bra{u_{\bm{k}}^{\lambda}} \hat{v}_{\mu} \ket{u_{\bm{k}+\bm{q}}^{\lambda'}} \bra{u_{\bm{k}+\bm{q}}^{\lambda'}} \hat{v}_{\nu} \ket{u_{\bm{k}}^{\lambda}} }{\hbar\omega + \ii\hbar(\Gamma^{\lambda} + \Gamma^{\lambda'}) + \epsilon_{\bm{k}}^{\lambda} - \epsilon_{\bm{k}+\bm{q}}^{\lambda'} }
\frac{n_{F}(\xi_{\bm{k}}^{\lambda} + \ii\hbar\Gamma^{\lambda}) - n_{F}(\xi_{\bm{k}+\bm{q}}^{\lambda'} - \ii\hbar\Gamma^{\lambda'})}{\ii\hbar(\Gamma^{\lambda} + \Gamma^{\lambda'}) + \epsilon_{\bm{k}}^{\lambda} - \epsilon_{\bm{k}+\bm{q}}^{\lambda'}},
\end{eqnarray}
where also the electronic eigenvalues $u_{\bm{k}}^{\lambda}$ are functions of $\hbar\Gamma^{\lambda}$. Assuming that the dissipation rates are small, in a first order approximation, by using $\Gamma = \Gamma^{\lambda} + \Gamma^{\lambda'}$, we obtain the Kubo formula we are going to use
\begin{eqnarray}\label{General_Kubo_Formula}
\sigma_{\mu\nu}^{\rm{K}}(\qq, \Gamma, \mu, T) & = & - \ii e^{2}\hbar\sum_{\lambda,\lambda'}\int_{BZ}\frac{\dd^{d}\bm{k}}{(2\pi)^{d}}
\frac{\bra{u_{\bm{k}}^{\lambda}} \hat{v}_{\mu} \ket{u_{\bm{k}+\bm{q}}^{\lambda'}} \bra{u_{\bm{k}+\bm{q}}^{\lambda'}} \hat{v}_{\nu} \ket{u_{\bm{k}}^{\lambda}} }{\hbar(\omega + \ii\Gamma) + \epsilon_{\bm{k}}^{\lambda} - \epsilon_{\bm{k}+\bm{q}}^{\lambda'} }
\frac{n_{F}(\xi_{\bm{k}}^{\lambda}) - n_{F}(\xi_{\bm{k}+\bm{q}}^{\lambda'})}{ \epsilon_{\bm{k}}^{\lambda} - \epsilon_{\bm{k}+\bm{q}}^{\lambda'}},
\end{eqnarray}
\end{widetext}
where any possible dependence of $u_{\bm{k}}^{\lambda}$ on $\hbar\Gamma^{\lambda}$ has disappeared as well and the effect of the small electronic dissipation is given by the phenomenological dissipation parameter $\Gamma > 0$ in this approximation.

In \Eq{General_Kubo_Formula}, the dissipation rate is the inverse of the mean lifetime of the electronic quasiparticle $\Gamma = \tau^{-1}$ \cite{DasSarmaRMP2011}\cite{Gusynin2007b}\cite{Gusynin2009}. It appears as the imaginary part of $(\epsilon_{\bm{k}}^{\lambda} - \epsilon_{\bm{k}+\bm{q}}^{\lambda'})$, but it is not a bad approximation to consider it as a constant, therefore, $\ii\Gamma$ can be absorbed into a now complex $\omega\longmapsto q_{0} = (\omega + \ii\Gamma)$ \cite{yanagisawa2004}\cite{Szunyogh1999}. This formula for the electric conductivity (\Eq{General_Kubo_Formula}) derived from the Random Phase Approximation is entirely equivalent to the Kubo formula, which has been derived elsewhere \cite{Luttinger1968}\cite{non-local_Graphene_Lilia_Pablo}\cite{Falkovsky2007}\cite{Gusynin2006}\cite{ALLEN2006165}\cite{Szunyogh1999}\cite{Szunyogh2002}\cite{universe7070237}\cite{Mahan_Book}\cite{Lilia_full_graphene_Conductivity}\cite{Drosdoff2012}\cite{watson2023mathematical}. In the following sections, we are going to relate the expressions for the electric conductivity obtained in the different models we consider.

\section{Tensor decomposition}\label{Section_Tensor_decomposition}
In order to compare the results of the QFT model \cite{PRL_Mohideen} with the results of the Kubo formula \cite{non-local_Graphene_Lilia_Pablo}, we start observing that, in the former, the results for the non-local Polarization operator are written in terms of the component $\Pi_{00}$ and of the quantity $\Pi = q_{\surf}^{2}\Pi_{\text{tr}} - \theta_{z}^{2}\Pi_{00}$, where $\Pi_{\text{tr}}=\text{Tr}(\Pi)$, $\theta_{z} = \sqrt{ \tfrac{\xi^{2}}{c^{2}} + q_{\surf}^{2} }$ and $\omega = \ii\xi$. Correspondingly, the QFT model electric conductivity is expressed in terms of the components $\sigma_{00}$ and $\sigma = q_{\surf}^{2}\sigma_{\text{tr}} - \theta_{z}^{2}\sigma_{00}$. Differently, in the Kubo formula, the non-local electric conductivity tensor $\sigma_{ij}$ is expressed in terms of longitudinal ($\sigma_{L}$), transversal ($\sigma_{T}$), Hall ($\sigma_{H}$) and sinusoidal ($\sigma_{S}$) components as \cite{Fialkovsky2011}\cite{Bordag2021}\cite{Zeitlin1995}\cite{Dorey1992}

\begin{eqnarray}
\label{General_non_local_conductivity_form}
\sigma_{ij}(\qq, \mu, T) & = & \frac{q_{i}q_{j}}{q_{\surf}^{2}}\sigma_{L}(\omega, \tilde{q}, \mu, T)\nonumber\\
& & + 
\left( \delta_{ij} - \frac{q_{i}q_{j}}{q_{\surf}^{2}} \right)\sigma_{T}(\omega, \tilde{q}, \mu, T)\nonumber\\
& & + \epsilon_{ij}\sigma_{H}(\omega, \tilde{q}, \mu, T)\nonumber\\
& & + \Big( \frac{q_{i}q_{\ell}\epsilon_{\ell j} - \epsilon_{i\ell}q_{\ell}q_{j}}{q_{\surf}^{2}} \Big)\sigma_{S}(\omega, \tilde{q}, \mu, T),
\end{eqnarray}
A similar relation can be derived for the polarization operator $\Pi_{\mu\nu}$. Here, $\qq = \left(\tilde{q}_{0},\tilde{\bm{q}}_{\surf}\right)$, $\tilde{\bm{q}}_{\surf} = \left(\tilde{q}_{1},\tilde{q}_{2}\right)$, $\tilde{q}_{\surf}^{2} = \sqrt{\tilde{q}_{1}^{2} + \tilde{q}_{2}^{2}}$ (see \Tab{Notation}), $\delta_{ij}$ is the Kronecker delta function and $\epsilon_{ij}$ is the 2D Levi-Civita symbol. The sum over repeated indices is assumed. In general, $\sigma_{S}\neq 0$, but for graphene (and for any 2D Dirac material), it is zero \cite{non-local_Graphene_Lilia_Pablo}. In this section, we provide the explicit connection between the component $\{\sigma_{00},\sigma_{\text{tr}}\}$, and the components $\{\sigma_{L},\sigma_{T},\sigma_{H},\sigma_{S}\}$ (and between the components $\{\Pi_{00},\Pi_{\text{tr}}\}$, and the components $\{\Pi_{L},\Pi_{T},\Pi_{H},\Pi_{S}\}$), as a side result, we also provide the spatio-temporal generalization of \Eq{General_non_local_conductivity_form}.\\

To this end, we will use the transversality condition $q^{\mu}\Pi_{\mu\nu}(\qq) = 0$ and \Eq{General_non_local_conductivity_form} for $\Pi_{\mu\nu}$ instead for $\sigma_{\mu\nu}$ (we can define the same quantities for $\sigma_{\mu\nu}$, but $\Pi_{\mu\nu}$ is a tensor that relates two 4-vectors, while $\sigma_{\mu\nu}$ relates one 4-vector with a part of a second-order tensor). The transversality condition for the Polarization operator $\Pi_{\mu\nu}$ can be deduced from the application of the continuity equation ($\partial^{\mu}j_{\mu}(x_{\mu}) = 0$) for the charge 4-current (Note that, for 2D materials, covariant vectors are in $(2+1)$ dimensions, however, for the sake of clarity, we will call them 4-vectors as usual) inside the material together with the constitutive relation for linear currents [$j_{\mu}(\qq) = - \Pi_{\mu\nu}(\qq)A^{\nu}(\qq)$]:
\begin{eqnarray}
q^{\mu}j_{\mu}(\qq) = 
- q^{\mu}\Pi_{\mu\nu}(\qq)A^{\nu}(\qq) = 0
\,\Rightarrow\,
q^{\mu}\Pi_{\mu\nu}(\qq) = 0
\end{eqnarray}
for all $A_{\nu}$. Here, $q_{\mu}=(q_{0}, q_{1}, q_{2}) = (\tfrac{\omega}{c}, q_{1}, q_{2})$ (see \Tab{Notation}) is the momentum of the quasiparticle. In all the text that follows, we use the metric tensor $g_{\mu\nu} = \text{diag}\left\lbrace +1, -1, -1\right\rbrace$. Separating the temporal component of the 4-vectors, we get $q^{\mu}\Pi_{\mu\nu}(\qq) = q^{0}\Pi_{0\nu}(\qq) - q^{a}\Pi_{a\nu}(\qq) = 0$, therefore, we obtain ($a\in\{1,2\}$)
\begin{eqnarray}
\Pi_{0\nu}(\qq) = \frac{q^{a}\Pi_{a\nu}(\qq)}{q^{0}}.
\end{eqnarray}
Using \Eq{General_non_local_conductivity_form}, we obtain
\begin{eqnarray}
\begin{array}{rcl}
\Pi_{01}(\qq) & = & \dfrac{ q^{1}\Pi_{L}(\qq) + q^{2}(\Pi_{H}(\qq) - \Pi_{S}(\qq) )}{q^{0}},\\
\Pi_{02}(\qq) & = & \dfrac{ q^{2}\Pi_{L}(\qq) - q^{1}(\Pi_{H}(\qq) - \Pi_{S}(\qq) )}{q^{0}}.
\end{array}
\end{eqnarray}
Now we can use that the transversality condition is also fulfilled for the second index of the polarization tensor $\Pi_{\mu\nu}(\qq)q^{\nu} = 0$ as well, obtaining that
\begin{eqnarray}
\Pi_{\mu 0}(\qq) = \frac{\Pi_{\mu a}(\qq)q^{a}}{q^{0}},
\end{eqnarray}
Using again \Eq{General_non_local_conductivity_form}, we obtain
\begin{eqnarray}
\begin{array}{rcl}
\Pi_{10}(\qq) & = & \dfrac{ q^{1}\Pi_{L}(\qq) - q^{2}(\Pi_{H}(\qq) + \Pi_{S}(\qq) )}{q^{0}},\\
\Pi_{20}(\qq) & = & \dfrac{ q^{2}\Pi_{L}(\qq) + q^{1}(\Pi_{H}(\qq) + \Pi_{S}(\qq) )}{q^{0}}.
\end{array}
\end{eqnarray}
In general, $\Pi_{\mu\nu}$ is not symmetric because of $\Pi_{S}$ and of the purely antisymmetric term $\Pi_{H}$. From those results, we can now derive the $00$ component as
\begin{eqnarray}\label{sigma_00=f_sigma_L}
\Pi_{00}(\qq) = \frac{q^{a}\Pi_{a0}(\qq)}{q^{0}} = \frac{q^{a}\Pi_{ab}(\qq)q^{b}}{(q^{0})^{2}} = \frac{q_{\surf}^{2}}{q_{0}^{2}}\Pi_{L}(\qq).
\end{eqnarray}
where we have used $(q^{0})^{2} = q_{0}^{2}$. Now we have derived the full form of the polarization tensor; we obtain the trace as
\begin{eqnarray}\label{DefTrsigma}
\Tr{\Pi} = \Pi_{\text{tr}} = g^{\mu\nu}\Pi_{\mu\nu} = - \Pi_{T}(\qq) - \Pi_{L}(\qq)\dfrac{q_{z}^{2}}{q_{0}^{2}},
\end{eqnarray}
with $q_{z} = \sqrt{ q_{0}^{2} - q_{\surf}^{2} }$. From the expressions for $\Pi_{00}$ and $\Pi_{\text{tr}}$, we finally obtain that
\begin{eqnarray}\label{sigmaTr_sigma00}
\begin{array}{rcl}
\Pi_{L}(\qq) & = & \dfrac{q_{0}^{2}}{q_{\surf}^{2}}\Pi_{00}(\qq),\label{sigma_L=f_sigma_00}\\
\Pi_{T}(\qq) & = & - \Pi_{\text{tr}} - \dfrac{q_{z}^{2}}{q_{\surf}^{2}}\Pi_{00}(\qq).
\end{array}
\end{eqnarray}
Note that for imaginary frequencies
\begin{eqnarray}%\label{sigmaTr_sigma00}
- q_{\surf}^{2}\Pi_{T}(\qq) = q_{\surf}^{2}\Pi_{\text{tr}} - \theta_{z}^{2}\Pi_{00}(\qq) = \Pi,
\end{eqnarray}
where $\theta_{z} = \sqrt{ \tfrac{\xi^{2}}{c^{2}} + q_{\surf}^{2} }$. In conclusion, the electric conductivity and Polarization tensors can be decomposed into the sum of four components \cite{Dorey1992}\cite{Fialkovsky2016} as
\begin{eqnarray}\label{Tensor_Factorization_Pi_munu}
\Pi_{\mu\nu} = L_{\mu\nu}\Pi_{L} + T_{\mu\nu}\Pi_{T} + H_{\mu\nu}\Pi_{H} + S_{\mu\nu}\Pi_{S},
\end{eqnarray}
and
\begin{eqnarray}\label{Tensor_Factorization_sigma_munu}
\sigma_{\mu\nu}^{\rm{NR}} = L_{\mu\nu}\sigma_{L} + T_{\mu\nu}\sigma_{T} + H_{\mu\nu}\sigma_{H} + S_{\mu\nu}\sigma_{S},
\end{eqnarray}
with
\begin{eqnarray}
L_{\mu\nu} & = & \left( \delta_{\mu 0} - \frac{q_{0}q_{\mu}}{q_{z}^{2}} \right)\frac{q_{z}^{4}}{q_{\surf}^{2}q_{0}^{2}}\left( \delta_{\nu 0} - \frac{q_{0}q_{\nu}}{q_{z}^{2}} \right),\\
T_{\mu\nu} & = & \delta_{\mu}^{\phantom{\mu}i}\left( \delta_{ij} - \frac{q_{i}q_{j}}{q_{\surf}^{2}} \right)\delta_{\phantom{j}\nu}^{j},\\
H_{\mu\nu} & = & (-1)^{\delta_{0\mu}+\delta_{0\nu}}\epsilon_{\mu\nu\rho}\frac{q^{\rho}}{q_{0}},\\
S_{\mu\nu} & = & \dfrac{\bar{q}_{\mu}\bar{q}_{\rho}\epsilon_{\rho\nu} - \epsilon_{\mu\rho}\bar{q}_{\rho}\bar{q}_{\nu}}{q_{\surf}^{2}},
\end{eqnarray}
where
\begin{eqnarray}
\bar{q}_{\alpha} = q_{\alpha} - \frac{q_{z}^{2}}{q_{0}}\delta_{0\alpha}.
\end{eqnarray}
So, now we can compare the results of \cite{PRL_Mohideen} with the results obtained in \cite{non-local_Graphene_Lilia_Pablo}

\section{Kubo formula for graphene}\label{Section_non_local_Kubo}
In \cite{non-local_Graphene_Lilia_Pablo}, by using the Kubo formula (\Eq{General_Kubo_Formula}), it has been derived the spatial part of the 2D-electric conductivity tensor for the 2D Dirac cone given by \Eq{Eq_Hamiltonian_spinor} (see \Tab{Notation})
\begin{eqnarray}\label{Eq_Hamiltonian_spinor2}
\hat{H}_{s}^{\eta}(\bm{k}) = \eta\tilde{k}_{1}\tau_{1} + \tilde{k}_{2}\tau_{2} + \tau_{3}\Delta_{s}^{\eta},
\end{eqnarray}
where $m_{z} = \Delta_{s}^{\eta}$. Depending on the spin ($s$) and valley ($\eta$) indices, and on external and internal perturbations of the 2D material, each mass-gap $\Delta_{s}^{\eta}$ can take different values \cite{Ezawa2015}\cite{Ezawa2012PRL}\cite{Ezawa2012PRB}\cite{Ezawa2013PRL}\cite{Rodriguez-Lopez2017}, which are zero for suspended and unperturbed graphene sheets. The velocity vector operator is $\hat{v}_{i} = \partial_{k_{i}}\hat{H} = \hbar v_{F} (\eta\tau_{x}, \tau_{y})$ and the eigenstates of the Hamiltonian from \Eq{Eq_Hamiltonian_spinor} are
\begin{eqnarray}\label{Eigenfunctions}
\ket{u_{\bm{k}}^{\lambda}} = \frac{1}{\sqrt{2\epsilon_{\bm{k}}^{\lambda}(\epsilon_{\bm{k}}^{\lambda} + \Delta_{s}^{\eta}) }} 
\left(\begin{array}{c}
-(\Delta_{s}^{\eta} + \epsilon_{\bm{k}}^{\lambda})\dfrac{\tilde{k}_{2} + \ii\eta\tilde{k}_{1}}{\tilde{k}_{\surf}}\\
\tilde{k}_{\surf}
\end{array}\right),
\end{eqnarray}
with corresponding eigenenergies (compare with \Eq{Eq_Energy_BiSpinor})
\begin{eqnarray}\label{Eq_Energy_Hamiltonian_spinor}
\epsilon^{\lambda}_{\bm{k}} = \lambda\sqrt{ \tilde{k}_{\surf}^{2} + (\Delta_{s}^{\eta})^{2} }.
\end{eqnarray}
If we compare \Eq{Eq_Hamiltonian_spinor2} with the Hamiltonian of graphene given in \Eq{Dirac_H}, we see that we can rewrite the four-spinor Hamiltonian given in \Eq{Eq_Hamiltonian_spinor2} into the sum of two two-spinors Hamiltonians of the form of \Eq{Dirac_H}, one from the second and third rows and columns, and the other from the first and fourth, as indicated here
\begin{eqnarray}\label{Dirac_Decomposition}
\hat{\mathcal{H}}_{s}^{D}(\bm{k}) = \left(\begin{array}{c V{3} cc V{3} c}
\tilde{k}_{0} + m & 0 & 0 & \tilde{k}_{1} - \ii\tilde{k}_{2} \\
\hlineB{4}
 0 & \tilde{k}_{0} + m & \tilde{k}_{1} + \ii\tilde{k}_{2} & 0 \\
 0 & \tilde{k}_{1} - \ii\tilde{k}_{2} & \tilde{k}_{0} - m & 0 \\
\hlineB{4}
\tilde{k}_{1} + \ii\tilde{k}_{2} & 0 & 0 & \tilde{k}_{0} - m
\end{array}\right).
\end{eqnarray}
Then, the electric conductivity $\sigma_{p}^{\rm{K}}(\qq, \Delta_{s}^{\eta})$ for each Dirac cone defined by the two-spinor hamiltonian of \Eq{Eq_Hamiltonian_spinor2} can be obtained from the results of \cite{non-local_Graphene_Lilia_Pablo}, also given in Appendix \ref{NonLocalKuboAppendix}. To obtain each electric conductivity term of Graphene $\sigma_{p}(\qq)$ (where $p\in\{L,T,H,S\}$), we must sum the contribution of the four Dirac cones (two cones due to the factorization of \Eq{Dirac_H} shown in \Eq{Dirac_Decomposition}, each one counted two times because of the spin degeneracy $g_{s}=2$), taking into account their respective Dirac masses $\Delta_{s}^{\eta}$ (which can be positive or negative):
\begin{eqnarray}
\sigma_{p}^{\rm{K}}(\qq) = \sum_{\eta=\pm}\sum_{s=\pm}\sigma_{p}^{\rm{K}}(\qq, \Delta_{s}^{\eta}).
\end{eqnarray}
From the decomposition shown in \Eq{Dirac_Decomposition} of $\hat{\mathcal{H}}_{s}^{D}(\bm{k})$ into two 2-spinor Hamiltonians $\hat{H}_{s}^{\eta}(\bm{k})$ of the form of \Eq{Eq_Hamiltonian_spinor2} with Dirac masses $\Delta_{s}^{\eta} = \eta m$ (rotate the 2-spinor Hamiltonian obtained from the first and fourth rows and columns $\pi/2$ rads), the Chern number of the studied model of graphene is
\begin{eqnarray}
C = \sum_{\eta=\pm}\sum_{s=\pm}\sgn(\eta m) = 0.
\end{eqnarray}
Therefore, graphene with the induced mass studied here is topologically trivial, and there is not any Hall electric conductivity ($\sigma_{H}^{\rm{K}} = 0$). This analysis is consistent with the velocity-velocity correlators given in \cite{non-local_Graphene_Lilia_Pablo}, and in \Eq{Z/N=<vv>/(k0-xi)(k0+q0-xi)}.

Due to the requirements of causality and realism, $\sigma_{\mu\nu}^{\rm{K}}$ does not have poles for $\omega$ in the upper complex plane, and \Eq{General_Kubo_Formula} is valid for all complex-values frequency with positive imaginary part simply promoting $\omega\in\mathbb{R}$ as a complex variable $\omega\in\mathbb{C}^{+}$. In \cite{non-local_Graphene_Lilia_Pablo}, it was proven that the spatial components $\sigma_{ij}^{\rm{K}}$ of the electric conductivity tensor can be conveniently expressed by separating between longitudinal $\sigma_{L}^{\rm{K}}$, transverse $\sigma_{T}^{\rm{K}}$, and Hall $\sigma_{H}^{\rm{K}}$ contributions \cite{Fialkovsky2011}\cite{non-local_Graphene_Lilia_Pablo}\cite{Zeitlin1995}\cite{Dorey1992}  (\Eq{General_non_local_conductivity_form}. The explicit analytical form of those three functions for real and complex frequencies in the zero temperature limit can be found in \cite{non-local_Graphene_Lilia_Pablo} and in the appendix \ref{NonLocalKuboAppendix}. To obtain similar results for finite temperatures, we should apply the Maldague formula \cite{Giuliani}\cite{Maldague}
\begin{eqnarray}\label{Maldague_Formula}
\sigma_{ij}^{\rm{K}}(\qq, \mu, T) = \int_{-\infty}^{\infty}d\text{M}\dfrac{\sigma_{ij}^{\rm{K}}(\qq, \text{M}, 0)}{4 k_{\rm B}T\cosh^{2}\left(\frac{\text{M}-\mu}{2 k_{\rm B}T}\right)} ,
\end{eqnarray}
where $\sigma_{ij}^{\rm{K}}(\qq,\mu,0)$ is the zero-temperature electric conductivity result. This is the more general formula for the linear non-local electric conductivity based on the linearized tight-binding model with a constant dissipation time parameter $\tau = \Gamma^{-1}$, and it is completely equivalent to \Eq{General_Kubo_Formula}. To go beyond this result, the full tight-binding model of graphene should be used \cite{Link2016}\cite{Lilia_full_graphene_Conductivity} instead of the linear approximation, a deep more detailed study of the effects of the different interactions in electronic quasiparticle spectrum \cite{Ostrovsky2006}\cite{Kashuba2008}\cite{Fritz2008}\cite{TeberPhDThesis2018} or more detailed ab-initio models \cite{Link2016}\cite{PhysRevB.103.125421}\cite{Arxivwang2024photon} should be considered.

\section{Local limit of the Kubo electric conductivity}\label{Section_yes_local_Kubo}
A very useful limiting case of the general non-local Kubo electric conductivity of \Eq{General_Kubo_Formula} is its local limit $\bm{q}_{\surf}\to\bm{0}$. Remarkably, a simple local electric conductivity expression valid for all temperature $T$, chemical potential $\mu$, and mass gap $\Delta$ has been derived \cite{non-local_Graphene_Lilia_Pablo}\cite{Rodriguez-Lopez2017} by applying \Eq{Maldague_Formula} to 
\begin{eqnarray}\label{local_sigma_realw}
\sigma_{xx}^{\rm{L}}(\omega,\bm{0}, \mu, 0) & = & \ii\frac{\sigma_{0}}{\pi} 
\Bigg[ \frac{\mu^{2} - \Delta^{2}}{\abs{\mu}}\frac{1}{\Omega}\Theta\left(\abs{\mu} - \abs{\Delta}\right) \nonumber\\
&+&  \frac{\Delta^{2}}{M\Omega} - \frac{\Omega^{2}+4\Delta^2}{2\ii\Omega^2}   \tan^{-1}\left(\frac{\ii\Omega}{2M}\right) \Bigg], \cr
\sigma_{xy}^{\rm{L}} (\omega,\bm{0}, \mu, 0) &=&  \frac{2\sigma_{0}}{\pi}\frac{\eta\Delta}{\ii\Omega}\tan^{-1}\left(\frac{\ii\Omega}{2M}\right),
\end{eqnarray}
where $\sigma_{0}=\alpha c/4$ is the universal electric conductivity of graphene ($\alpha = \frac{e^{2}}{\hbar c}$ is the fine structure constant), $\Omega = \hbar\omega + \ii\hbar\Gamma$, $M = \text{Max}\left[\abs{\Delta}, \abs{\mu}\right]$ and $\Theta(x)$ is the Heaviside Theta function. These results are per Dirac cone and are consistent with those found by other researchers \cite{Fialkovsky2011}\cite{Fialkovsky2012}\cite{Klimchitskaya2018}\cite{Gusynin2007b}\cite{Falkovsky2007}\cite{Gusynin2006}\cite{Gusynin2009}\cite{Fialkovsky2008}\cite{Bordag2009}\cite{Ludwig1994}\cite{MacDonald2006}\cite{Falkovsky2007b}\cite{PhysRevB.88.045442}\cite{WangKong2010}\cite{WangKong2011}. The first term in $\sigma_{xx}^{\rm{L}}$ corresponds to intraband transitions. The last two terms correspond to interband transitions. Note that, in the local limit $\bm{q}_{\surf}=\bm{0}$ one obtains $\sigma_{xx}^{\rm{L}}(\omega,\bm{0})=\sigma_{yy}^{\rm{L}}(\omega,\bm{0})=\sigma_{L}^{\rm{L}}(\omega,\bm{0})=\sigma_{T}^{\rm{L}}(\omega,\bm{0})$, and $\sigma_{xy}^{\rm{L}}(\omega,\bm{0})=-\sigma_{yx}^{\rm{L}}(\omega,\bm{0})=\sigma_{H}^{\rm{L}}(\omega, \bm{0})$. 

This model serves to model the local electric conductivity of graphene with mass and any other 2D Dirac cones, like the surface states of a 3D topological insulator \cite{PabloTI} and Chern Insulators \cite{PabloCI}. By summing the contribution of several different 2D Dirac cones with $\Delta\neq 0$, non-trivial topological states with non-zero Chern number can be studied \cite{Rodriguez-Lopez2017}\cite{PabloCI}.

In this section, we show that this local limit, in the specific case of $\Delta=0$, exactly reproduces the well-known electric conductivity expression derived by Falkovsky \cite{Fialkovsky2008}.
To this end, we need to calculate the $\Delta\to0$ limit of the sum of the contribution of 4 Dirac cones, then
\begin{eqnarray}\label{local_sigma_Delta0_realw}
\sigma_{xx}^{\rm{L}}(\omega,\bm{q}_{\surf} = \bm{0},T=0,\Delta=0) & = & \sigma^{\text{intra}}_{xx}(\omega) + \sigma^{\text{inter}}_{xx}(\omega),\nonumber\\
\sigma^{\text{intra}}_{xx}(\omega) & = & \dfrac{\alpha c}{\pi}\dfrac{\abs{\mu}}{-\ii\Omega},\label{local_intra_T=0}\nonumber\\
\sigma^{\text{inter}}_{xx}(\omega) & = & \dfrac{\alpha c}{2\pi}\tan^{-1}\left(\dfrac{- \ii\Omega}{2\abs{\mu}}\right),\nonumber\\
\sigma_{xy}^{\rm{L}}(\omega,\bm{q}_{\surf} = \bm{0},T=0,\Delta=0) & = & 0.
\end{eqnarray}
By applying the Maldague formula (\Eq{Maldague_Formula}) to this result with $\Delta = 0$, in the non-dissipative limit for the interband term and for finite temperatures, the well-known result of Falkovsky \cite{Gusynin2009}\cite{Fialkovsky2008} is obtained as
\begin{eqnarray}\label{Falkovsky_conductivity}
\sigma^{\rm{F}}_{xx}(\omega, \Gamma, T, \mu) & = & \phantom{+} \sigma^{\rm{F},\text{intra}}_{xx}(\omega, \Gamma, T, \mu)\nonumber\\
& & + \sigma^{\rm{F},\text{inter}}_{xx}(\omega, 0, T, \mu),
\end{eqnarray}
\begin{eqnarray}
\sigma^{\rm{F},\text{intra}}_{xx}(\omega, \Gamma, T, \mu) & = & \dfrac{1}{\pi\hbar}\dfrac{2\ii\alpha c k_{\rm B}T}{\omega + \ii\Gamma}
\ln\left[2\cosh\left(\frac{\mu}{2k_{\rm B}T}\right)\right],\nonumber\\
\sigma^{\rm{F},\text{inter}}_{xx}(\omega, 0, T, \mu) & = & \dfrac{\alpha c}{4}G\left(\Abs{\dfrac{\hbar\omega}{2}}\right)\\
& + & \ii \dfrac{\alpha c}{4}\dfrac{4\hbar\omega}{\pi}\int_{0}^{\infty}\dd\xi\frac{G(\xi) - G\left(\Abs{\tfrac{\hbar\omega}{2}}\right)}{(\hbar\omega)^{2} - 4\xi^{2}},\nonumber
\end{eqnarray}
with
\begin{eqnarray}\label{Def_G}
G\left(\epsilon\right)
& = & \dfrac{\sinh\left(\beta\epsilon\right)}{\cosh\left(\beta\mu\right) + \cosh\left(\beta\epsilon\right)}.
\end{eqnarray}
We have the absolute value in the preceding formula because we also handle negative real frequencies, while in \cite{Fialkovsky2008} the result for only positive frequencies was derived. The derivation of this result for real and imaginary frequencies is given in the appendix \ref{LocalKuboAppendix}.

\section{On the definition of the electric conductivity as available in current literature}\label{Section_QFT_Model}
In literature it exists QFT model for the non-local electric conductivity, also based on the Polarization operator of a $2D+1$ Dirac Hamiltonian \cite{Fialkovsky2016}\cite{Fialkovsky2012}\cite{Klimchitskaya2018}\cite{universe6090150}\cite{PRL_Mohideen}\cite{Bordag2015}\cite{Bimonte2017}, but predicting a different result for electric conductivity if compared with the QFT model we just derived in \Sect{Section_non_local_Kubo}. That previous QFT model has been extensively used in the context of the Casimir effect, compared to experimental results \cite{PRL_Mohideen}, and it has been proposed as intrinsically more fundamental than the Kubo one since based on ''first principles'' and being ''not phenomenological'' \cite{Mostdispersionrelations,MostepanenkoReflConfPol}. Due to a claimed coherence with the Casimir-Lifshitz theory, has also been suggested as model to modify the well-known dielectric function of metals itself \cite{MostepanenkoReflConfPol}. This section shows that the previous QFT model has several problems. First, it does not include unavoidable losses from the inelastic interactions of electronic quasiparticles with different objects always present in real samples, such as phonons, scattering centers, lattice dislocations, and non-linear interactions, as discussed in Subsec.~\ref{Effect_interactions}. To add such losses at this level of theory, it is enough to introduce a constant parameter $\Gamma=\tau^{-1}$ as the inverse of the mean lifetime of the electron quasiparticle, which is an experimentally measured quantity (See Subsect. \ref{Effect_interactions}). 
In addition to that problem, the electric conductivity derived in the QFT model predicts nonphysical features like an intrinsic plasma behavior that cannot be cured even by adding losses. We explain the origin of that pathology and show that standard regularization of this model (as done in \Sect{Section_EM_response}) makes the QFT model identical to the Kubo model we derived in the previous section \Sect{Section_non_local_Kubo}.\\
The starting point is the linear relationship of the induced electric current $J_{\mu}(\omega, \bm{q})$ with the potential vector $A^{\nu}(\omega, \bm{q})$ given in \Eq{JdependsonA}
\begin{eqnarray}
\mean{ J_{\mu}(\omega, \bm{q}) } = - \Pi_{\mu\nu}(\omega,\bm{q})A^{\nu}(\omega, \bm{q}).
\end{eqnarray}
To connect this result with the microscopic Ohm's Law \cite{Bordag2015b}\cite{Klimchitskaya2018}
\begin{eqnarray}
\mean{ J_{\mu}(\omega,\bm{q}) } = \sigma_{\mu\nu}(\qq)E^{\nu}(\omega,\bm{q}),
\end{eqnarray}
where $\sigma_{\mu\nu}(\qq)$ is the electric conductivity operator, and $E^{\mu}(\omega,\bm{q}) = \ii\omega A^{\mu}(\omega,\bm{q})$ is the electric field, the relation between $\Pi_{\mu\nu}(\qq)$ and $\sigma_{\mu\nu}(\qq)$ is assumed to be \cite{Fialkovsky2016}\cite{Fialkovsky2012}\cite{Bordag2015b}\cite{Klimchitskaya2018}\cite{universe6090150}\cite{Klimchitskaya2016}\cite{Klimchitskaya2016b}\cite{Klimchitskaya2017}
\begin{eqnarray}\label{Relation_Polarization_conductivity}
\sigma_{\mu\nu}^{\rm{NR}}(\qq) = \dfrac{\Pi_{\mu\nu}(\qq)}{-\ii\omega},
\end{eqnarray}
with the Polarization operator defined in \Eq{def:Polarization_Operator} using the covariant action and covariant Dirac Hamiltonian given in \Eqs{Weyl_covariant_Dirac_Action} and \Eq{Weyl_H}.
Note that, contrary to the result obtained in \Eq{j_eq10_maintext} and in \Sect{KubosigmaderivationAppendix}, there is no additional regularization term $\dlim_{\omega\to0}\Pi_{\mu\nu}(\omega,\bm{q}_{\surf})$ \eq{True_Relation_Polarization_conductivity_maintext}. Remarkably, we show that this absence of this regularization term makes the corresponding transversal electric conductivity critically different from the result of the Kubo formalism.

%with the Polarization operator obtained in \Eq{General_Polarization_Operator}, is that, instead of using the expression of the Green function as an expansion on eigenfunctions given in \Eq{Electric_Green0_function}, 

It is worth stressing that, by looking at the Polarization operator, the main formal difference is that the Kubo expression uses the Green function as an expansion on eigenfunctions given in \Eq{Electric_Green0_function}. In contrast, the QFT model uses a covariant form of the Green function of the Dirac Hamiltonian \Eq{Weyl_covariant_Dirac_Green_function}. This is the reason why apparently the results obtained in \cite{Fialkovsky2011}\cite{Fialkovsky2016}\cite{Fialkovsky2012}\cite{universe6090150}\cite{PRL_Mohideen}\cite{Bordag2015}\cite{Bimonte2017}\cite{Klimchitskaya2020}\cite{Bordag2009} look completely different to the ones obtained by the use of the Kubo formula. 

In this section: in subsection \ref{sectens} (i), we show that the two Polarization tensors used in the Kubo and in the available QFT theory, respectively, are indeed exactly identical\NEW{;} (ii) we derive explicitly the Polarization operator $\Pi_{\mu\nu}$ of the available QFT theory and express it in terms of Longitudinal and Transversal parts, and in subsections \ref{subsect_analitical_diff_Kubo_QFT} and \ref{Section_Numerical_comparison} we analyze, analytically and numerically, the differences between the Kubo electric conductivity of the electric conductivity of the available QFT result. We show that when the regularization term $\dlim_{\omega\to0}\Pi_{\mu\nu}(\omega,\bm{q}_{\surf})$ is zero (which is the case of the local limit and for the non-local longitudinal component) the two electric conductivities are identical, while when $\dlim_{\omega\to0}\Pi_{\mu\nu}(\omega,\bm{q}_{\surf})$ is different from zero (it is the case for the non-local transversal part) the available polarization-based electric conductivity is drastically different from the Kubo one, hence it shows pathological plasma-like behavior.

\subsection{Comparison of the Polarization tensors \label{sectens}}
Here, we calculate the QFT Polarization operator and compare it with the one used in the Kubo formalism.

By using the general definition of the current operator (\Eq{Current_Operator_definition}), we obtain that
\begin{eqnarray}%\label{Current_Operator_Weyl_covariant}
\hat{J}^{\mu}(\kk)
& = & - \dfrac{\partial\hat{\mathcal{H}}_{s}^{W}(k_{\alpha} + e A_{\alpha}(\kk))}{\partial A_{\mu}(\kk)}\nonumber\\
& = & - e( \gamma^{0}, v_{F}\gamma^{1}, v_{F}\gamma^{2} )\nonumber\\
& = & - \frac{e}{\hbar}( \tilde{\gamma}^{0}, \tilde{\gamma}^{1}, \tilde{\gamma}^{2} ).
\end{eqnarray}
Inserting this result into \Eq{def:Polarization_Operator}, and using the covariant Green function (\Eq{Weyl_covariant_Dirac_Green_function}), which is diagonal in spin, we get for the QFT Polarization:
\begin{eqnarray}\label{def:Covariant_Polarization_Operator}
\Pi_{\mu\nu}(\qq) = g_{s}\frac{e^{2}}{\hbar^{2}}\dfrac{-\ii}{\hbar}\int_{\kk}\Tr{ \mathcal{G}_{0}^{W}(\kk)\tilde{\gamma}_{\mu}\mathcal{G}_{0}^{W}(\kk+\qq)\tilde{\gamma}_{\nu} },
\end{eqnarray}
where $g_{s}=2$ is the spin degeneration, $e^{2} = \alpha c\hbar$, and $\tilde{\gamma}_{\mu} = \hbar(\gamma_{0}, v_{F}\bm{\gamma})$ (see \Tab{Notation}). After carrying out the trace, one obtains
\begin{eqnarray}\label{def:Covariant_Polarization_Operator_definitions}
\Pi_{\mu\nu}(\qq) = g_{s}e^{2}\dfrac{-\ii}{\hbar}\int_{k}\dfrac{Z_{\mu\nu}(K_{\alpha},q_{\alpha})}{\left[ K^{\rho}K_{\rho} - m^{2} \right]\left[ S^{\zeta}S_{\zeta} - m^{2} \right]},
\end{eqnarray}
where $S_{\zeta} = K_{\zeta} + Q_{\zeta}$ (see \Tab{Notation}), and $Z_{\mu\nu}(K_{\alpha},q_{\alpha})$ is obtained as \cite{Bordag2015}
\begin{widetext}
\begin{eqnarray}\label{Numerator_Polarization_Operator_Mostepanenko}
Z_{\mu\nu}(K_{\alpha},q_{\alpha}) = 4\left(\begin{array}{ccc}
\tilde{k}_{\mu}\tilde{s}_{\mu} + \tilde{k}_{1}\tilde{s}_{1} + \tilde{k}_{2}\tilde{s}_{2} + m^{2} & - v_{F}(\tilde{k}_{\mu}\tilde{s}_{1} + \tilde{k}_{1}\tilde{s}_{\mu}) & 
- v_{F}(\tilde{k}_{\mu}\tilde{s}_{2} + \tilde{k}_{2}\tilde{s}_{\mu}) \\
- v_{F}(\tilde{k}_{\mu}\tilde{s}_{1} + \tilde{k}_{1}\tilde{s}_{\mu}) & v_{F}^{2}\left( \tilde{k}_{\mu}\tilde{s}_{\mu} + \tilde{k}_{1}\tilde{s}_{1} - \tilde{k}_{2}\tilde{s}_{2} - m^{2} \right) & v_{F}^{2}(\tilde{k}_{1}\tilde{s}_{2} + \tilde{k}_{2}\tilde{s}_{1}) \\
- v_{F}(\tilde{k}_{\mu}\tilde{s}_{2} + \tilde{k}_{2}\tilde{s}_{\mu}) & v_{F}^{2}(\tilde{k}_{1}\tilde{s}_{2} + \tilde{k}_{2}\tilde{s}_{1}) & v_{F}^{2}\left(\tilde{k}_{\mu}\tilde{s}_{\mu} + \tilde{k}_{2}\tilde{s}_{2} - \tilde{k}_{1}\tilde{s}_{1} - m^{2}\right)
\end{array}
\right),
\end{eqnarray}
where $\tilde{k}_{\mu} = \tilde{k}_{0} + \mu$, $\tilde{s}_{\mu} = \tilde{s}_{0} + \mu$, $\tilde{k}_{i} = \hbar v_{F}k_{i}$ and $\tilde{s}_{i} = \hbar v_{F}s_{i}$ (see \Tab{Notation}). From this result, it is easy to obtain \cite{Pyatkovskiy_2009}\cite{Bordag2015b}
\begin{eqnarray}\label{Z/N=<vv>/(k0-xi)(k0+q0-xi)}
\dfrac{Z_{\mu\nu}(K_{\alpha},q_{\alpha})}{N(K_{\alpha},q_{\alpha})}
& = & \sum_{\lambda,\lambda'}\left(\begin{array}{ccc}
1 + \frac{\tilde{k}_{1}\tilde{s}_{1} + \tilde{k}_{2}\tilde{s}_{2} + m^{2}}{\epsilon_{\bm{k}}^{\lambda}\epsilon_{\bm{s}}^{\lambda'}} & v_{F}\left(\frac{\tilde{k}_{1}}{\epsilon_{\bm{k}}^{\lambda}} + \frac{\tilde{s}_{1}}{\epsilon_{\bm{s}}^{\lambda'}}\right) & v_{F}\left(\frac{\tilde{k}_{2}}{\epsilon_{\bm{k}}^{\lambda}} + \frac{\tilde{s}_{2}}{\epsilon_{\bm{s}}^{\lambda'}}\right)\\
v_{F}\left(\frac{\tilde{k}_{1}}{\epsilon_{\bm{k}}^{\lambda}} + \frac{\tilde{s}_{1}}{\epsilon_{\bm{s}}^{\lambda'}}\right) & v_{F}^{2}\left[ 1 + \frac{\tilde{k}_{1}\tilde{s}_{1} - \tilde{k}_{2}\tilde{s}_{2} - m^{2}}{\epsilon_{\bm{k}}^{\lambda}\epsilon_{\bm{s}}^{\lambda'}}\right] & v_{F}^{2}\frac{\tilde{k}_{1}\tilde{s}_{2} + \tilde{k}_{2}\tilde{s}_{1}}{\epsilon_{\bm{k}}^{\lambda}\epsilon_{\bm{s}}^{\lambda'}}\\
v_{F}\left(\frac{\tilde{k}_{2}}{\epsilon_{\bm{k}}^{\lambda}} + \frac{\tilde{s}_{2}}{\epsilon_{\bm{s}}^{\lambda'}}\right) & v_{F}^{2}\frac{\tilde{k}_{1}\tilde{s}_{2} + \tilde{k}_{2}\tilde{s}_{1}}{\epsilon_{\bm{k}}^{\lambda}\epsilon_{\bm{s}}^{\lambda'}} & v_{F}^{2}\left[ 1 + \frac{\tilde{k}_{2}\tilde{s}_{2} - \tilde{k}_{1}\tilde{s}_{1} - m^{2}}{\epsilon_{\bm{k}}^{\lambda}\epsilon_{\bm{s}}^{\lambda'}}\right]
\end{array}\right)
\frac{1}{\tilde{k}_{0} + \xi_{\bm{k}}^{\lambda}}\frac{1}{\tilde{s}_{0} + \xi_{\bm{s}}^{\lambda'}}\nonumber\\
& = & 2\sum_{\lambda,\lambda'}
\frac{\bra{u_{\bm{k}}^{\lambda}} \hat{v}_{\mu} \ket{u_{\bm{k}+\bm{q}}^{\lambda'}} \bra{u_{\bm{k}+\bm{q}}^{\lambda'}} \hat{v}_{\nu} \ket{u_{\bm{k}}^{\lambda}}}{\left[ \tilde{k}_{0} -\xi_{\bm{k}}^{\lambda}\right]\left[\tilde{s}_{0} - \xi_{\bm{s}}^{\lambda'}\right]},
\end{eqnarray}
\end{widetext}
where $N(K_{\alpha},q_{\alpha}) = \left[ K^{\rho}K_{\rho} - m^{2} \right]\left[ S^{\zeta}S_{\zeta} - m^{2} \right]$ is the denominator of \Eq{def:Covariant_Polarization_Operator_definitions}, $\ket{u_{\bm{k}}^{\lambda}}$ are the eigenfunctions of the spinor Hamiltonian \eqref{Eq_Hamiltonian_spinor} with $\Delta\to m$ defined in \Eq{Eigenfunctions} and $\xi_{\bm{k}}^{\lambda}$ are their corresponding eigenvalues, defined in \Eq{Eigensystem} with \Eq{Eq_Energy_Hamiltonian_spinor}. The product of the velocity correlators in \Eq{Z/N=<vv>/(k0-xi)(k0+q0-xi)} is also identical to the one obtained in \cite{non-local_Graphene_Lilia_Pablo}, with $m$ instead of $\Delta$. This explicitly prove that that, even if the QFT formalism is derived from the covariant Green's function, the final Polarization operator for graphene is exactly identical to the one obtained with the Kubo formalism (for the $\Gamma\to0$ limit), which use the Green's function \Eq{Electric_Green0_function}. 

While the two methods use exactly the same Polarization operator $\Pi_{\mu\nu}(\qq)$, they differ in their definition of the electric conductivity: The Luttinger formula given in \Eq{True_Relation_Polarization_conductivity_maintext} and \Eq{General_Kubo_Formula} derived from the Kubo formalism is different from the non regularized electric conductivity \Eq{Relation_Polarization_conductivity} using \Eq{def:Covariant_Polarization_Operator_definitions} of the available QFT model. As a consequence, the Kubo electric conductivity derived from \Eq{General_Kubo_Formula} is regular for small frequencies with a finite DC electric conductivity, while the QFT electric conductivity obtained from \Eq{Relation_Polarization_conductivity} and \Eq{def:Covariant_Polarization_Operator_definitions} predicts an infinite DC electric conductivity. This divergent behavior comes from the infinite dissipation time used in the QFT model and a spurious plasma behavior originating in the interband transition of the non-regularized transversal electric conductivity part. These pathological behaviors can be solved by introducing losses and using the regularized expression for the electric conductivity.

In the rest of this subsection, we show that the expression \Eq{def:Covariant_Polarization_Operator_definitions} of the QFT Polarization tensor (we just shown being identical to the Kubo one) can be presented in exactly the same form as it appears in literature (the complete explicit calculation is done in Appendix \ref{Appendix_Derivation_Polarization_Mostepanenko}).

By applying the Matsubara formalism directly to the expression \Eq{def:Covariant_Polarization_Operator_definitions} and using \Eq{Numerator_Polarization_Operator_Mostepanenko} instead of \Eq{Z/N=<vv>/(k0-xi)(k0+q0-xi)}, we obtain the expression for the Polarization operator shown in \cite{Bordag2015} (see \Eq{def:Covariant_Polarization_Operator_definitions_Appendix} in Appendix \ref{Appendix_Derivation_Polarization_Mostepanenko}):
\begin{eqnarray}\label{Obtained_Polarization_Mostepanenko}
\Pi_{\mu\nu}(\qq) & = & g_{s}e^{2}\dint_{\bm{k}}\left[ 1 - N_{\mu}(\epsilon_{\bm{k}})\right]\nonumber\\& & \times
\sum_{\lambda=\pm}\dfrac{Z_{\mu\nu}(\epsilon_{\bm{k}}^{\lambda},\tilde{\bm{k}}_{\surf},\tilde{q}_{0},\tilde{\bm{q}}_{\surf})}{2\epsilon_{\bm{k}}\left( (\tilde{q}_{0} + \epsilon_{\bm{k}}^{\lambda})^{2} - \epsilon_{\bm{s}}^{2} \right)},
\end{eqnarray}
where we have defined (using \Eq{Def_G}):
\begin{eqnarray}\label{Def_NF_mu}
N_{\mu}(\epsilon) & = & \sum_{\eta=\pm}n_{F}(\epsilon + \eta\mu) = 1 - G(\epsilon).
\end{eqnarray}
Following the notation of \cite{Bordag2015}, the Polarization operator given in \Eq{def:Covariant_Polarization_Operator_definitions} can be written as
\begin{eqnarray}
\Pi_{\mu\nu}(\qq) = \Pi^{(0)}_{\mu\nu}(\qq) + \Delta_{T}\Pi_{\mu\nu}(\qq).
\end{eqnarray}
By construction, $\Pi^{(0)}_{\mu\nu}(\qq)$ is independent of the temperature and of the chemical potential $\mu$ \cite{Bordag2015}, and it can be interpreted as the interband contribution with $\mu=k_{B}T=0\eV$. Note that $\Pi^{(0)}_{\mu\nu}(\qq)$, as shown in \Eq{Obtained_Polarization_Mostepanenko}, has an ultraviolet divergence, which can be removed with a Pauli-Villars subtraction scheme \cite{Fialkovsky2011}, by a $1/N$ expansion \cite{Pisarski1984,Dorey1992}, by an explicit regularization of the ultraviolet divergent integral (Eq.~(50) of \cite{bordag2024}) or by solving the regular integral given in \Eq{Def_Polarization} \cite{Bordag2015b}\cite{non-local_Graphene_Lilia_Pablo}\cite{Wunsch_2006} instead (in this case this divergence is absent from the very beginning).
In \Eq{Pi_Split_in_L_T} of appendix \ref{Appendix_Derivation_Polarization_Mostepanenko}, we show that the polarization operator derived from \Eq{Obtained_Polarization_Mostepanenko} can be written in terms of its longitudinal and transversal parts as
\begin{eqnarray}\label{Pi_Split_in_L_T_main_text}
\Pi_{ij}(\qq) & = & \Pi_{L}(\qq)\frac{\tilde{q}_{i}\tilde{q}_{j}}{\tilde{q}_{\surf}^{2}} + \Pi_{T}(\qq)\left( \delta_{ij} - \frac{\tilde{q}_{i}\tilde{q}_{j}}{\tilde{q}_{\surf}^{2}} \right).
\end{eqnarray}
The longitudinal Polarization is obtained as (\Eq{App_Final_Pi_L} in Appendix \ref{Appendix_Derivation_Polarization_Mostepanenko})
\begin{eqnarray}\label{Final_Pi_L}
&\Pi_{L}(\qq)
= 2g_{s}e^{2}v_{F}^{2}\dfrac{\tilde{q}_{0}^{2}}{\tilde{q}_{\surf}^{2}}\dint_{\bm{k}}\dfrac{1}{2\epsilon_{\bm{k}}}\left[ 1 - N_{\mu}(\epsilon_{\bm{k}})\right]\nonumber
\\& \times
\dsum_{\lambda=\pm}\left[ 
1 + \dfrac{M_{00}(\tilde{q}_{0},\tilde{k}_{\surf},\tilde{q}_{\surf})}{ Q(\tilde{q}_{0},\tilde{k}_{\surf},\tilde{q}_{\surf}) + 2\tilde{\bm{k}}\cdot\tilde{\bm{q}} }\right],
\end{eqnarray}
with $\tilde{q}_{z}^{2} = \tilde{q}_{0}^{2} - \tilde{q}_{\surf}^{2}$ (see \Tab{Notation})
\begin{eqnarray}\label{Def_M_00_text}
M_{00}(\tilde{q}_{0},\tilde{\bm{k}}_{\surf},\tilde{\bm{q}}_{\surf}) & = & - \tilde{q}_{z}^{2} + 4\tilde{q}_{0}\epsilon_{\bm{k}}^{\lambda} + 4\epsilon_{\bm{k}}^{2},\\
Q(\tilde{q}_{0},\tilde{k}_{\surf},\tilde{q}_{\surf}) & = & - \tilde{q}_{z}^{2} - 2\tilde{q}_{0}\epsilon_{\bm{k}}^{\lambda},
\end{eqnarray}
while the transversal Polarization is (\Eq{App_Final_Pi_T} in Appendix \ref{Appendix_Derivation_Polarization_Mostepanenko})
\begin{eqnarray}\label{Final_Pi_T}
&\Pi_{T}(\qq)
= 2g_{s}e^{2}v_{F}^{2}\dfrac{\tilde{q}_{0}^{2}}{\tilde{q}_{\surf}^{2}}\dint_{\bm{k}}\dfrac{1}{2\epsilon_{\bm{k}}}\left[ 1 - N_{\mu}(\epsilon_{\bm{k}})\right]\nonumber\\
& \times
\dsum_{\lambda=\pm}\left[ 
1 + \dfrac{M_{00}(\tilde{q}_{0},\tilde{k}_{\surf},\tilde{q}_{\surf}) - 4\frac{\tilde{q}_{\surf}^{2}}{q_{0}^{2}}\left( \tilde{k}_{\surf}^{2} + \tilde{q}_{0}\epsilon_{\bm{k}}^{\lambda} \right)}{ Q(\tilde{q}_{0},\tilde{k}_{\surf},\tilde{q}_{\surf}) + 2\tilde{\bm{k}}_{\surf}\cdot\tilde{\bm{q}}_{\surf} }\right].
\end{eqnarray}
Here $\tilde{q}_{z} = \sqrt{ \tilde{q}_{0}^{2} - \tilde{q}_{\surf}^{2} }$ and $\delta = \frac{2\abs{m}}{-\ii\tilde{q}_{z}}$ for any complex frequency $\omega\in\mathbb{C}^{+}$ ($\tilde{q}_{0} = \hbar(\omega + \ii\Gamma)$), and
\begin{eqnarray}
\Psi(x) = 2\left[ x + \left( 1 - x^{2} \right)\tan^{-1}\left(\dfrac{1}{x}\right) \right].
\end{eqnarray}
The ultraviolet regularization of $\Pi^{(0)}_{\mu\nu}(\qq)$ in \Eq{Final_Pi_L} and \Eq{Final_Pi_T} lead to \cite{Klimchitskaya2016}\cite{PRL_Mohideen}\cite{Bordag2015}\cite{Bimonte2017}
\begin{eqnarray}
\Pi_{L}^{(0)}(\qq) = - g_{s}\ii\dfrac{\alpha c}{8\pi\hbar}\dfrac{\tilde{q}_{0}^{2}}{\tilde{q}_{z}}\Psi\left(\frac{2m}{-\ii\tilde{q}_{z}}\right),
\end{eqnarray}
\begin{eqnarray}\label{Final_Pi0_T}
\Pi_{T}^{(0)}(\qq) = - g_{s}\ii\dfrac{\alpha c}{8\pi\hbar}\tilde{q}_{z}\Psi\left(\frac{2m}{-\ii\tilde{q}_{z}}\right),
\end{eqnarray}
where $\alpha = \tfrac{e^{2}}{\hbar c}$. In the particular case of imaginary frequencies, we have $\tilde{q}_{0} = \ii\hbar\xi = \ii\Xi$, $\tilde{q}_{z} = \ii\sqrt{ \Xi^{2} + \tilde{q}_{\surf}^{2} } = \ii\tilde{\theta}_{z}$, $\gamma = \frac{\Xi}{\tilde{\theta}_{z}}$ and $\delta = \frac{2\abs{m}}{\tilde{\theta}_{z}}$  (see \Tab{Notation}). It is show in the appendix \ref{Appendix_Derivation_Polarization_Mostepanenko} that the integrals of $\Pi_{L}(\qq)$ given \Eq{Final_Pi_L} and of $\Pi_{T}(\qq)$ given in \Eq{Final_Pi_T} can be reduced to
\begin{eqnarray}\label{Final_Pi0_L0Im}
\Pi_{L}^{(0)}(\qq) = g_{s}\dfrac{\alpha c}{8\pi\hbar}\dfrac{\Xi^{2}}{\tilde{\theta}_{z}}\Psi(\delta),
\end{eqnarray}
\begin{eqnarray}\label{Delta_T_PI_00Imv6}
&\Delta_{T}\Pi_{L}(\qq)
= g_{s}\dfrac{\alpha c}{2\pi\hbar}\dfrac{\Xi^{2}}{\tilde{q}_{\surf}^{2}}\tilde{\theta}_{z}
\dint_{\delta}^{\infty}\dd u N_{\mu}\Big(\tilde{\theta}_{z}\frac{u}{2}\Big)\nonumber\\
& \times\left[ 
 1 - \Real{\dfrac{ 1 - u^{2} - 2\ii\gamma u}{\sqrt{ 1 - u^{2} - 2\ii\gamma u + (1 - \gamma^{2})\delta^{2} } }}\right],
\end{eqnarray}
\begin{eqnarray}\label{Final_Pi0_T0}
\Pi_{T}^{(0)}(\qq) = g_{s}\dfrac{\alpha c}{8\pi\hbar}\tilde{\theta}_{z}\Psi(\delta),
\end{eqnarray}
\begin{eqnarray}\label{Delta_T_PI_T}
&\Delta_{T}\Pi_{T}(\qq)
= - g_{s}\dfrac{\alpha c}{2\pi\hbar}\dfrac{\Xi^{2}}{\tilde{q}_{\surf}^{2}}\tilde{\theta}_{z}
\dint_{\delta}^{\infty}\dd u N_{\mu}\Big(\tilde{\theta}_{z}\frac{u}{2}\Big)\nonumber\\
& \times\left[ 
 1 - \Real{\dfrac{ \left( 1 + \ii\gamma^{-1}u\right)^{2} + \left( 
\gamma^{-2} - 1 \right)\delta^{2}}{\sqrt{ 1 - u^{2} - 2\ii\gamma u + (1 - \gamma^{2})\delta^{2} } }}\right].
\end{eqnarray}
These results exactly coincide with the results published in \cite{Klimchitskaya2016}\cite{Klimchitskaya2017}\cite{PRL_Mohideen}\cite{Bordag2015}\cite{Bimonte2017} in their appropriate limits. The associated electric conductivity can be derived from those formulas by using the definition
\begin{eqnarray}\label{Relation_Polarization_conductivityb}
\sigma_{\mu\nu}^{\rm{NR}}(\qq) = \dfrac{\Pi_{\mu\nu}(\qq)}{-\ii\omega} = \hbar\dfrac{\Pi_{\mu\nu}(\qq)}{-\ii\tilde{q}_{0}},
\end{eqnarray}

\subsection{Differences between the Kubo and the non-regularized QFT electric conductivity expressions}\label{subsect_analitical_diff_Kubo_QFT}

As we have just shown, the only difference between the Kubo and polarization-based model is the absence of the regularization term $\dlim_{\omega\to0}\Pi_{\mu\nu}(\omega,\bm{q})$ in the latter (once losses are also added to the QFT model). In order to compare the two electric conductivity models we need to explore the values of this term for the longitudinal $\Pi_{L}(\omega,\bm{q})$ and transversal $\Pi_{T}(\omega,\bm{q})$ components of the Polarization operator, both in the non-local and local regime $\bm{q}\to\bm{0}$.

The longitudinal Polarization $\Pi_{L}(\omega,\bm{q})$ is explicitly given in \Eq{UVRegularized_App_Final_Pi_L}, its zero frequency limit is zero $\dlim_{\omega\to0}\Pi_{L}(\omega,\bm{q})=0$ in both the local and non-local regions. In particular, it goes to zero as $\mathcal{O}(\omega^{2})$. We can conclude that the longitudinal part of the Kubo and the polarization-based electric conductivities are identical in the local and pure non-local regions.  

Concerning the transversal polarization $\Pi_{T}(\omega,\bm{q})$, its explicit expression is given in \Eq{UVRegularized_App_Final_Pi_T}. We can see that in the local regime, its zero frequency limit is zero $\dlim_{\omega\to0}\Pi_{T}(\omega,\bm{q}=\bm{0})=0$, while in the purely non-local regime $\bm{q}\neq\bm{0}$ this is not the case and $\dlim_{\omega\to0}\Pi_{T}(\omega,\bm{q}\neq\bm{0})\neq 0$. We can conclude that the transversal part of the Kubo and the polarization-based electric conductivities are identical in the local region. At the same time, they are different in the purely nonlocal region.

Hence, let us investigate the transversal electric conductivity in detail.

\begin{widetext}

\begin{figure}[H]
\centering
\includegraphics[width=0.325\linewidth]{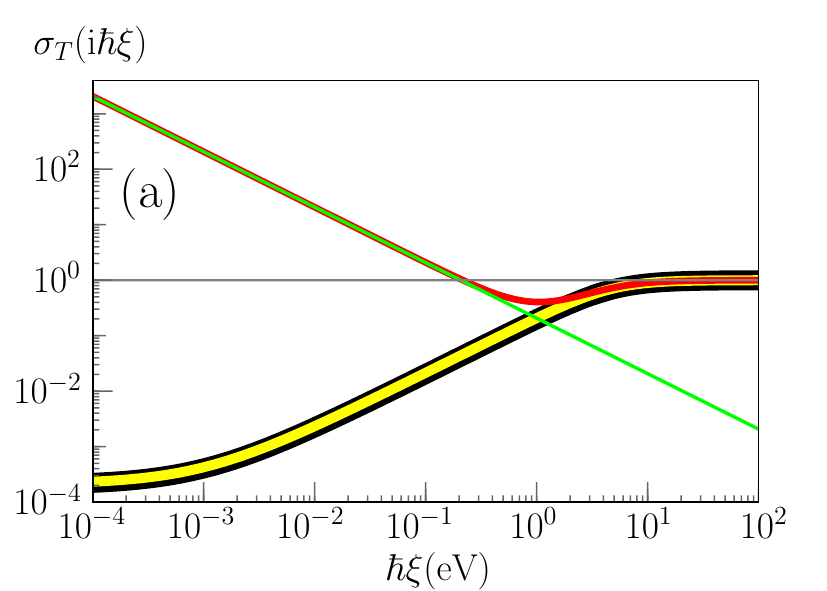}
\includegraphics[width=0.325\linewidth]{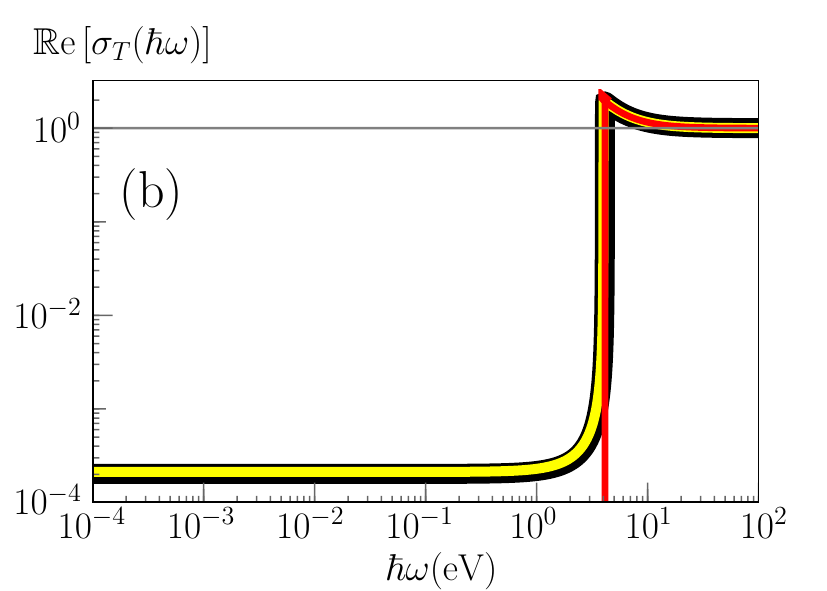}
\includegraphics[width=0.325\linewidth]{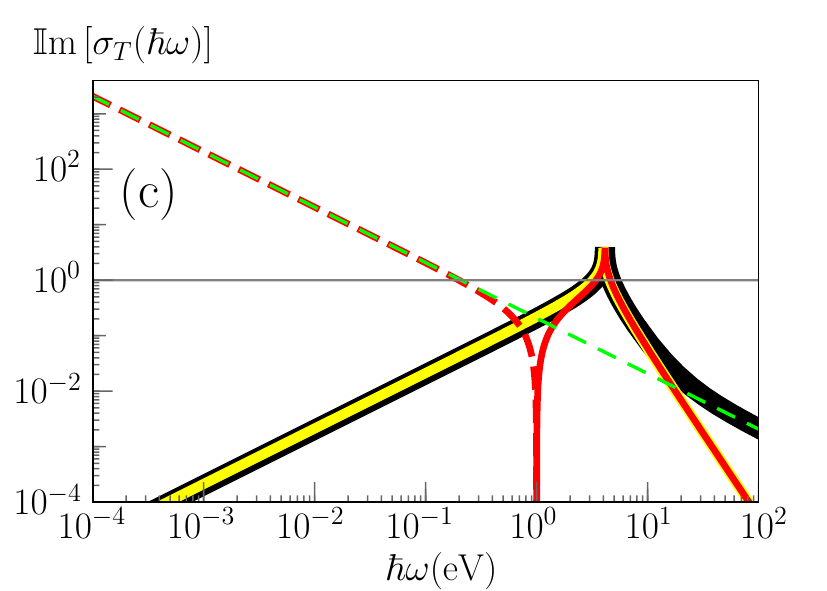}
\caption{ (Color online) Double logarithmic plots of the real part of the transversal electric conductivity $\sigma_{T}$ (in units of the universal electric conductivity of graphene $\sigma_{0} = \frac{\alpha c}{4}$) as a function of the imaginary frequency $\hbar\omega = \ii\hbar\xi$ in panel (a), the real part for real frequency $\hbar\omega$ in panel (b) and the imaginary part for real frequencies in (c), for the case with $k_{B}T = \mu = 0$, $\Delta = 2\eV$, $\tilde{q}_{\surf} = 1\eV$ and dissipation rate $\hbar\Gamma = 10^{-3}\eV$. The thick black curve is the non-local electric conductivity derived from the Kubo formula $\sigma_{T,0}^{\rm{K}}(\omega, \tilde{q}_{\surf})$, given in this limit by \Eq{Kubo_sigmaT_interband_mu0_T0}, the red curve is the non-local electric conductivity $\sigma_{T}^{(0),\rm{NR}}(\omega, \tilde{q}_{\surf})$ shown in \Eq{QFT_sigmaT_interband_mu0_T0}, the yellow curve is the local electric conductivity $\sigma_{T,0}^{\rm{L}}(\omega, \bm{0})$ shown in \Eq{local_sigma_realw} and the green curve is the $\hbar\xi\to0$ divergence of \Eq{QFT_sigmaT_interband_mu0_T0}, given in \Eq{smallhxi_divergence_QFT_sigmaT_interband_mu0_T0}. The dashed curves in panel (c) represent the positive imaginary parts of the electric conductivity, while the full curves represent the negative values.}
\label{fig:sigmaT_tq=1eV_T_0_mu_0}
\end{figure}

\end{widetext}

Taking into account that $\Pi_{T}^{(0)}(\qq)$ corresponds to the $\mu = k_{B}T = 0\eV$ case, it represents the interband electric conductivity. Therefore, the real part of this electric conductivity must be zero when $\hbar\Gamma\to0$ (an electron in the valence band has to jump to a hole place in the conduction band to conduct; therefore, a finite gap exists as long as valence and conduction bands do not touch). In \cite{non-local_Graphene_Lilia_Pablo} it was found that the interband electric conductivity can be written as
\begin{eqnarray}
\lim_{\hbar\Gamma\to0}\Real{\sigma_{T}^{\rm{K}}(\qq)} = f(\qq)\Theta\left(\hbar\omega - \sqrt{4\Delta^{2} + \tilde{q}_{\surf}^{2}}\right).
\end{eqnarray}
However, from the expression published in \cite{Bordag2015}, we find that
\begin{eqnarray}\label{QFT_sigmaT_interband_mu0_T0}
\sigma_{T}^{(0),\rm{NR}}(\omega, \tilde{q}_{\surf}) = \frac{\alpha c}{4\pi}\dfrac{\sqrt{ \tilde{q}_{\surf}^{2} - \hbar^{2}\omega^{2} }}{-\ii\hbar\omega}\Psi\left(\frac{2\Delta}{\sqrt{ \tilde{q}_{\surf}^{2} - \hbar^{2}\omega^{2} }}\right)
\end{eqnarray}
diverges at small frequencies as
\begin{eqnarray}\label{smallhxi_divergence_QFT_sigmaT_interband_mu0_T0}
\sigma_{T}^{(0),\rm{NR}}(\omega, \tilde{q}_{\surf}) = \frac{\alpha c}{4\pi}\dfrac{\tilde{q}_{\surf}}{-\ii\hbar\omega}\Psi\left(\frac{2\Delta}{\tilde{q}_{\surf}}\right) + \mathcal{O}\left[\hbar\omega\right].
\end{eqnarray}
with $\tilde{q}_{\surf} = \hbar v_{F}q_{\surf}$. Note that $\sigma_{T}^{(0),\rm{NR}}(\omega, \tilde{q}_{\surf})$ behaves as a plasma model without conduction electrons or holes. This result of the polarization-based theory would imply a dissipation-less electric current generated by an electric field, which is clearly unacceptable in normal materials. This is an explicit example of the necessity to regularize the electric conductivity operator in such a way the condition $\dlim_{\omega\to0}\Pi_{\mu\nu}(\qq)A^{\nu}(\omega,\bm{q}) = 0$ $\forall A^{\nu}(0,\bm{q})$ (i.e. for $E_{\nu}(\omega,\bm{q}) = 0$) is fulfilled. Therefore, we use the $\mu=0$ interband transversal electric conductivity derived in \cite{non-local_Graphene_Lilia_Pablo} and given by
\begin{eqnarray}\label{Kubo_sigmaT_interband_mu0_T0}
\sigma_{T,0}^{\rm{K}}(\qq) = \frac{\alpha c}{4\pi}\frac{1}{-\ii\tilde{q}_{0}}
\left[ \tilde{\theta}_{z}\Psi(\delta) - \tilde{q}_{\surf}\Psi(x) \right],
\end{eqnarray}
with $\tilde{\theta}_{z} = \sqrt{ \tilde{q}_{\surf}^{2} - \tilde{q}_{0}^{2} }$, $\delta = \frac{2\abs{\Delta}}{\tilde{\theta}_{z}}$ and $x = \frac{2\abs{\Delta}}{\tilde{q}_{\surf}}$ (see \Tab{Notation}). Note that this expression corresponds to the explicit elimination of the $\omega\to0$ limit to the Polarization operator $\dlim_{\omega\to0}\Pi_{\mu\nu}(\omega,\bm{q}_{\surf})A^{\nu}(\omega,\bm{q}_{\surf}) = 0$ for constant static $A^{\nu}(0,\bm{q}_{\surf})$ \cite{Annett:730995}\cite{Khalilov2015}. This is the difference between using the polarization-based expression of the electric conductivity operator given in \Eq{QFTconductivitydef} and the regularized one used in \Eq{True_Relation_Polarization_conductivity_maintext}. In \Fig{fig:sigmaT_tq=1eV_T_0_mu_0} can be observed, for real ($\hbar\omega$) and imaginary ($\hbar\omega = \ii\hbar\xi$) frequencies the divergence of \Eq{QFT_sigmaT_interband_mu0_T0} and the convergence of \Eq{Kubo_sigmaT_interband_mu0_T0} to the local limit given in \Eq{local_sigma_realw}. For imaginary frequencies (\Fig{fig:sigmaT_tq=1eV_T_0_mu_0}a) and for the imaginary part of the electric conductivity for real frequencies (\Fig{fig:sigmaT_tq=1eV_T_0_mu_0}c), the appearance of the plasma-like peak is evident, for the real part of the electric conductivity at real frequencies, the plasma-like peak is a Dirac delta. It cannot be observed in the figure (\Fig{fig:sigmaT_tq=1eV_T_0_mu_0}b). Interestingly, when $\tilde{q}_{\surf}\to0$, the plasma divergence disappears. %In \Sect{Section_Numerical_comparison}, we will see that the same problem appears with the thermal contribution of the Polarization operator of the QFT model.

Here is clearly shown the importance of using the regularized electric conductivity that requires the subtraction of the diverging term $\dlim_{\omega\to0}\Pi_{\mu\nu}(\omega,\bm{q})$ for electric field induced currents. In the previous QFT model, the lack of regularization at zero frequency opens the possibility of obtaining spurious non-physical plasma electric conductivities without dissipation ($\Gamma=0$) in graphene. This result is not cured even if dissipation ($\Gamma > 0$) is artificially added to the model. This plasma-divergence happens for the transversal electric conductivity $\sigma_{T}^{(0),\rm{NR}}(\xi, \tilde{q}_{\surf})$ in \Eq{QFT_sigmaT_interband_mu0_T0}, while the longitudinal electric conductivity is saved from this divergence because it scales with $\omega^{2}$ at small $\omega$ (See \Eq{sigma_L=f_sigma_00}) hence the term to subtract in the regularized prescription is zero.
It is worth noting that this divergence disappears in the local limit, so none of the models studied here have this non-physical dissipation-less plasma current in their local limit. Finally, by construction, the non-local Kubo model does not have this divergence.

\subsection{Numerical comparison}\label{Section_Numerical_comparison}
\begin{figure*}%[H]
\includegraphics[width=0.24\linewidth]{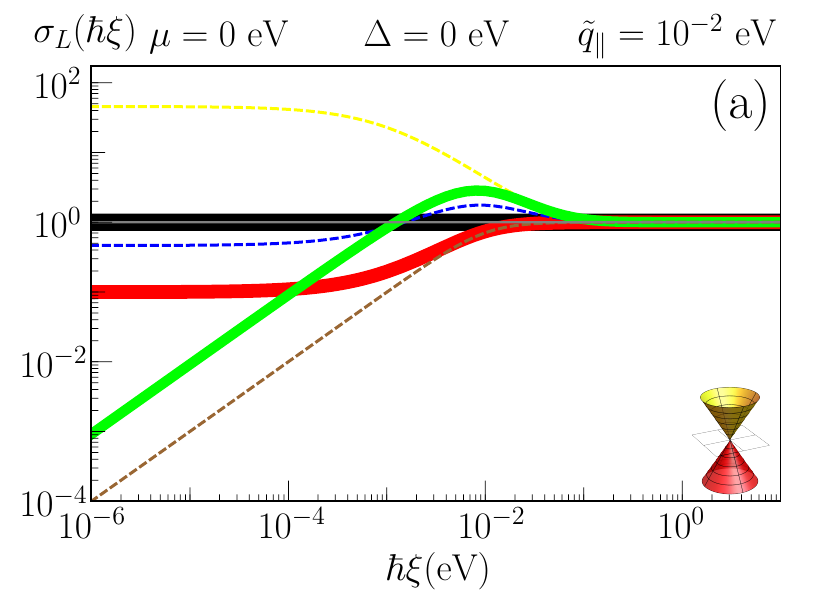}
\includegraphics[width=0.24\linewidth]{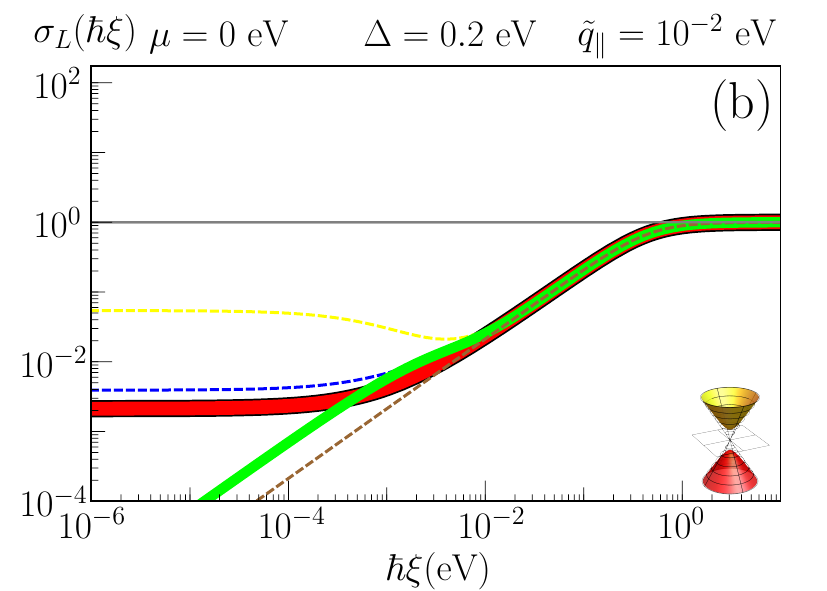}
\includegraphics[width=0.24\linewidth]{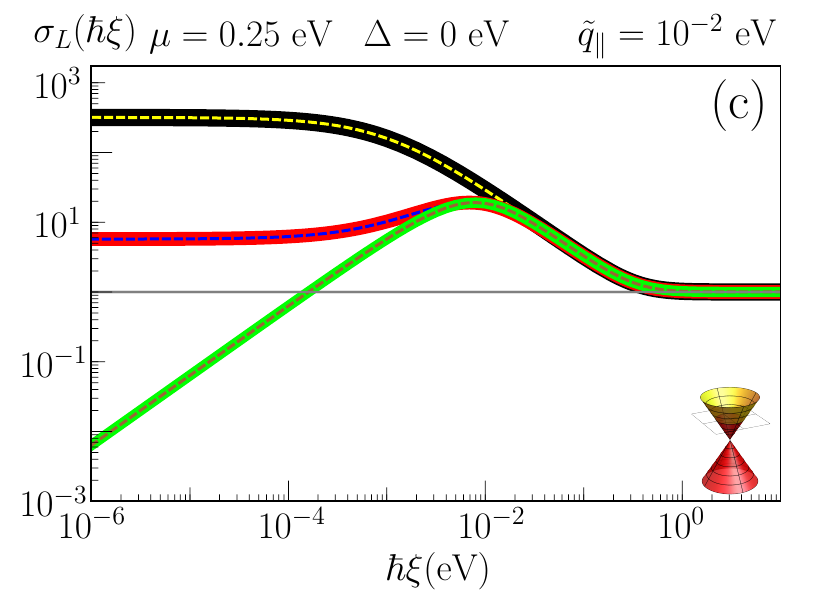}
\includegraphics[width=0.24\linewidth]{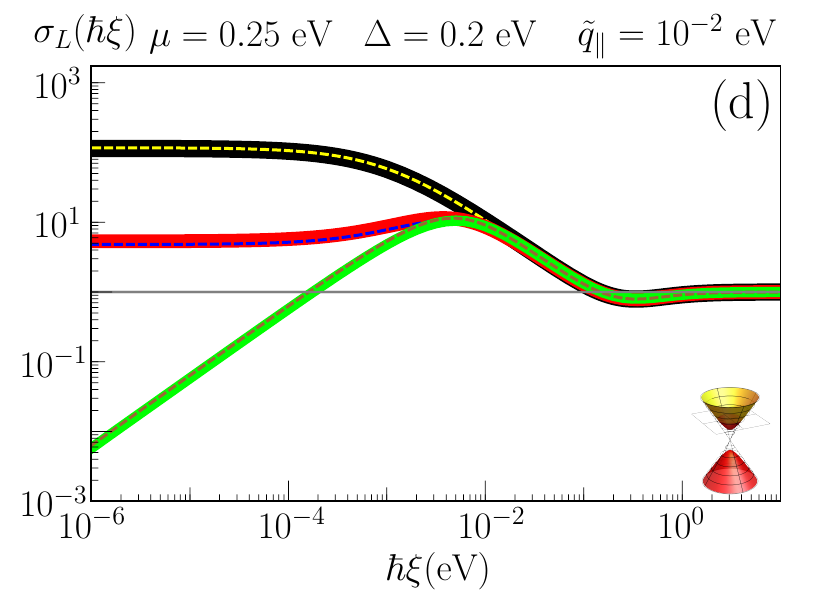}\\
\includegraphics[width=0.24\linewidth]{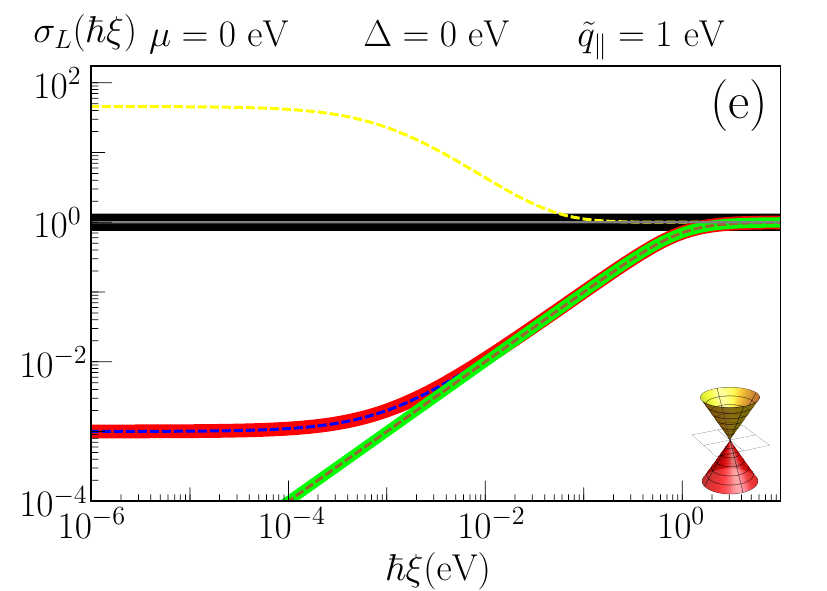}
\includegraphics[width=0.24\linewidth]{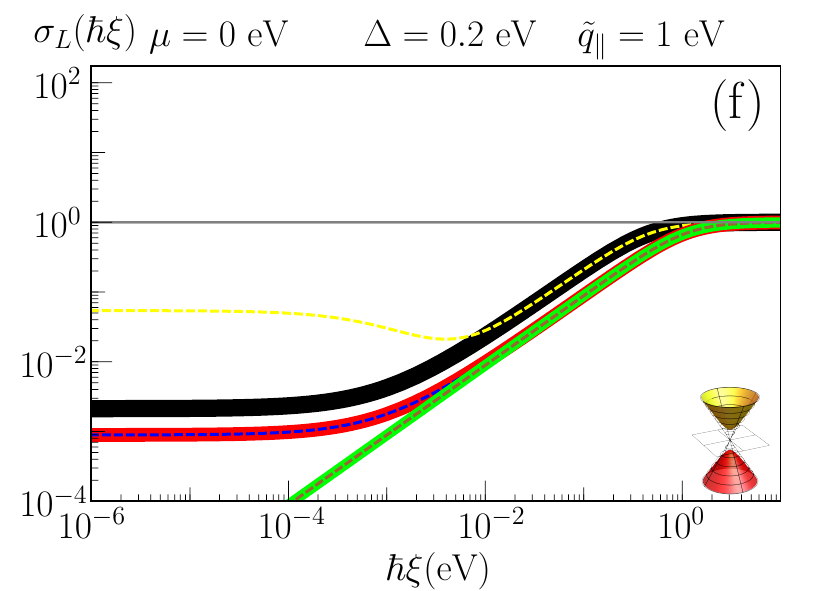}
\includegraphics[width=0.24\linewidth]{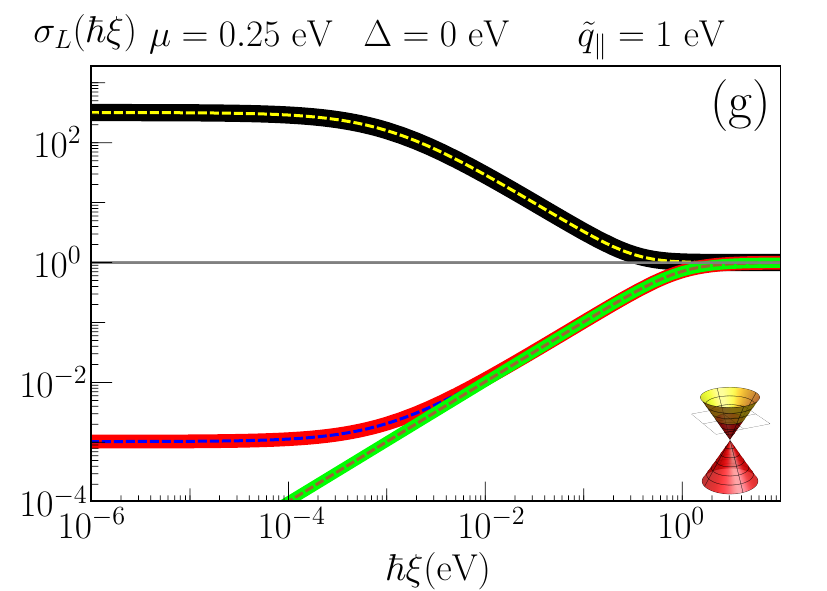}
\includegraphics[width=0.24\linewidth]{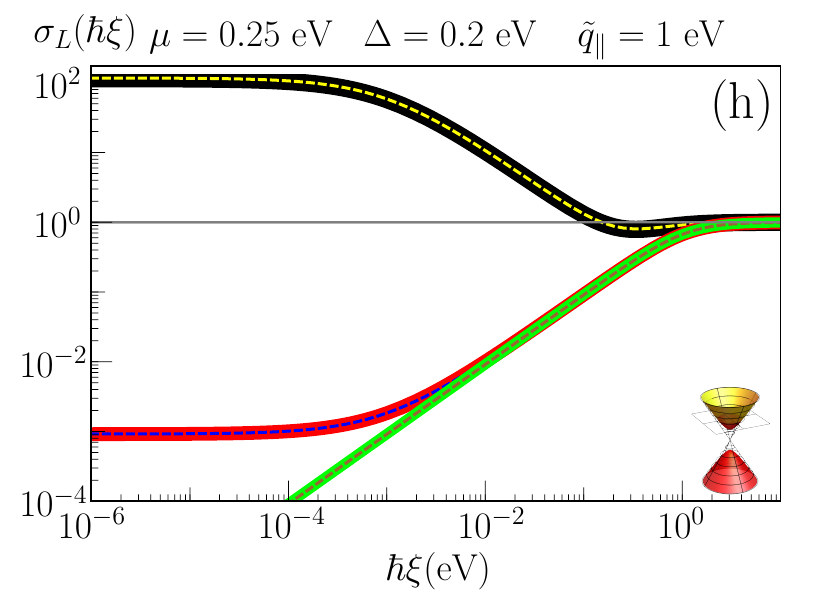}
\caption{ (Color online) Double logarithmic plots of the longitudinal electric conductivity $\sigma_{L}(\ii\hbar\xi)$ (in units of the universal electric conductivity of graphene $\sigma_{0} = \frac{\alpha c}{4}$) as a function of the imaginary frequency $\hbar\omega = \ii\hbar\xi$. The thick black and the yellow dashed curves are the local electric conductivity $\sigma_{xx}^{\rm{F}}(\omega,\Gamma,T,\mu)$ ($\tilde{q}_{\surf}=0$) at $T=0\KK$ and $T=300\KK$ respectively given by \Eq{Falkovsky_conductivity}. The thick red and the blue dashed curves are the non-local electric conductivities derived from the Kubo formula $\sigma_{L}^{\rm{K}}(\omega,\tilde{q}_{\surf},\Gamma,T,\mu)$ (\Eq{Sigma_L_0} and \Eq{Sigma_L_mu} and \Eq{Maldague_Formula}) with $\tilde{q}_{\surf}\neq0$, $\hbar\Gamma = 10^{-3}\eV$ at $T=0\KK$ and $T=300\KK$ respectively. The thick green and brown dashed curves are the non-local electric conductivity derived from the QFT model $\sigma_{L}^{\rm{NR}}(\omega,\tilde{q}_{\surf},T,\mu)$ (\Eq{QFT_Sigma_L}) with $\tilde{q}\neq0$ at $T=300\KK$ and $T=0\KK$ respectively, finally, the thin gray line is the universal electric conductivity of graphene with $\sigma(\ii\hbar\xi) = \sigma_{0} = \frac{\alpha c}{4}$. The chemical potential ($\mu$), Dirac mass ($\Delta$) and momentum ($\tilde{q}_{\surf}$) are specified in each panel.}
\label{fig_sigmaL}
\end{figure*}

\begin{figure*}%[H]
\includegraphics[width=0.24\linewidth]{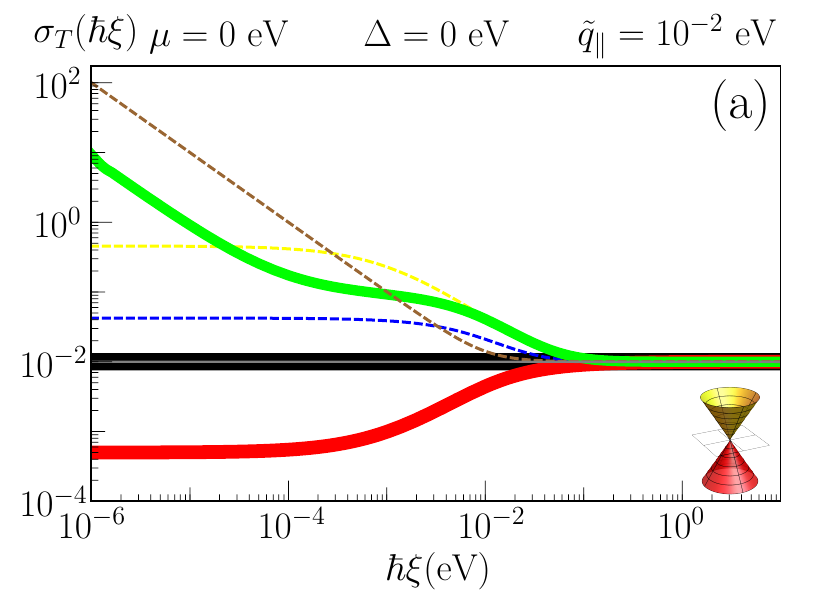}
\includegraphics[width=0.24\linewidth]{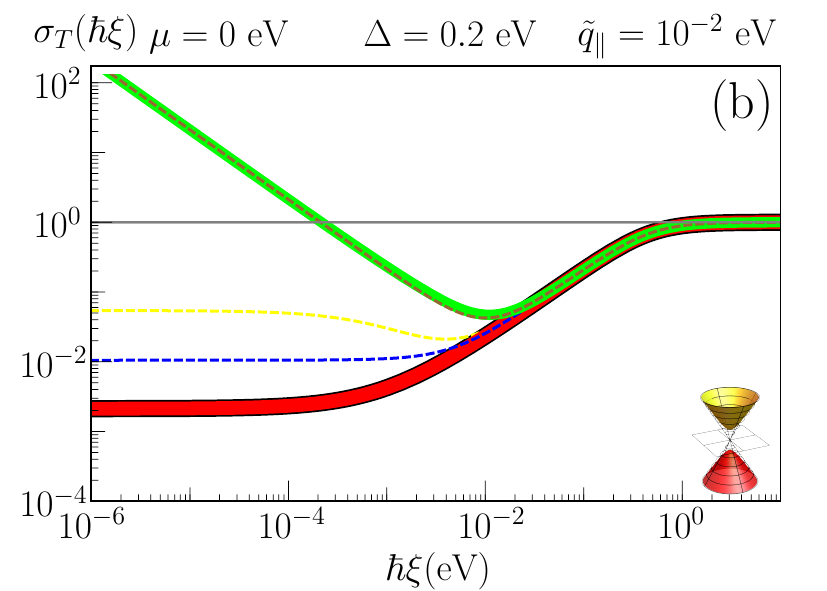}
\includegraphics[width=0.24\linewidth]{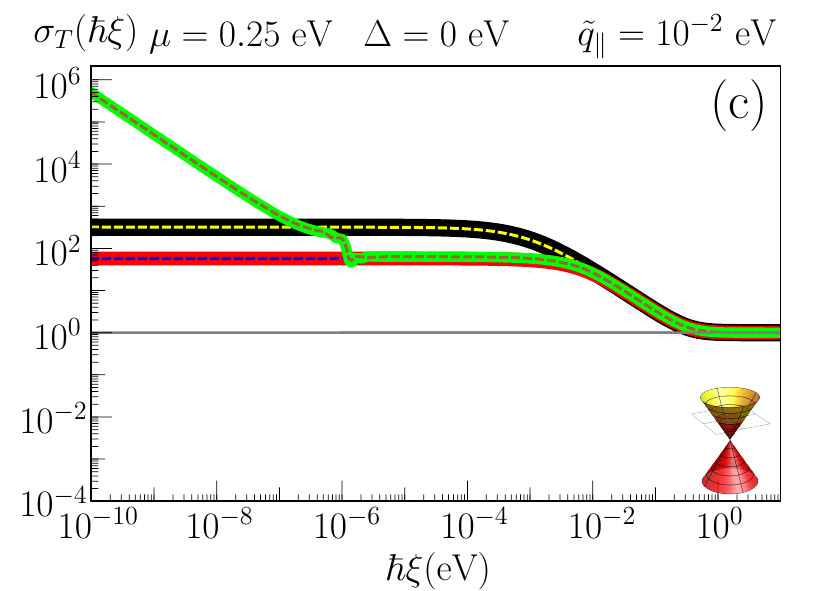}
\includegraphics[width=0.24\linewidth]{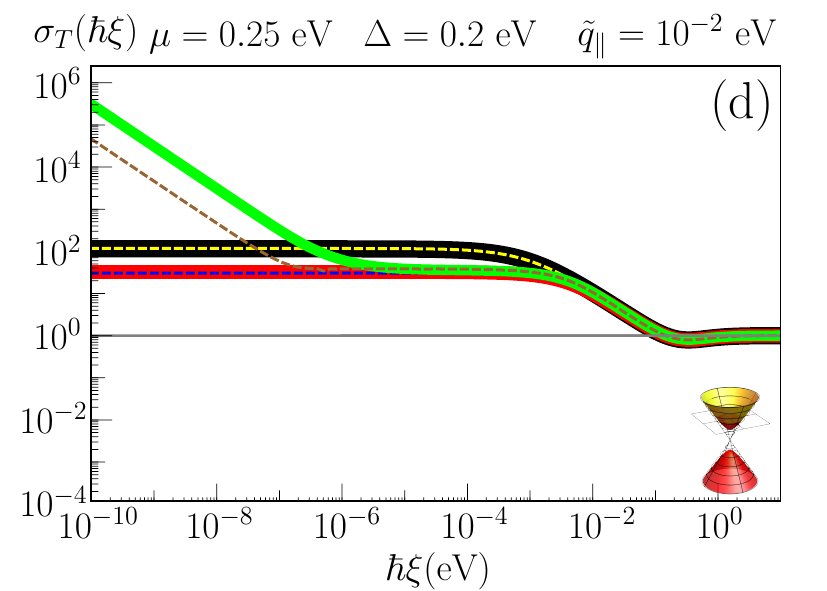}\\
\includegraphics[width=0.24\linewidth]{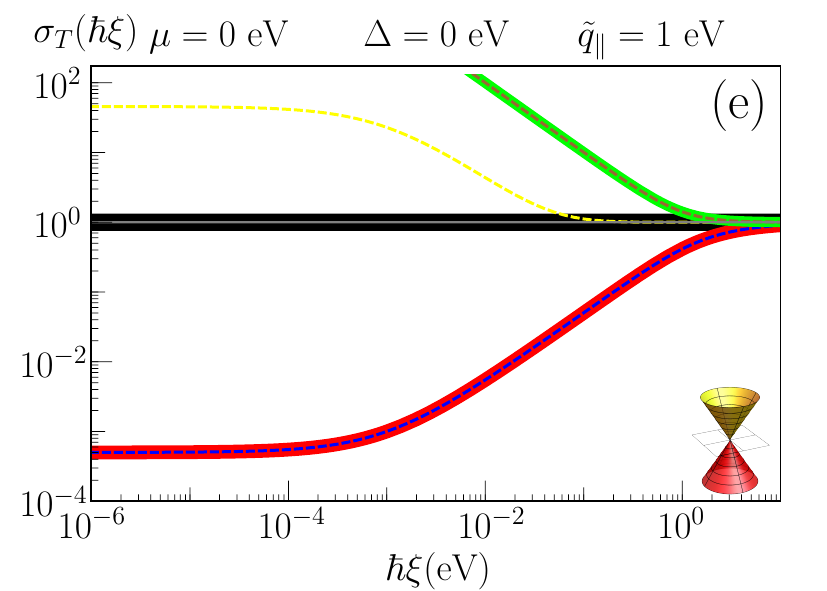}
\includegraphics[width=0.24\linewidth]{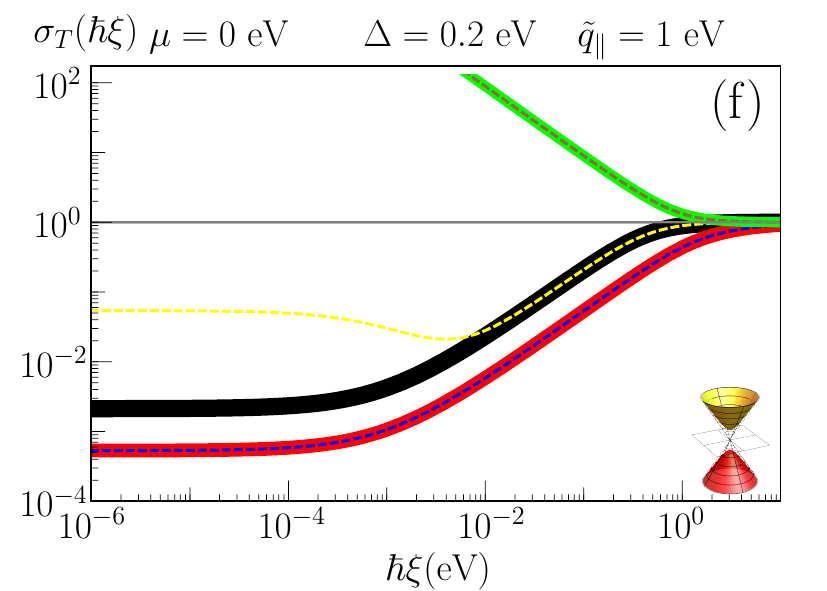}
\includegraphics[width=0.24\linewidth]{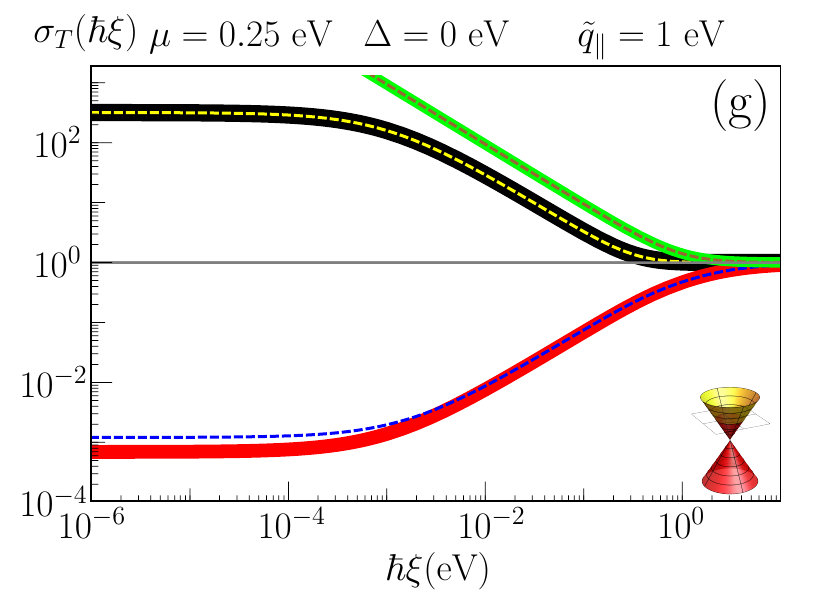}
\includegraphics[width=0.24\linewidth]{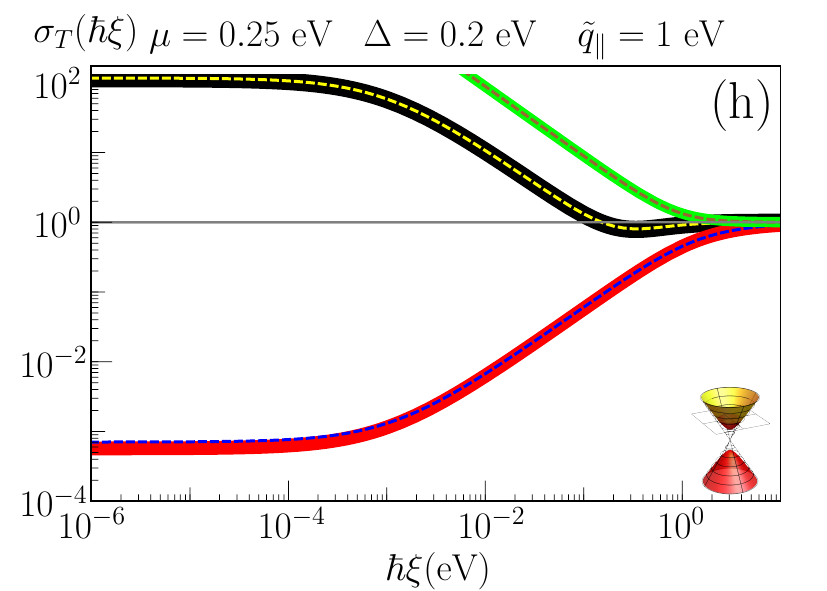}
\caption{ (Color online) Double logarithmic plots of the transversal electric conductivity $\sigma_{T}(\ii\hbar\xi)$ (in units of the universal electric conductivity of graphene $\sigma_{0} = \frac{\alpha c}{4}$) as a function of the imaginary frequency $\hbar\omega = \ii\hbar\xi$. The thick black and the yellow dashed curves are the local electric conductivity $\sigma_{xx}^{\rm{F}}(\omega,\Gamma,T,\mu)$ ($\tilde{q}_{\surf}=0$) at $T=0\KK$ and $T=300\KK$ respectively given by \Eq{Falkovsky_conductivity}. The thick red and the blue dashed curves are the non-local electric conductivities derived from the Kubo formula $\sigma_{T}^{\rm{K}}(\omega,\tilde{q}_{\surf},\Gamma,T,\mu)$ (\Eq{Sigma_T_0} and \Eq{Sigma_T_mu} and \Eq{Maldague_Formula}) with $\tilde{q}_{\surf}\neq0$, $\hbar\Gamma = 10^{-3}\eV$ at $T=0\KK$ and $T=300\KK$ respectively. The thick green and brown dashed curves are the non-local electric conductivity derived from the QFT model $\sigma_{T}^{\rm{NR}}(\omega,\tilde{q}_{\surf},T,\mu)$ (\Eq{QFT_Sigma_T}) with $\tilde{q}_{\surf}\neq0$ at $T=300\KK$ and $T=0\KK$ respectively, finally, the thin gray line is the universal electric conductivity of graphene with $\sigma(\ii\hbar\xi) = \sigma_{0} = \frac{\alpha c}{4}$. The chemical potential ($\mu$), Dirac mass ($\Delta$) and momentum ($\tilde{q}_{\surf}$) are specified in each panel.}
\label{fig_sigmaT}
\end{figure*}

In the following, we numerically compare the longitudinal (\Fig{fig_sigmaL}) and transversal (\Fig{fig_sigmaT}) electric conductivities derived from the three models we have studied here. We compared them for different temperatures $T$, chemical potentials $\mu$, mass gaps $\Delta$ and momentum $\tilde{q}$. In \Fig{fig_sigmaL} and \Fig{fig_sigmaT}, the local limit ($\tilde{q}_{\surf}=0$) at $T=0\KK$ and $T=300\KK$ are represented by the thick black and the yellow dashed curves respectively. The non-local electric conductivities derived from the Kubo formula with $\tilde{q}_{\surf}\neq 0\eV$ and $\hbar\Gamma = 10^{-3}\eV$ at $T=0\KK$ and $T=300\KK$ are represented by the thick red and the blue dashed curves respectively. The non-local electric conductivity derived from the QFT model with $\tilde{q}_{\surf}\neq 0\eV$ at $T=0\KK$ and $T=300\KK$ are represented by the dashed brown and thick green curve, respectively. We study the non-locality for $\tilde{q}_{\surf}=10^{-2}\eV$ \Fig{fig:sigmaT_tq=1eV_T_0_mu_0}(a-d) and for $\tilde{q}_{\surf}=1\eV$ \Fig{fig:sigmaT_tq=1eV_T_0_mu_0}(e-h).

As shown in \Fig{fig_sigmaL}, the non-local longitudinal electric conductivities derived from the Kubo formula and the QFT model almost coincide. If we artificially add a non-zero dissipation rate $\hbar\Gamma$ to the QFT model (which we remember is exactly zero for this model \cite{Bordag2015}), the two curves would superimpose. This fact remarks the contribution the electronic quasi-particle dissipation has in the electric conductivity, mainly for frequencies $\Abs{\omega} \lesssim \Gamma$.

As can be observed in \Fig{fig_sigmaT}, the results for the non-local transversal electric conductivities derived from the Kubo formula and the previous QFT model are very different. The main problem here is that a spurious asymptote proportional to $(\hbar\xi)^{-1}$ appears for very small imaginary frequencies. This difference is explained because $\sigma_{T}^{\rm{NR}}(\qq)$ is not regularized as imposed by the Kubo formula to fulfill $\dlim_{\omega\to0}\Pi_{\mu\nu}(\omega,\bm{q})A^{\nu}(\omega,\bm{q}) = 0$.

%\newpage

\section{Conclusions}\label{Section_Conclusions}
In this article, we have shown a detailed derivation of the Polarization $\Pi_{\mu\nu}$ and electric conductivity $\sigma_{\mu\nu}$ tensor for graphene close to the Dirac point in the continuous limit. We have used the Kubo formula \cite{non-local_Graphene_Lilia_Pablo} ($\sigma^{\rm{K}}$), a Quantum Field Theory based (QFT) model (which approaches the electronic quasiparticles as $(2+1)D$ Dirac electrons) \cite{Bordag2015} ($\sigma^{\rm{NR}}$) and a local model \cite{Fialkovsky2008} ($\sigma^{\rm{F}}$).

The more general result is obtained with the Kubo formula $\sigma^{\rm{K}}$. This result is valid for any complex frequency (with positive imaginary part) $\omega\in\mathbb{C}^{+}$, constant dissipation rate $\Gamma$, chemical potential $\mu$ and Dirac mass $m$ as a closed analytical formula at zero temperature $T$. The non-zero temperature results can be obtained after integration using the Maldague formula (\Eq{Maldague_Formula}).

We obtain that the local limit of $\sigma^{\rm{K}}$ coincides with the local limit of $\sigma^{\rm{NR}}$ when $\Gamma=0$. In addition to that, when $m=0$, $\sigma^{\rm{K}}$ converges to $\sigma^{\rm{F}}$, and when $m=0$ and $\Gamma=0$, $\sigma^{\rm{NR}}$ also converges to $\sigma^{\rm{F}}$.

We have derived the Polarization (and, therefore, the electric conductivity) as in the QFT model, we obtain again the results published elsewhere, and we find that the longitudinal electric conductivity derived from the Kubo formula and from the QFT model almost coincide (\Fig{fig_sigmaL}), with any difference explained by the different regularization strategies used. However, there is no such coincidence with the transversal electric conductivity (\Fig{fig_sigmaT}).

There are several differences between the two formalisms; the first one is that in the Kubo formalism, the effect of losses is taken into account. As a result, the intraband conductivity in the Kubo formula is described by a Drude model (in both longitudinal and transversal terms), while, in the QFT model, those intraband conductivities are described by the dissipation-less plasma model. In addition to that, the main difference comes from the regularization of the Polarization operator used. In the case of the Kubo formula, the expression of the Polarization operator is regularized by imposing $\dlim_{\omega\to0}\Pi^{\mu\nu}(\omega,\bm{q}_{\surf})A_{\nu}(\omega,\bm{q}_{\surf}) = 0$ for all $A_{\nu}(\omega,\bm{q}_{\surf})$, as shown in \Eq{j_eq10} of \Sect{KubosigmaderivationAppendix}. On the contrary, in the derivation of the QFT model, as the assumed constitutive relation (\Eq{JdependsonA}) does not impose any regularization term, the Longitudinal electric conductivity derived from the QFT model coincides with the result derived from the Kubo formula, but this is not the case of the Transversal electric conductivity, for which the QFT result fulfills $\dlim_{\omega\to0}\Pi_{T}(\omega,\bm{q}_{\surf}) \neq 0$. As a consequence, a transversal plasma electric conductivity is obtained in the QFT model, which would imply a dissipation-less electric current generated by an electric field that is not corrected by the addition of losses; this is clearly not acceptable in normal materials.\\
We have shown that the use of Ohm's Law as the constitutive relation between the electric current $J_{\mu}$ and the electric field $E^{\nu}$ (\Eq{OhmLaw}) instead of the assumed linear relationship of the electric current with the potential vector $A^{\nu}$ (\Eq{QFTconductivitydef}) in the Kubo formula leads to different models for the electric conductivity of 2D materials described by the Dirac Hamiltonian, like graphene. This difference can be traced to a regularization term that must be applied to the electric conductivity tensor. Once this regularization term is considered, the Kubo and QFT model exactly coincide for all parameters of the model.

\begin{acknowledgments}
P. R.-L. acknowledges support from Ministerio de Ciencia, Innovaci\'on y Universidades (Spain), Agencia Estatal de Investigaci\'on, under project NAUTILUS (PID2022-139524NB-I00), from AYUDA PUENTE, URJC, from QuantUM program of the University of Montpellier and the hospitality of the Theory of Light-Matter and Quantum Phenomena group at the Laboratoire Charles Coulomb, University of Montpellier, where part of this work was done.
J.-S.W. acknowledges support from MOE FRC tier 1 grant A-8000990-00-00.
M.A. acknowledges the QuantUM program of the University of Montpellier, the grant ”CAT”, No. A-HKUST604/20, from the ANR/RGC Joint Research Scheme sponsored by the French National Research Agency (ANR) and the Research Grants Council (RGC) of the Hong Kong Special Administrative Region.
The authors acknowledge Prof. Carsten Henkel his insightful comments and constructive suggestions, which have contributed to enhance the clarity and rigor of this manuscript.
%P.R.-L., and M.A. acknowledge the QuantUM program of the University of Montpellier
\end{acknowledgments}

%\appendix

%\section{A little more on appendixes}
%\subsection{\label{app:subsec}A subsection in an appendix}

\appendix
\begin{widetext}
\section{Non-local Kubo electric conductivity expressions}\label{NonLocalKuboAppendix}
The analytical formulas for the non-local electric conductivities of 2D Dirac cones have been derived in \cite{non-local_Graphene_Lilia_Pablo} from \Eq{General_Kubo_Formula}. Those formulas are naturally divided into two parts, one independent of the chemical potential $\mu$ and another term for which the chemical potential is accounted for. Namely,
$\sigma_{p}^{\rm{K}}(\qq) =  \sigma_{p,0}^{\rm{K}}(\qq) + \Theta(\abs{\mu} - \abs{\Delta})\sigma_{p,1}^{\rm{K}}(\qq)$, where $\Theta$ is the Heaviside step function and $p\in\{L, T, H\}$. Using
\begin{eqnarray}
\Psi(x) = 2\left[ x + \left( 1 - x^{2} \right)\tan^{-1}\left(\dfrac{1}{x}\right) \right],
\end{eqnarray}

\begin{eqnarray}
\label{Sigma_L_0}
\sigma_{L,0}^{\rm{K}}(\qq) & = &  \frac{\sigma_{0}}{2\pi}
\frac{ - \ii \tilde{q}_{0}}{\tilde{\theta}_{z}^{2}}\left[ 2\abs{\Delta} + \frac{ \tilde{\theta}_{z}^{2} - 4\Delta^{2}}{\tilde{\theta}_{z}}\tan^{-1}\left(\dfrac{\tilde{\theta}_{z}}{2\abs{\Delta}}\right)\right]
= \frac{\sigma_{0}}{4\pi}
\frac{ - \ii \tilde{q}_{0}}{\tilde{\theta}_{z}}\Psi(\delta),\\
\label{Sigma_L_mu}
\sigma_{L,1}^{\rm{K}}(\qq) & = &  \frac{\sigma_{0}}{2\pi}
\frac{-\ii\tilde{q}_{0}}{\tilde{q}_{\surf}^{2}}
\left[
4(\abs{\mu} - \abs{\Delta} ) + \frac{1}{2\tilde{\theta}_{z}}\left( \mathcal{F}_{1} + \left( R^{2} - \tilde{q}_{\surf}^{2} \right)\mathcal{F}_{2} \right)
+ \frac{\tilde{q}_{\surf}}{q_{0}^{2}}\Theta\left(\tilde{q}_{\surf}^{2} - 4\left( \mu^{2} - \Delta^{2} \right)\right)\mathcal{F}_{3}\right], \\
\label{Sigma_T_0}
\sigma_{T,0}^{\rm{K}}(\qq) & = & \frac{\sigma_{0}}{2\pi}\frac{1}{-\ii\tilde{q}_{0}}
\left[ \frac{\tilde{\theta}_{z}^{2} - 4\Delta^{2} }{\tilde{\theta}_{z}} \tan^{-1}\left(\frac{\tilde{\theta}_{z}}{2\abs{\Delta}}\right) - \frac{ \tilde{q}_{\surf}^{2} - 4\Delta^{2}}{\tilde{q}_{\surf}} \tan^{-1}\left(\frac{\tilde{q}_{\surf}}{2\abs{\Delta}}\right)
\right] = \frac{\sigma_{0}}{4\pi}\frac{1}{-\ii\tilde{q}_{0}}
\left[ \tilde{\theta}_{z}\Psi(\delta) - \tilde{q}_{\surf}\Psi(x) \right],\\
\label{Sigma_T_mu}
\sigma_{T,1}^{\rm{K}}(\qq) & = & \frac{\sigma_{0}}{2\pi} \frac{\ii}{q_{0}}
\!\!\left[
4\frac{\tilde{q}_{0}^{2}}{\tilde{q}_{\surf}^{2}}4( \abs{\mu} - \abs{\Delta} ) + 2\abs{\Delta} - 
\frac{1}{2\tilde{\theta}_{z} }\left( \frac{\tilde{\theta}_{z}^{2}}{\tilde{q}_{\surf}^{2}} \mathcal{F}_{1} + \left( \tilde{\theta}_{z}^{2} - 4\Delta^{2} \right)\mathcal{F}_{2} \right) \right. \nonumber \\
&& \left. ~~~~~~~~ + \frac{1}{\tilde{q}_{\surf}}\left( \mathcal{F}_{4}\Theta\left( \tilde{q}_{\surf}^{2} - 4\left( \mu^{2} - \Delta^{2} \right) \right) + \mathcal{F}_{5}\Theta\left( 4\left( \mu^{2} - \Delta^{2} \right) - \tilde{q}_{\surf}^{2} \right)\right)
\right],\\
\label{Sigma_D_0}
\sigma_{H,0}^{\rm{K}}(\qq) & = & \frac{2\sigma_{0}}{\pi}\frac{\eta\Delta}{\tilde{\theta}_{z}}\tan^{-1}\left(\frac{\tilde{\theta}_{z}}{2\abs{\Delta}}\right), \\
\sigma_{H,1}^{\rm{K}}(\qq) & = & - \frac{\sigma_{0}}{\pi}\frac{\eta\Delta}{\tilde{\theta}_{z}} 
\left[
\tan^{-1}\left( \frac{\tilde{q}_{0} - 2\abs{\Delta}}{\sqrt{ R^{2} - (\tilde{q}_{0} - 2\abs{\Delta})^{2} }} \right) 
- \tan^{-1}\left( \frac{\tilde{q}_{0} - 2\abs{\mu}}{\sqrt{ R^{2} - (\tilde{q}_{0} - 2\abs{\mu})^{2} }} \right) \right. \cr
&&
\left.
~~~~~~~~~~~~~ + 
\ii\log\left( \frac{\tilde{q}_{0} + 2\abs{\mu}    + \sqrt{ ( \tilde{q}_{0} + 2\abs{\mu}    )^{2} - R^{2} } }{\tilde{q}_{0} + 2\abs{\Delta} + \sqrt{ ( \tilde{q}_{0} + 2\abs{\Delta} )^{2} - R^{2} } } \right) \right],\\
\label{Sigma_D_0b}
\sigma_{S,0}^{\rm{K}}(\qq) & = & 0, \\
\sigma_{S,1}^{\rm{K}}(\qq) & = & 0
\label{Sigma_D_mu}
\end{eqnarray}
where $\sigma_{0} = \frac{\alpha c}{4} = \frac{e^{2}}{4\hbar}$, $\tilde{q}_{0} = \hbar\Omega  = \hbar\omega + \ii\hbar\Gamma$ ($\Gamma=\tau^{-1}$ accounts for the relaxation time), $\tilde{q}_{\surf}=  \hbar v_{F} q$, $\tilde{\theta}_{z} = \sqrt{ \tilde{q}_{\surf}^{2} - \tilde{q}_{0}^{2} }$, $\gamma = \frac{\Xi}{\tilde{\theta}_{z}}$, $\delta = \frac{2\abs{\Delta}}{\tilde{\theta}_{z}}$, $R = \tilde{q}_{\surf}\sqrt{ 1 + \delta^{2}}$ and $x = \frac{2\abs{\Delta}}{\tilde{q}_{\surf}}$. It is important to note that only the modulus of the Dirac mass enter into the expressions for $\sigma_{L}^{\rm{K}}$ and $\sigma_{T}^{\rm{K}}$, while $\sigma_{H}^{\rm{K}}$ has an additional dependency on the sign of the gaps through the combination $\eta \Delta$. The auxiliary functions $\{\mathcal{F}_{n}\}_{n=1}^{5}$ are the following
\begin{eqnarray}
\mathcal{F}_{1} & = & 
\phantom{+\ii} (\tilde{q}_{0} - 2\abs{\mu})\sqrt{ R^{2} - ( \tilde{q}_{0} - 2\abs{\mu} )^{2} } - \phantom{\ii}  (\tilde{q}_{0} - 2\abs{\Delta})\sqrt{ R^{2} - ( \tilde{q}_{0} - 2\abs{\Delta} )^{2} }
\nonumber\\
& & + \ii( \tilde{q}_{0} + 2\abs{\mu} )\sqrt{ ( \tilde{q}_{0} + 2\abs{\mu} )^{2} - R^{2} }
- \ii( \tilde{q}_{0} + 2\abs{\Delta} )\sqrt{ ( \tilde{q}_{0} + 2\abs{\Delta} )^{2} - R^{2} },
\\
\mathcal{F}_{2} & = & 
\phantom{-\ii} \tan^{-1}\left(\frac{ \tilde{q}_{0} -2\abs{\Delta} }{\sqrt{ R^{2} - ( \tilde{q}_{0} - 2\abs{\Delta} )^{2} } }\right)
- \tan^{-1}\left(\frac{ \tilde{q}_{0} -2\abs{\mu} }{\sqrt{ R^{2} - ( \tilde{q}_{0} - 2\abs{\mu} )^{2} } }\right)\nonumber\\
& & - \ii\log\left( \tilde{q}_{0} + 2\abs{\Delta} +\sqrt{ ( \tilde{q}_{0} + 2\abs{\Delta} )^{2} - R^{2} } \right)
+ \ii\log\left( \tilde{q}_{0} + 2\abs{\mu} +\sqrt{ ( \tilde{q}_{0} + 2\abs{\mu} )^{2} - R^{2} } \right),
\\
\mathcal{F}_{3} & = & \phantom{+} 2\mu\left( \sqrt{ \tilde{q}_{\surf}^{2} - 4\left( \mu^{2} - \Delta^{2} \right) } + \ii\sqrt{ 4\left( \mu^{2} - \Delta^{2} \right) - \tilde{q}_{\surf}^{2} }\right)\nonumber\\
& & + \left(4\Delta^{2} - \tilde{q}_{\surf}^{2}\right)\left[\begin{array}{l}
   \ii\log( 2\abs{\Delta} + \ii\tilde{q}_{\surf} )
 - \ii\log\left( 2\abs{\mu} + \sqrt{ 4\left( \mu^{2} - \Delta^{2} \right) - \tilde{q}_{\surf}^{2} }\right)\\
\end{array}\right]\nonumber\\
& & + \left(4\Delta^{2} - \tilde{q}_{\surf}^{2}\right)\left[\begin{array}{l}
    \tan^{-1}\left(\frac{2\abs{\mu}}{\sqrt{ \tilde{q}_{\surf}^{2} - 4(\mu^{2} - \Delta^{2}) }}\right)
 -  \tan^{-1}\left(\frac{2\abs{\Delta}}{\tilde{q}_{\surf}}\right)
\end{array}\right],
\\
\mathcal{F}_{4} & = & - 2\abs{\mu}\sqrt{ \tilde{q}_{\surf}^{2} - 4\left( \mu^{2} - \Delta^{2} \right) }
 + \left( \tilde{q}_{\surf}^{2} - 4\Delta^{2} \right) \left[ \tan^{-1}\left(\frac{2\abs{\mu}}{\sqrt{ \tilde{q}_{\surf}^{2} - 4(\mu^{2} - \Delta^{2}) }}\right) - \tan^{-1}\left(\frac{2\abs{\Delta}}{\tilde{q}_{\surf}}\right)\right],
\end{eqnarray}
\begin{eqnarray}
\mathcal{F}_{5} & = & \left( \tilde{q}_{\surf}^{2} - 4\Delta^{2} \right)\left[ \frac{\pi}{2} - \tan^{-1}\left(\frac{2\abs{\Delta}}{\tilde{q}_{\surf}}\right)\right].
\end{eqnarray}
Those results are valid for all frequencies $\omega\in\mathbb{C}^{+}$, and we remind that they are the electric conductivity per Dirac cone, to obtain the electric conductivity of Graphene, we must to sum the contribution of each 4 Dirac cones, having into account their respective (signed) Dirac masses by using
\begin{eqnarray}
\sigma_{p}^{\rm{K}}(\qq) = \sum_{\eta=\pm}\sum_{s=\pm}\sigma_{p}^{\rm{K}}(\qq, \Delta_{s}^{\eta}).
\end{eqnarray}

\section{Local Kubo electric conductivity expressions}\label{LocalKuboAppendix}
There is a special case in the local limit (when we apply the $\bm{q}_{\surf}\to\bm{0}$ limit to the Kubo formula \eqref{General_Kubo_Formula}) when results valid for all temperatures can be obtained. The local limit of the electric conductivities of one massive Dirac cone is
\begin{eqnarray} 
\sigma_{xx}(\omega,\bm{0}, \mu, \Delta) & = & \ii\frac{\sigma_{0}}{\pi} 
\Bigg[ \frac{\mu^{2} - \Delta^{2}}{\abs{\mu}}\frac{1}{\Omega}\Theta\left(\abs{\mu} - \abs{\Delta}\right)
+ \frac{\Delta^{2}}{M\Omega} - \frac{\Omega^2+4\Delta^2}{2 i \Omega^2}   \tan^{-1}\left(\frac{i\Omega}{2M}\right) \Bigg], \cr
\sigma_{xy} (\omega,\bm{0}, \mu, \Delta) & = &  \frac{2 \sigma_{0}}{\pi}\frac{\eta\Delta}{i\Omega}\tan^{-1}\left(\frac{i\Omega}{2M}\right),
\end{eqnarray}
where $\Omega = \hbar\omega + \ii\hbar\Gamma$, $\Gamma = \tau^{-1}$ and $M = \text{Max}\left[\abs{\Delta}, \abs{\mu}\right]$. These results are per Dirac cone, and they are consistent with the ones found by other researchers \cite{Fialkovsky2011}\cite{Fialkovsky2012}\cite{Falkovsky2007}\cite{Fialkovsky2008}\cite{Bordag2009}\cite{PhysRevB.88.045442}\cite{WangKong2010}\cite{WangKong2011}. The first term in $\sigma_{xx}$ corresponds to intra-band transitions, and the last two terms to inter-band transitions. Note that, in the local limit $\bm{q}_{\surf}\to\bm{0}$ one obtains $\sigma_{xx}(\omega,\bm{0})=\sigma_{yy}(\omega,\bm{0})=\sigma_{L}(\omega,\bm{0})=\sigma_{T}(\omega,\bm{0})$, and $\sigma_{xy}(\omega,\bm{0})=-\sigma_{yx}(\omega,\bm{0})=\sigma_H(\omega,\bm{0})$.

From those results, we obtain the electric conductivity at zero temperature of a Dirac cone with $\Delta=0$ mass gap as
\begin{eqnarray}\label{Local_Delta0_T0}
\sigma_{xx}(\omega,\bm{q}_{\surf} = \bm{0},T=0,\Delta=0) & = & \sigma^{\text{intra}}_{xx}(\omega) + \sigma^{\text{inter}}_{xx}(\omega),\nonumber\\
\sigma^{\text{intra}}_{xx}(\omega) & = & \dfrac{\alpha c}{\pi}\dfrac{\abs{\mu}}{-\ii\Omega},\label{local_intra_T=0v2}\\
\sigma^{\text{inter}}_{xx}(\omega) & = & \dfrac{\alpha c}{2\pi}\tan^{-1}\left(\dfrac{- \ii\Omega}{2\abs{\mu}}\right),\label{local_inter_T=0v2}\\
\sigma_{xy}(\omega,\bm{q}_{\surf} = \bm{0},T=0,\Delta=0) & = & 0.
\end{eqnarray}

From this result with $\Delta = 0$, in the non-dissipation limit for the interband term and for finite temperatures, by using the Maldague formula, the well-known result of Falkovsky is obtained as
\begin{eqnarray}
\sigma^{\rm{F}}_{xx}(\omega, \Gamma, T, \mu) & = & \sigma^{\rm{F},\text{intra}}_{xx}(\omega, \Gamma, T, \mu) + \sigma^{\rm{F},\text{inter}}_{xx}(\omega, 0, T, \mu),\nonumber\\
\sigma^{\rm{F},\text{intra}}_{xx}(\omega, \Gamma, T, \mu) & = & \dfrac{1}{\pi\hbar}\dfrac{2\ii\alpha c k_{\rm B}T}{\omega + \ii\Gamma}
\ln\left[2\cosh\left(\frac{\mu}{2k_{\rm B}T}\right)\right],\nonumber\\
\sigma^{\rm{F},\text{inter}}_{xx}(\omega, 0, T, \mu) & = & \dfrac{\alpha c}{4}G\left(\Abs{\dfrac{\hbar\omega}{2}}\right)%\nonumber\\& &
+ \ii \dfrac{\alpha c}{4}\dfrac{4\hbar\omega}{\pi}\int_{0}^{\infty}\dd\xi\frac{G(\xi) - G\left(\Abs{\dfrac{\hbar\omega}{2}}\right)}{(\hbar\omega)^{2} - 4\xi^{2}}.
\end{eqnarray}
with
\begin{eqnarray}
G\left(\epsilon\right)
= n_{F}(-\epsilon + \mu) - n_{F}(+\epsilon + \mu)
= \dfrac{\sinh\left(\beta\epsilon\right)}{\cosh\left(\beta\mu\right) + \cosh\left(\beta\epsilon\right)}.
\end{eqnarray}
The intraband term is obtained by the use of the Maldague formula (\Eq{Maldague_Formula}) for \Eq{local_intra_T=0v2}, as 
\begin{eqnarray}
\sigma^{\text{intra}}_{xx}(\omega, T)
& = & \dfrac{\alpha c}{\pi}\dfrac{1}{-\ii\Omega}
\frac{2}{\beta}\ln\left[2\cosh\left(\frac{\beta\mu}{2}\right)\right]%\nonumber\\& = &
= \dfrac{\alpha c}{\pi\hbar}\dfrac{2\ii k_{\rm B}T}{\omega + \ii\Gamma}
\ln\left[2\cosh\left(\frac{\mu}{2k_{\rm B}T}\right)\right].
\end{eqnarray}
To derive the local interband electric conductivity at finite temperatures, we first need to apply the zeroth dissipation limit $\tau\to\infty$ ($\Gamma\to0$) of the real part $\omega\in\mathbb{R}$ of the interband electric conductivity of graphene, as
\begin{eqnarray}
& &\lim_{\tau\to\infty}\Real{\sigma^{\text{inter}}_{xx}(\omega)}
 = \lim_{\tau\to\infty}\dfrac{\alpha c}{2\pi}\Real{\tan^{-1}\left(\dfrac{- \ii\Omega}{2\abs{\mu}}\right)}%\nonumber\\& &
= \dfrac{\alpha c}{4}\Big[\Theta\left( \hbar\omega - 2\abs{\mu} \right) + \Theta\left( - \hbar\omega - 2\abs{\mu} \right)\Big],
\end{eqnarray}
which coincides with the result shown in \cite{Fialkovsky2008} in the $T\to0$ limit for $\omega>0$. Note that $\Real{\sigma^{\text{inter}}_{xx}(-\omega)} = + \Real{\sigma^{\text{inter}}_{xx}(\omega)}$ is an even function in $\omega$.\\
Applying the Maldague formula (\Eq{Maldague_Formula}) to \Eq{local_inter_T=0v2}, we obtain
\begin{eqnarray}
& & \Real{\sigma^{\text{inter}}_{xx}(\omega,\mu, T)}
= \dfrac{\alpha c}{4}
\left(  \dfrac{1}{e^{\beta\left(\mu - \abs{\frac{\hbar\omega}{2}}\right)} + 1} - \dfrac{1}{e^{\beta\left(\mu + \abs{\frac{\hbar\omega}{2}}\right)} + 1} \right)%\nonumber\\& = &
= \dfrac{\alpha c}{4}\dfrac{\sinh\left(\beta\abs{\frac{\hbar\omega}{2}}\right)}{\cosh\left(\beta\mu\right) + \cosh\left(\beta\abs{\frac{\hbar\omega}{2}}\right)} = \dfrac{\alpha c}{4}G\left(\Abs{\dfrac{\hbar\omega}{2}}\right).%\nonumber\\& = & .
\end{eqnarray}
We have the absolute value in the formula above because we also handle negative real frequencies, while in \cite{Fialkovsky2008} the result for only positive frequencies was derived.
Applying a regularized version of the Kramers-Krönig relationships that avoid the use of the principal part of a function, that read as \cite{dresselhaus2018solid}\cite{jackson_classical_1999}\cite{jones1973theoretical}:
\begin{eqnarray}
\sigma_{R}(x) = \frac{2}{\pi}\int_{0}^{\infty}\dd\omega\frac{\omega\sigma_{I}(\omega) - x\sigma_{I}(x)}{\omega^{2} - x^{2}},%\nonumber\\
\hspace{2cm}
\sigma_{I}(x) = \frac{2}{\pi}x\int_{0}^{\infty}\dd\omega\frac{\sigma_{R}(x) - \sigma_{R}(\omega)}{\omega^{2} - x^{2}},
\end{eqnarray}
it is immediate that
\begin{eqnarray}\label{Imag_Inter_Falkovsky}
\Imag{\sigma_{xx}^{\text{inter}}(\omega,T)}%\nonumber\\% = G_{I}\left(\dfrac{\hbar\omega}{2}\right)\nonumber\\& = &
= \dfrac{\alpha c}{4}\dfrac{4\hbar\omega}{\pi}\int_{0}^{\infty}\dd\xi\frac{G(\xi) - G\left(\Abs{\dfrac{\hbar\omega}{2}}\right)}{(\hbar\omega)^{2} - 4\xi^{2}},
\end{eqnarray}
which coincides with the imaginary part of the interband term of Falkovsky \cite{PhysRevB.93.115427}.\\
Finally, by using the Kramers-Krönig relation to find the real part of the electric conductivity at imaginary frequencies $\omega = \ii\xi$ \cite{Lilia_full_graphene_Conductivity}
\begin{eqnarray}
\Real{\sigma_{ij}(\ii\xi)} & = & \frac{2}{\pi}\int_{0}^{\infty}\dd\omega\frac{\xi}{\omega^{2} + \xi^{2}}\Real{\sigma_{ij}(\omega)} = \frac{2}{\pi}\int_{0}^{\infty}\dd\omega\frac{\omega}{\omega^{2} + \xi^{2}}\Imag{\sigma_{ij}(\omega)},
\end{eqnarray}
we obtain the interband electric conductivity for imaginary frequencies as \cite{MAntezza17}
\begin{eqnarray}
\sigma_{xx}^{\rm{F},\text{inter}}
(\ii\hbar\xi) & = & \dfrac{\alpha c}{4}\frac{2}{\pi}\int_{0}^{\infty}\dd\omega\frac{\xi}{\omega^{2} + \xi^{2}}G\left(\frac{\hbar\omega}{2}\right).
\end{eqnarray}
Note this result is entirely equivalent to the use of the Maldague formula to $\sigma_{xx}^{\rm{F},\text{inter}}$ given in \Eq{Local_Delta0_T0}, and by making the substitution $\xi\rightarrow\xi + \Gamma$, we can automatically add the constant dissipation to this interband electric conductivity.

\section{Derivation of the Polarization and electric conductivity from the QFT model previously used in literature }\label{Appendix_Derivation_Polarization_Mostepanenko}
In this appendix, we will show how to derive the results for the Polarization operator given in \cite{PRL_Mohideen}\cite{Bordag2015}\cite{Bimonte2017} presented in \Sect{Section_QFT_Model} from \Eq{def:Covariant_Polarization_Operator_definitions}.
\begin{eqnarray}\label{def:Covariant_Polarization_Operator_definitions_Appendix}
\Pi_{\mu\nu}(\qq) = g_{s}e^{2}\dfrac{-\ii}{\hbar}\int_{k}\dfrac{Z_{\mu\nu}(K_{\alpha},q_{\alpha})}{\left[ K^{\rho}K_{\rho} - m^{2} \right]\left[ S^{\zeta}S_{\zeta} - m^{2} \right]},
\end{eqnarray}
We are going to use the following notation: $S_{\zeta} = K_{\zeta} + q_{\zeta}$, $K_{\rho} = (\tilde{k}_{0} + \mu, \tilde{\bm{k}}_{\surf}) = (\ii\hbar\omega_{n} + \mu, \hbar v_{F}\bm{k}_{\surf})$, $q_{\alpha} = (\tilde{q}_{0}, \tilde{\bm{q}}_{\surf}) = (\ii\hbar\omega_{m}, \hbar v_{F}\bm{q}_{\surf})$ (see \Tab{Notation}), $Z_{\mu\nu}(K_{\alpha},q_{\alpha})$ is given in \Eq{Numerator_Polarization_Operator_Mostepanenko}.

We apply the Matsubara formalism directly to the expression of \Eq{def:Covariant_Polarization_Operator_definitions_Appendix} to obtain the expression shown in \cite{Bordag2015}. Remembering that, for fermions we have $\tilde{k}_{0} = \ii\hbar\omega_{n} = \ii\frac{2\pi}{\beta}\left(n + \frac{1}{2}\right)$ $\forall n\in\mathbb{Z}$ and $\tilde{q}_{0} = \ii\hbar\omega_{m} = \ii\frac{2\pi}{\beta}\left( m + \frac{1}{2}\right)$ $\forall m\in\mathbb{Z}$, we obtain
\begin{eqnarray}\label{Matsubara_Sum_Mostepanenko}
\dfrac{\Pi_{\mu\nu}(\qq)}{g_{s}e^{2}}
& = & \dfrac{-\ii}{\hbar}\dint_{BZ}\dfrac{\dd^{2}\bm{k}_{\surf}}{(2\pi)^{2}}\int_{-\infty}^{\infty}\dfrac{\dd \tilde{k}_{0}}{2\pi}\dfrac{Z_{\mu\nu}(\tilde{k}_{0}+\mu,\tilde{\bm{k}}_{\surf},\tilde{q}_{0},\tilde{\bm{q}}_{\surf})}{\left[ K^{\mu}K_{\mu} - m^{2} \right]\left[ S^{\mu}S_{\mu} - m^{2} \right]}\nonumber\\
& = & \dfrac{-\ii}{\hbar}\dint_{\bm{k}}\int_{-\infty}^{\infty}\dfrac{\dd \tilde{k}_{0}}{2\pi}\dfrac{Z_{\mu\nu}(\tilde{k}_{0} + \mu,\tilde{\bm{k}}_{\surf},\tilde{q}_{0},\tilde{\bm{q}}_{\surf})}{\left[ (\tilde{k}_{0} + \mu)^{2} - \epsilon_{\bm{k}}^{2} \right]\left[ (\tilde{k}_{0} + \tilde{q}_{0} + \mu)^{2} - \epsilon_{\bm{k}+\bm{q}}^{2} \right]}\nonumber\\
& = & \frac{-1}{\beta}\dint_{\bm{k}}\sum_{n\in\mathbb{Z}}^{\text{Fermi}}\dfrac{Z_{\mu\nu}(\ii\hbar\omega_{n} + \mu,\tilde{\bm{k}}_{\surf},\ii\hbar\omega_{m},\tilde{\bm{q}}_{\surf})}{\left[ (\ii\hbar\omega_{n} + \mu)^{2} - \epsilon_{\bm{k}}^{2} \right]\left[ (\ii\hbar\omega_{n} + \ii\hbar\omega_{m} + \mu)^{2} - \epsilon_{\bm{s}}^{2} \right]}\nonumber\\
& = & \dint_{\bm{k}}\left[ \dfrac{Z_{\mu\nu}(- \ii\hbar\omega_{m} -\epsilon_{\bm{s}},\tilde{\bm{k}}_{\surf},\ii\hbar\omega_{m},\tilde{\bm{q}}_{\surf})}{2\epsilon_{\bm{s}}\left( ( \ii\hbar\omega_{m} + \epsilon_{\bm{s}})^{2} - \epsilon_{\bm{k}}^{2} \right)} + \dfrac{Z_{\mu\nu}(-\epsilon_{\bm{k}},\tilde{\bm{k}}_{\surf},\ii\hbar\omega_{m},\tilde{\bm{q}}_{\surf})}{2\epsilon_{\bm{k}}( ( \ii\hbar\omega_{m} - \epsilon_{\bm{k}})^{2} - \epsilon_{\bm{s}}^{2} )} \right]\nonumber\\
& & - \dint_{\bm{k}}\sum_{\lambda=\pm}\dfrac{n_{F}(\epsilon_{\bm{s}} - \lambda\mu)Z_{\mu\nu}( - \ii\hbar\omega_{m} +\epsilon_{\bm{s}}^{\lambda},\tilde{\bm{k}}_{\surf},  \ii\hbar\omega_{m},\tilde{\bm{q}}_{\surf})}{2\epsilon_{\bm{s}}\left( ( \ii\hbar\omega_{m} - \epsilon_{\bm{s}}^{\lambda})^{2} - \epsilon_{\bm{k}}^{2} \right)} %\nonumber\\& &
- \dint_{\bm{k}}\sum_{\lambda=\pm}\dfrac{n_{F}(\epsilon_{\bm{k}} - \lambda\mu)Z_{\mu\nu}(\epsilon_{\bm{k}}^{\lambda},\tilde{\bm{k}}_{\surf}, \ii\hbar\omega_{m},\tilde{\bm{q}}_{\surf})}{2\epsilon_{\bm{k}}\left( ( \ii\hbar\omega_{m} + \epsilon_{\bm{k}}^{\lambda})^{2} - \epsilon_{\bm{s}}^{2} \right)},
\end{eqnarray}
where we have used that $n_{F}(-\ii\hbar\omega_{m} + \epsilon_{\bm{s}} - \lambda\mu) = n_{F}(\epsilon_{\bm{s}} - \lambda\mu)$ and $\bm{s}_{\surf} = \bm{k}_{\surf} + \bm{q}_{\surf}$ (see \Tab{Notation}). Next, we analytically expand the Matsubara sum given in \Eq{Matsubara_Sum_Mostepanenko} to the whole upper complex plane by applying the formal change $\ii\omega_{m} \to \omega\in\mathbb{C}^{+}$ into the definition of the Polarization operator given in \Eq{def:Covariant_Polarization_Operator_definitions_Appendix}, using $\tilde{q}_{0} = \hbar\omega$, we get
\begin{eqnarray}\label{Matsubara_Sum_Mostepanenko2}
\dfrac{\Pi_{\mu\nu}(\qq)}{g_{s}e^{2}}
& = & \dint_{\bm{k}}\left[ \dfrac{Z_{\mu\nu}(-\tilde{q}_{0}-\epsilon_{\bm{s}},\tilde{\bm{k}}_{\surf},\tilde{q}_{0},\tilde{\bm{q}}_{\surf})}{2\epsilon_{\bm{s}}\left( ( \tilde{q}_{0} + \epsilon_{\bm{s}})^{2} - \epsilon_{\bm{k}}^{2} \right)} + \dfrac{Z_{\mu\nu}(-\epsilon_{\bm{k}},\tilde{\bm{k}}_{\surf},\tilde{q}_{0},\tilde{\bm{q}}_{\surf})}{2\epsilon_{\bm{k}}( ( \tilde{q}_{0} - \epsilon_{\bm{k}})^{2} - \epsilon_{\bm{s}}^{2} )} \right]\nonumber\\
& & - \dint_{\bm{k}}\sum_{\lambda=\pm}\dfrac{n_{F}(\epsilon_{\bm{s}} - \lambda\mu)Z_{\mu\nu}( - \tilde{q}_{0} + \epsilon_{\bm{s}}^{\lambda},\tilde{\bm{k}}_{\surf},\tilde{q}_{0},\tilde{\bm{q}}_{\surf})}{2\epsilon_{\bm{s}}\left( ( \tilde{q}_{0} - \epsilon_{\bm{s}}^{\lambda})^{2} - \epsilon_{\bm{k}}^{2} \right)} %\nonumber\\& &
- \dint_{\bm{k}}\sum_{\lambda=\pm}\dfrac{n_{F}(\epsilon_{\bm{k}} - \lambda\mu)Z_{\mu\nu}(\epsilon_{\bm{k}}^{\lambda},\tilde{\bm{k}}_{\surf},\tilde{q}_{0},\tilde{\bm{q}}_{\surf})}{2\epsilon_{\bm{k}}\left( ( \tilde{q}_{0} + \epsilon_{\bm{k}}^{\lambda})^{2} - \epsilon_{\bm{s}}^{2} \right)}.
\end{eqnarray}
The term in brackets corresponds to the integrand of the $T=0$ limit, while the second and third terms correspond to the correction due to the temperature. Therefore, following the notation of \cite{Bordag2015}, the Polarization operator given in \Eq{def:Covariant_Polarization_Operator_definitions_Appendix} can be written as
\begin{eqnarray}
\Pi_{\mu\nu}(\qq) = \Pi^{(0)}_{\mu\nu}(\qq) + \Delta_{T}\Pi_{\mu\nu}(\qq),
\end{eqnarray}
by construction, $\Pi^{(0)}_{\mu\nu}(\qq)$ is independent of temperature and of the chemical potential $\mu$ \cite{Bordag2015}, therefore, it corresponds to the interband electric conductivity with $\mu = k_{B}T = 0\eV$. On the other hand, to simplify $\Delta_{T}\Pi_{\mu\nu}(\qq)$, we apply the change of variables $\bm{k}_{\surf} \to - (\bm{k}_{\surf} + \bm{q}_{\surf})$, we also make use of the symmetry of the relation of dispersion $\epsilon_{\bm{k}} = \epsilon_{-\bm{k}}$ and we transform the dummy variable $\lambda\to -\lambda$ to the first summand of $\Delta_{T}\Pi_{\mu\nu}(q)$ to obtain
\begin{eqnarray}
\dint_{\bm{k}}\sum_{\lambda=\pm}\dfrac{n_{F}(\epsilon_{\bm{s}} - \lambda\mu)Z_{\mu\nu}(-\tilde{q}_{0}+\epsilon_{\bm{s}}^{\lambda},\tilde{\bm{k}}_{\surf},\tilde{q}_{0},\tilde{\bm{q}}_{\surf})}{2\epsilon_{\bm{s}}\left( (\tilde{q}_{0} - \epsilon_{\bm{s}}^{\lambda})^{2} - \epsilon_{\bm{k}}^{2} \right)}%\nonumber\\
& = & \dint_{\bm{k}}\sum_{\lambda=\pm}\dfrac{n_{F}(\epsilon_{-\bm{k}} + \lambda\mu)Z_{\mu\nu}(-\tilde{q}_{0}-\epsilon_{-\bm{k}}^{\lambda},-\tilde{\bm{s}}_{\surf},\tilde{q}_{0},\tilde{\bm{q}}_{\surf})}{2\epsilon_{-\bm{k}}\left( (\tilde{q}_{0} + \epsilon_{-\bm{k}}^{\lambda})^{2} - \epsilon_{-\bm{s}}^{2} \right)}\nonumber\\
& = & \dint_{\bm{k}}\sum_{\lambda=\pm}\dfrac{n_{F}(\epsilon_{\bm{k}} + \lambda\mu)Z_{\mu\nu}(-\epsilon_{\bm{k}}^{\lambda}-\tilde{q}_{0},-\tilde{\bm{s}}_{\surf},\tilde{q}_{0},\tilde{\bm{q}}_{\surf})}{2\epsilon_{\bm{k}}\left( (\tilde{q}_{0} + \epsilon_{\bm{k}}^{\lambda})^{2} - \epsilon_{\bm{s}}^{2} \right)}\nonumber\\
& = & \dint_{\bm{k}}\sum_{\lambda=\pm}\dfrac{n_{F}(\epsilon_{\bm{k}} + \lambda\mu)Z_{\mu\nu}(\epsilon_{\bm{k}}^{\lambda},\tilde{\bm{k}}_{\surf},\tilde{q}_{0},\tilde{\bm{q}}_{\surf})}{2\epsilon_{\bm{k}}\left( (\tilde{q}_{0} + \epsilon_{\bm{k}}^{\lambda})^{2} - \epsilon_{\bm{s}}^{2} \right)}.
\end{eqnarray}
where we have used that
\begin{eqnarray}\label{Z_symmetry}
Z_{\mu\nu}(-\epsilon_{\bm{k}}^{\lambda}-\tilde{q}_{0},-\tilde{\bm{k}}_{\surf} - \tilde{\bm{q}}_{\surf},\tilde{q}_{0},\tilde{\bm{q}}_{\surf}) = Z_{\mu\nu}(\epsilon_{\bm{k}}^{\lambda},\tilde{\bm{k}}_{\surf},\tilde{q}_{0},\tilde{\bm{q}}_{\surf}).
\end{eqnarray}
Joining all together, and using 
\begin{eqnarray}\label{Def_NF_mu_app}
N_{\mu}(\epsilon) & = & \sum_{\eta=\pm}n_{F}(\epsilon + \eta\mu),
\end{eqnarray}
which is \Eq{Def_NF_mu} of the main text, we simplify $\Delta_{T}\Pi_{\mu\nu}(\qq)$ into
\begin{eqnarray}\label{Delta_T_PI_mu_nu}
\Delta_{T}\Pi_{\mu\nu}(\qq)
= - g_{s}e^{2}\dint_{\bm{k}}N_{\mu}(\epsilon_{\bm{k}})\sum_{\lambda=\pm}\dfrac{Z_{\mu\nu}(\epsilon_{\bm{k}}^{\lambda},\tilde{\bm{k}}_{\surf},\tilde{q}_{0},\tilde{\bm{q}}_{\surf})}{2\epsilon_{\bm{k}}\left( (\tilde{q}_{0} + \epsilon_{\bm{k}}^{\lambda})^{2} - \epsilon_{\bm{s}}^{2} \right)}.
\end{eqnarray}
The same change of variables $\bm{k}_{\surf} \to - (\bm{k}_{\surf} + \bm{q}_{\surf})$ can be applied to the first summand of $\Pi^{0}_{\mu\nu}(\qq)$, obtaining
\begin{eqnarray}
%L_{3} & = & 
\dint_{\bm{k}}\dfrac{Z_{\mu\nu}(-\tilde{q}_{0}-\epsilon_{\bm{s}},\tilde{\bm{k}}_{\surf},\tilde{q}_{0},\tilde{\bm{q}}_{\surf})}{2\epsilon_{\bm{s}}\left( (\tilde{q}_{0} + \epsilon_{\bm{s}})^{2} - \epsilon_{\bm{k}}^{2} \right)}%\nonumber\\
& = & \dint_{\bm{k}}\dfrac{Z_{\mu\nu}(-\tilde{q}_{0}-\epsilon_{-\bm{k}},-\tilde{\bm{s}}_{\surf},\tilde{q}_{0},\tilde{\bm{q}}_{\surf})}{2\epsilon_{-\bm{k}}\left( (\tilde{q}_{0} + \epsilon_{-\bm{k}})^{2} - \epsilon_{-\bm{s}}^{2} \right)}\nonumber\\
& = & \dint_{\bm{k}}\dfrac{Z_{\mu\nu}(-\epsilon_{\bm{k}}-\tilde{q}_{0},-\tilde{\bm{k}}_{\surf} -\tilde{\bm{q}}_{\surf},\tilde{q}_{0},\tilde{\bm{q}}_{\surf})}{2\epsilon_{\bm{k}}\left( (\tilde{q}_{0} + \epsilon_{\bm{k}})^{2} - \epsilon_{\bm{s}}^{2} \right)}\nonumber\\
& = & \dint_{\bm{k}}\dfrac{Z_{\mu\nu}(\epsilon_{\bm{k}},\tilde{\bm{k}}_{\surf},\tilde{q}_{0},\tilde{\bm{q}}_{\surf})}{2\epsilon_{\bm{k}}\left( (\tilde{q}_{0} + \epsilon_{\bm{k}})^{2} - \epsilon_{\bm{s}}^{2} \right)},
\end{eqnarray}
where we have used the symmetry shown in \Eq{Z_symmetry}, then we have
\begin{eqnarray}
\Pi^{(0)}_{\mu\nu}(\qq)
& = & g_{s}e^{2}\dint_{\bm{k}}%\nonumber\\& &
\left[ \dfrac{Z_{\mu\nu}(-\tilde{q}_{0}-\epsilon_{\bm{s}},\tilde{\bm{k}}_{\surf},\tilde{q}_{0},\tilde{\bm{q}}_{\surf})}{2\epsilon_{\bm{s}}\left( (\tilde{q}_{0} + \epsilon_{\bm{s}})^{2} - \epsilon_{\bm{k}}^{2} \right)} + \dfrac{Z_{\mu\nu}(-\epsilon_{\bm{k}},\tilde{\bm{k}}_{\surf},\tilde{q}_{0},\tilde{\bm{q}}_{\surf})}{2\epsilon_{\bm{k}}( (\tilde{q}_{0} - \epsilon_{\bm{k}})^{2} - \epsilon_{\bm{s}}^{2} )} \right]\nonumber\\
& = & g_{s}e^{2}\dint_{\bm{k}}%\nonumber\\& &
\left[ \dfrac{Z_{\mu\nu}(\epsilon_{\bm{k}},\tilde{\bm{k}}_{\surf},\tilde{q}_{0},\tilde{\bm{q}}_{\surf})}{2\epsilon_{\bm{k}}\left( (\tilde{q}_{0} + \epsilon_{\bm{k}})^{2} - \epsilon_{\bm{s}}^{2} \right)} + \dfrac{Z_{\mu\nu}(-\epsilon_{\bm{k}},\tilde{\bm{k}}_{\surf},\tilde{q}_{0},\tilde{\bm{q}}_{\surf})}{2\epsilon_{\bm{k}}( (\tilde{q}_{0} - \epsilon_{\bm{k}})^{2} - \epsilon_{\bm{s}}^{2} )} \right]\nonumber\\
& = & g_{s}e^{2}\dint_{\bm{k}}\sum_{\lambda=\pm}\dfrac{Z_{\mu\nu}(\epsilon_{\bm{k}}^{\lambda},\tilde{\bm{k}}_{\surf},\tilde{q}_{0},\tilde{\bm{q}}_{\surf})}{2\epsilon_{\bm{k}}\left( (\tilde{q}_{0} + \epsilon_{\bm{k}}^{\lambda})^{2} - \epsilon_{\bm{s}}^{2} \right)}.
\end{eqnarray}
Finally, the full Polarization operator can be written as
\begin{eqnarray}
\Pi_{\mu\nu}(\qq) & = & g_{s}e^{2}\dint_{\bm{k}}\left[ 1 - N_{\mu}(\epsilon_{\bm{k}})\right]%\nonumber\\& & \times
\sum_{\lambda=\pm}\dfrac{Z_{\mu\nu}(\epsilon_{\bm{k}}^{\lambda},\tilde{\bm{k}}_{\surf},\tilde{q}_{0},\tilde{\bm{q}}_{\surf})}{2\epsilon_{\bm{k}}\left( (\tilde{q}_{0} + \epsilon_{\bm{k}}^{\lambda})^{2} - \epsilon_{\bm{s}}^{2} \right)},
\end{eqnarray}
which is the result shown in \Eq{Obtained_Polarization_Mostepanenko}. In addition to that, from \Eq{Numerator_Polarization_Operator_Mostepanenko}, it can be seen that the spatial part of the Polarization operator can be split as \cite{non-local_Graphene_Lilia_Pablo}
\begin{eqnarray}
Z_{ij}(\kk,\qq) & = & v_{F}^{2}\Big[Z_{L}(\kk,\qq)\frac{\tilde{q}_{i}\tilde{q}_{j}}{\tilde{q}_{\surf}^{2}} + Z_{T}(\kk,\qq)\left( \delta_{ij} - \frac{\tilde{q}_{i}\tilde{q}_{j}}{\tilde{q}_{\surf}^{2}} \right)%\nonumber\\& &
+ S_{ij}Z_{S}(\kk,\qq)\Big],
\end{eqnarray}
with
\begin{eqnarray}
Z_{00}(\kk,\qq) & = & 4\left[ \tilde{k}_{0}\tilde{s}_{0} + \tilde{k}_{\surf}^{2} + \tilde{k}_{\surf}\tilde{q}_{\surf}\cos(\varphi) + m^{2} \right],\nonumber\\
Z_{T}(\kk,\qq) & = & 4\left[ \tilde{k}_{0}\tilde{s}_{0} - \left( \tilde{k}_{\surf}\tilde{q}_{\surf}\cos(\varphi) + \tilde{k}_{\surf}^{2}\cos(2\varphi) - m^{2}\right)\right],\nonumber\\
Z_{L}(\kk,\qq) & = & 4\left[ \tilde{k}_{0}\tilde{s}_{0} + \left( \tilde{k}_{\surf}\tilde{q}_{\surf}\cos(\varphi) + \tilde{k}_{\surf}^{2}\cos(2\varphi) - m^{2}\right)\right],\nonumber\\
Z_{S}(\kk,\qq) & = & 4\left[ \tilde{k}_{\surf}\tilde{q}_{\surf}\sin(\varphi) + \tilde{k}_{\surf}^{2}\sin(2\varphi)\right],
\end{eqnarray}
where $\tilde{k}_{0} = \epsilon_{\bm{k}}^{\lambda}$, $\tilde{s}_{0} = \tilde{k}_{0} + \tilde{q}_{0}$, $\tilde{k}_{\surf} = \hbar v_{F}k_{\surf}$ and $\tilde{q}_{\surf} = \hbar v_{F}q_{\surf}$ (see \Tab{Notation}). Note that there is no Hall term because this model is topologically trivial ($C=0$). It is immediate to see that $\Pi_{S}(\kk,\qq) = 0$. Then, we can separate the polarization operator into longitudinal and transverse parts as (see \Eq{Pi_Split_in_L_T_main_text})
\begin{eqnarray}\label{Pi_Split_in_L_T}
\Pi_{ij}(\qq) & = & \Pi_{L}(\qq)\frac{\tilde{q}_{i}\tilde{q}_{j}}{\tilde{q}_{\surf}^{2}} + \Pi_{T}(\qq)\left( \delta_{ij} - \frac{\tilde{q}_{i}\tilde{q}_{j}}{\tilde{q}_{\surf}^{2}} \right),
\end{eqnarray}
Each term of the Polarization operator can be written, using $p\in\{00, L, T\}$ as
\begin{eqnarray}\label{General_Pi_P}
\Pi_{p}(\qq)
= 2g_{s}e^{2}v_{p}^{2}\dint_{\bm{k}}\dfrac{1}{2\epsilon_{\bm{k}}}\left[ 1 - N_{\mu}(\epsilon_{\bm{k}})\right]%\nonumber\\& \times
\dsum_{\lambda=\pm}\left[ 
 1 + \dfrac{M_{p}(\tilde{q}_{0},\tilde{\bm{k}}_{\surf},\tilde{\bm{q}}_{\surf})}{ Q(\tilde{q}_{0},\tilde{k}_{\surf},\tilde{q}_{\surf}) + 2\tilde{\bm{k}}_{\surf}\cdot\tilde{\bm{q}}_{\surf} }\right],
\end{eqnarray}
where $v_{00}=1$, $v_{L} = v_{T} = v_{F}$ and
\begin{eqnarray}
Q(\tilde{q}_{0},\tilde{k}_{\surf},\tilde{q}_{\surf}) & = & - \tilde{q}_{z}^{2} - 2\tilde{q}_{0}\epsilon_{\bm{k}}^{\lambda},\\
M_{p}(\tilde{q}_{0},\tilde{\bm{k}}_{\surf},\tilde{\bm{q}}_{\surf}) & = & \dfrac{1}{2}Z_{p}(\epsilon_{\bm{k}}^{\lambda},\tilde{\bm{k}}_{\surf},\tilde{q}_{0},\tilde{\bm{q}}_{\surf})%\nonumber\\& &
- Q(\tilde{q}_{0},\tilde{k}_{\surf},\tilde{q}_{\surf}) - 2\tilde{\bm{k}}_{\surf}\cdot\tilde{\bm{q}}_{\surf},
\end{eqnarray}
with $\tilde{q}_{z}^{2} = \tilde{q}_{0}^{2} - \tilde{q}_{\surf}^{2}$ (see \Tab{Notation}). In particular, we obtain
\begin{eqnarray}
M_{00}(\tilde{q}_{0},\tilde{\bm{k}}_{\surf},\tilde{\bm{q}}_{\surf}) & = & - \tilde{q}_{z}^{2} + 4\tilde{q}_{0}\epsilon_{\bm{k}}^{\lambda} + 4\epsilon_{\bm{k}}^{2},\label{Def_M_00}\\
M_{L}(\tilde{q}_{0},\tilde{\bm{k}}_{\surf},\tilde{\bm{q}}_{\surf}) & = & - \tilde{q}_{z}^{2} + 4\tilde{q}_{0}\epsilon_{\bm{k}}^{\lambda} + 4\tilde{k}_{\surf}^{2}\cos^{2}(\varphi),\label{Def_M_L}\\
M_{T}(\tilde{q}_{0},\tilde{\bm{k}}_{\surf},\tilde{\bm{q}}_{\surf}) & = & - \tilde{q}_{z}^{2} + 4\tilde{q}_{0}\epsilon_{\bm{k}}^{\lambda} + 4\tilde{k}_{\surf}^{2}\sin^{2}(\varphi) - 4\tilde{k}_{\surf}\tilde{q}_{\surf}\cos(\varphi),\label{Def_M_T}
\end{eqnarray}
note that $M_{00}(\tilde{q}_{0},\tilde{\bm{k}}_{\surf},\tilde{\bm{q}}_{\surf})$ is not a function of $\varphi$, so we can write $M_{00}(\tilde{q}_{0},\tilde{k}_{\surf},\tilde{q}_{\surf})$ instead.

\subsection{Temporal Polarization}
For the case $\Pi_{\mu\nu}(\qq) = \Pi_{00}(\qq)$, using \Eq{General_Pi_P} with \Eq{Def_M_00}
\begin{eqnarray}\label{App_Final_Pi_00}
\Pi_{00}(\qq)
= 2g_{s}e^{2}\dint_{\bm{k}}\dfrac{1}{2\epsilon_{\bm{k}}}\left[ 1 - N_{\mu}(\epsilon_{\bm{k}})\right]%\nonumber\\& \times
\dsum_{\lambda=\pm}\left[ 
1 + \dfrac{M_{00}(\tilde{q}_{0},\tilde{k}_{\surf},\tilde{q}_{\surf})}{ Q(\tilde{q}_{0},\tilde{k}_{\surf},\tilde{q}_{\surf}) + 2\tilde{\bm{k}}\cdot\tilde{\bm{q}} }\right],
\end{eqnarray}
where we have used that $\tilde{q}_{\surf} = \hbar v_{F}q_{\surf}$ and $\tilde{q}_{0} = \hbar q_{0}$ (see \Tab{Notation}). Removing the UV divergence of $\Pi_{00}^{(0)}$ \cite{Klimchitskaya2016}\cite{PRL_Mohideen}\cite{Bordag2015}\cite{Bimonte2017}, we get a finite result valid for all complex frequencies:
\begin{eqnarray}\label{UVRegularized_App_Final_Pi_00}
\Pi_{00}(\qq)
= - \ii\dfrac{g_{s}\alpha c}{8\pi\hbar}\dfrac{\tilde{q}_{\surf}^{2}}{v_{F}^{2}\tilde{q}_{z}}\Psi(\delta) + 2g_{s}e^{2}\dint_{\bm{k}}\dfrac{1}{2\epsilon_{\bm{k}}}N_{\mu}(\epsilon_{\bm{k}})%\nonumber\\& \times
\dsum_{\lambda=\pm}\left[ 
1 + \dfrac{M_{00}(\tilde{q}_{0},\tilde{k}_{\surf},\tilde{q}_{\surf})}{ Q(\tilde{q}_{0},\tilde{k}_{\surf},\tilde{q}_{\surf}) + 2\tilde{\bm{k}}\cdot\tilde{\bm{q}} }\right].
\end{eqnarray}
For imaginary frequencies $\tilde{q}_{0} = \ii\Xi = \ii\hbar\xi$, we can transform \Eq{App_Final_Pi_00} into
\begin{eqnarray}\label{Delta_T_PI_00}
&\Pi_{00}(\qq)
 = 2g_{s}e^{2}\dint_{0}^{k_{M}}\dfrac{\dd k_{\surf}}{2\pi}\dfrac{k_{\surf}}{2\epsilon_{\bm{k}}}\left[1-N_{\mu}(\epsilon_{\bm{k}})\right]%\nonumber\\& \times
\dsum_{\lambda=\pm}\dint_{0}^{2\pi}\dfrac{\dd\varphi}{2\pi}\left[ 
 1 + \dfrac{M_{00}(\ii\Xi,\tilde{k}_{\surf},\tilde{q}_{\surf})}{ Q(\ii\Xi,\tilde{k}_{\surf},\tilde{q}_{\surf}) + 2\tilde{\bm{k}}_{\surf}\cdot\tilde{\bm{q}}_{\surf} }\right],
\end{eqnarray}
where $\tilde{\bm{k}}_{\surf} = \hbar v_{F}\bm{k}_{\surf}$,  $\tilde{\bm{q}}_{\surf} = \hbar v_{F}\bm{q}_{\surf}$ (see \Tab{Notation}) and $k_{M}$ plays the role of an upper cut-off in frequencies, and 
\begin{eqnarray}
Q(\ii\Xi,\tilde{k}_{\surf},\tilde{q}_{\surf}) & = & \phantom{+}\Xi^{2} + \tilde{q}_{\surf}^{2} - 2\ii\Xi\epsilon_{\bm{k}}^{\lambda},\nonumber\\
M_{00}(\ii\Xi,\tilde{\bm{k}}_{\surf},\tilde{\bm{q}}_{\surf}) & = & - \Xi^{2} - \tilde{q}_{\surf}^{2} + 4\ii\Xi\epsilon_{\bm{k}}^{\lambda} + 4\epsilon_{\bm{k}}^{2}.
\end{eqnarray}
By using $\tilde{\bm{k}}_{\surf}\cdot\tilde{\bm{q}}_{\surf} = \tilde{k}_{\surf}\tilde{q}_{\surf}\cos(\varphi)$ with $\tilde{k}_{\surf} = \hbar v_{F}k_{\surf}$ and $\tilde{q}_{\surf} = \hbar v_{F}q_{\surf}$, the angular integral can be carried out as
\begin{eqnarray}\label{Angular_Integral}
\int_{0}^{2\pi}\dfrac{\dd\varphi}{2\pi}\dfrac{1}{Q + a\cos(\varphi)} = \dfrac{1}{\sqrt{ Q^{2} - a^{2} }},
\end{eqnarray}
Therefore, $\Pi_{00}(\qq)$ can be written as
\begin{eqnarray}\label{Delta_T_PI_00v2}
&\Pi_{00}(\qq)
= 2g_{s}e^{2}\dint_{0}^{k_{M}}\dfrac{\dd k_{\surf}}{2\pi}\dfrac{k_{\surf}}{2\epsilon_{\bm{k}}}\left[1 - N_{\mu}(\epsilon_{\bm{k}})\right]%\nonumber\\& \times
\left[ 
2 + \dsum_{\lambda=\pm}\dfrac{M_{00}(\ii\lambda\Xi,\tilde{k}_{\surf},\tilde{q}_{\surf})}{ N(\ii\lambda\Xi,\tilde{k}_{\surf},\tilde{q}_{\surf}) }\right].
\end{eqnarray}
with
\begin{eqnarray}
Q(\ii\lambda\Xi,\tilde{k}_{\surf},\tilde{q}_{\surf}) & = & \Xi^{2} + \tilde{q}_{\surf}^{2} - 2\ii\lambda\Xi\epsilon_{\bm{k}},\nonumber\\
M_{00}(\ii\lambda\Xi,\tilde{k}_{\surf},\tilde{q}_{\surf}) & = & - \Xi^{2} - \tilde{q}_{\surf}^{2} + 4\ii\lambda\Xi\epsilon_{\bm{k}} + 4\epsilon_{\bm{k}}^{2},\nonumber\\
N(\ii\lambda\Xi,\tilde{k}_{\surf},\tilde{q}_{\surf}) & = & \sqrt{ \left[Q(\ii\lambda\Xi,\tilde{k}_{\surf},\tilde{q}_{\surf})\right]^{2} - (2\tilde{k}_{\surf}\tilde{q}_{\surf})^{2} }.
\end{eqnarray}
Note that the ratio can be equivalently represented as
\begin{eqnarray}\label{Transformation_to_real_part}
\dsum_{\lambda=\pm}\dfrac{M_{00}(\ii\lambda\Xi,\tilde{k}_{\surf},\tilde{q}_{\surf})}{ N(\ii\lambda\Xi,\tilde{k}_{\surf},\tilde{q}_{\surf}) } = 2\Real{\dfrac{M_{00}(\ii\Xi,\tilde{k}_{\surf},\tilde{q}_{\surf})}{ N(\ii\Xi,\tilde{k}_{\surf},\tilde{q}_{\surf}) }},
\end{eqnarray}
therefore, we obtain
\begin{eqnarray}\label{Delta_T_PI_00v3}
&\Pi_{00}(\qq)
= 2g_{s}e^{2}\dint_{0}^{k_{M}}\dfrac{\dd k_{\surf}}{2\pi}\dfrac{k_{\surf}}{\epsilon_{\bm{k}}}\left[1 - N_{\mu}(\epsilon_{\bm{k}})\right]%\nonumber\\& \times
\left[ 1 + \Real{\dfrac{M_{00}(\ii\Xi,\tilde{k}_{\surf},\tilde{q}_{\surf})}{ N(\ii\Xi,\tilde{k}_{\surf},\tilde{q}_{\surf}) }}\right],
\end{eqnarray}
The next step is to change the integration variable from $k_{\surf}$ to $\epsilon = \sqrt{(\hbar v_{F}k_{\surf})^{2} + m^{2}}$, therefore, we get
\begin{eqnarray}\label{Delta_T_PI_00v4}
&\Pi_{00}(\qq)
= 2g_{s}e^{2}\dint_{m}^{\epsilon_{M}}\dfrac{\dd\epsilon}{2\pi \hbar^{2}v_{F}^{2}}\left[1 - N_{\mu}(\epsilon)\right]%\nonumber\\& \times
\left[ 
1 + \Real{\dfrac{ - \Xi^{2} - \tilde{q}_{\surf}^{2} + 4\ii\Xi\epsilon + 4\epsilon^{2}}{ \sqrt{ \left[\Xi^{2} + \tilde{q}_{\surf}^{2} - 2\ii\Xi\epsilon\right]^{2} - (2\tilde{q}_{\surf})^{2}(\epsilon^{2} - m^{2}) } }}\right].
\end{eqnarray}
It is followed by the change of variable into non-dimensional energy $u$, using $\tilde{\theta}_{z} = \sqrt{ \Xi^{2} + \tilde{q}_{\surf}^{2} }$ (see \Tab{Notation}), we apply the change of variable $\epsilon = \tilde{\theta}_{z}\frac{u}{2}$, obtaining
\begin{eqnarray}\label{Delta_T_PI_00v5}
&\Pi_{00}(\qq)
= 2g_{s}e^{2}\dint_{\frac{2m}{\tilde{\theta}_{z}}}^{\frac{2\epsilon_{M}}{\tilde{\theta}_{z}}}\dfrac{\tilde{\theta}_{z}\dd u}{4\pi \hbar^{2}v_{F}^{2}}\left[ 1 - N_{\mu}\Big(\tilde{\theta}_{z}\frac{u}{2}\Big) \right]%\nonumber\\& \times
\left[ 
1 + \Real{\dfrac{ - \tilde{\theta}_{z}^{2} + 2\ii\Xi \tilde{\theta}_{z}u + \tilde{\theta}_{z}^{2}u^{2}}{ \sqrt{ \left[ \tilde{\theta}_{z}^{2} - \ii\Xi \tilde{\theta}_{z}u\right]^{2} - (\tilde{\theta}_{z}^{2} - \Xi^{2})(\tilde{\theta}_{z}^{2}u^{2} - 4 m^{2}) } }}\right],
\end{eqnarray}
Taking a common factor of $\tilde{\theta}_{z}>0$, defining the nondimensional parameters $\gamma = \frac{\Xi}{\tilde{\theta}_{z}}$ and $\delta = \frac{2\abs{m}}{\tilde{\theta}_{z}}$, and applying the limit $\epsilon_{M}\to\infty$ we obtain
\begin{eqnarray}\label{Delta_T_PI_00v7}
&\Pi_{00}(\qq)
= \dfrac{g_{s}e^{2}}{2\pi}\dfrac{\tilde{\theta}_{z}}{\hbar^{2}v_{F}^{2}}
\dint_{\delta}^{\infty}\dd u \left[ 1 - N_{\mu}\Big(\tilde{\theta}_{z}\frac{u}{2}\Big) \right]%\nonumber\\& \times
\left[ 
1 - \Real{\dfrac{ 1 - u^{2} - 2\ii\gamma u}{\sqrt{ 1 - u^{2} - 2\ii\gamma u + (1 - \gamma^{2})\delta^{2} } }}\right],
\end{eqnarray}
Following \cite{Klimchitskaya2016}\cite{PRL_Mohideen}\cite{Bordag2015}\cite{Bimonte2017}, the divergent integral is regularized as
\begin{eqnarray}
\Pi_{00}(\qq) 
= - \dfrac{g_{s}\alpha c}{8\pi\hbar}\dfrac{\tilde{q}_{\surf}^{2}}{v_{F}^{2}\tilde{\theta}_{z}}\Psi(\delta) - g_{s}\dfrac{\alpha c}{2\pi\hbar}
\dfrac{\tilde{\theta}_{z}}{v_{F}^{2}}
\dint_{\delta}^{\infty}\dd u N_{\mu}\Big(\tilde{\theta}_{z}\frac{u}{2}\Big)
\left[ 
1 - \Real{\dfrac{ 1 - u^{2} - 2\ii\gamma u}{\sqrt{ 1 - u^{2} - 2\ii\gamma u + (1 - \gamma^{2})\delta^{2} } }}\right],
\end{eqnarray}
With this result, $\Pi_{00}(\qq)$ obtained from the QFT model is equivalent to the Polarization operator obtained from the non-local Kubo formula.

\subsection{Longitudinal Polarization}
For the case $\Pi_{\mu\nu}(\qq) = \Pi_{L}(\qq)$, using \Eq{General_Pi_P} and \Eq{Def_M_L}, we have
\begin{eqnarray}\label{Pi_Lv1}
\Pi_{L}(\qq)
= 2g_{s}e^{2}v_{F}^{2}\dint_{\bm{k}}\dfrac{1}{2\epsilon_{\bm{k}}}\left[ 1 - N_{\mu}(\epsilon_{\bm{k}})\right]%\nonumber\\& \times
\dsum_{\lambda=\pm}\left[ 
 1 + \dfrac{M_{L}(\tilde{q}_{0},\tilde{\bm{k}}_{\surf},\tilde{\bm{q}}_{\surf})}{ Q(\tilde{q}_{0},\tilde{k}_{\surf},\tilde{q}_{\surf}) + 2\tilde{\bm{k}}_{\surf}\cdot\tilde{\bm{q}}_{\surf} }\right],
\end{eqnarray}
Also, using the relation between the Longitudinal and Temporal terms of the Polarization operator derived from the transversality condition
\begin{eqnarray}\label{Relation_L_with_00}
\Pi_{L}(\qq) = \dfrac{q_{0}^{2}}{q_{\surf}^{2}}\Pi_{00}(\qq),
\end{eqnarray}
the Longitudinal Polarization can be obtained in terms of $\Pi_{00}(\qq)$ from \Eq{General_Pi_P}, \Eq{Def_M_00} and \Eq{Relation_L_with_00} as
\begin{eqnarray}\label{App_Final_Pi_L}
&\Pi_{L}(\qq)
= 2g_{s}e^{2}v_{F}^{2}\dfrac{\tilde{q}_{0}^{2}}{\tilde{q}_{\surf}^{2}}\dint_{\bm{k}}\dfrac{1}{2\epsilon_{\bm{k}}}\left[ 1 - N_{\mu}(\epsilon_{\bm{k}})\right]%\nonumber\\& \times
\dsum_{\lambda=\pm}\left[ 
1 + \dfrac{M_{00}(\tilde{q}_{0},\tilde{k}_{\surf},\tilde{q}_{\surf})}{ Q(\tilde{q}_{0},\tilde{k}_{\surf},\tilde{q}_{\surf}) + 2\tilde{\bm{k}}\cdot\tilde{\bm{q}} }\right],
\end{eqnarray}
where we have used that $\tilde{q}_{\surf} = \hbar v_{F}q_{\surf}$ and $\tilde{q}_{0} = \hbar q_{0}$ (see \Tab{Notation}). This is \Eq{Final_Pi_L} shown in \Sect{Section_QFT_Model}. Removing the UV divergence of $\Pi_{L}^{(0)}$ \cite{Klimchitskaya2016}\cite{PRL_Mohideen}\cite{Bordag2015}\cite{Bimonte2017}, we get a finite result valid for all complex frequencies:
\begin{eqnarray}\label{UVRegularized_App_Final_Pi_L}
&\Pi_{L}(\qq)
=  - \ii g_{s}\dfrac{\alpha c}{8\pi\hbar}\dfrac{\tilde{q}_{0}^{2}}{\tilde{q}_{z}}\Psi(\delta) + 2g_{s}e^{2}v_{F}^{2}\dfrac{\tilde{q}_{0}^{2}}{\tilde{q}_{\surf}^{2}}\dint_{\bm{k}}\dfrac{1}{2\epsilon_{\bm{k}}}N_{\mu}(\epsilon_{\bm{k}})%\nonumber\\& \times
\dsum_{\lambda=\pm}\left[ 
1 + \dfrac{M_{00}(\tilde{q}_{0},\tilde{k}_{\surf},\tilde{q}_{\surf})}{ Q(\tilde{q}_{0},\tilde{k}_{\surf},\tilde{q}_{\surf}) + 2\tilde{\bm{k}}\cdot\tilde{\bm{q}} }\right],
\end{eqnarray}
From \Eq{Pi_Lv1} and  \Eq{App_Final_Pi_L}, by using the definitions of $M_{L}$ and $M_{00}$ given in \Eq{Def_M_L} and \Eq{Def_M_00} respectively, we obtain the following equality ($\tilde{q}_{z} = \sqrt{ \tilde{q}_{0}^{2} - \tilde{q}_{\surf}^{2} }$)
\begin{eqnarray}
\dint_{\bm{k}}\dfrac{1}{2\epsilon_{\bm{k}}}\left[ 1 - N_{\mu}(\epsilon_{\bm{k}})\right]%\nonumber\\& \times
\dsum_{\lambda=\pm}\left[ 
1 + \dfrac{M_{L}(\tilde{q}_{0},\tilde{\bm{k}}_{\surf},\tilde{\bm{q}}_{\surf})}{ Q(\tilde{q}_{0},\tilde{k}_{\surf},\tilde{q}_{\surf}) + 2\tilde{\bm{k}}_{\surf}\cdot\tilde{\bm{q}}_{\surf} }\right]
 =
\dfrac{\tilde{q}_{0}^{2}}{\tilde{q}_{\surf}^{2}}\dint_{\bm{k}}\dfrac{1}{2\epsilon_{\bm{k}}}\left[ 1 - N_{\mu}(\epsilon_{\bm{k}})\right]%\nonumber\\& \times
\dsum_{\lambda=\pm}\left[ 
1 + \dfrac{M_{00}(\tilde{q}_{0},\tilde{k}_{\surf},\tilde{q}_{\surf})}{ Q(\tilde{q}_{0},\tilde{k}_{\surf},\tilde{q}_{\surf}) + 2\tilde{\bm{k}}\cdot\tilde{\bm{q}} }\right]
\end{eqnarray}
that leads to
\begin{eqnarray}\label{Simplify_Cos2}
\dint_{\bm{k}}\dfrac{1}{2\epsilon_{\bm{k}}}\left[ 1 - N_{\mu}(\epsilon_{\bm{k}})\right]\dsum_{\lambda=\pm} 
\dfrac{4\tilde{k}_{\surf}^{2}\cos^{2}(\varphi)}{ Q(\tilde{q}_{0},\tilde{k}_{\surf},\tilde{q}_{\surf}) + 2\tilde{\bm{k}}_{\surf}\cdot\tilde{\bm{q}}_{\surf} }\nonumber\\
= \dint_{\bm{k}}\dfrac{1}{2\epsilon_{\bm{k}}}\left[ 1 - N_{\mu}(\epsilon_{\bm{k}})\right]\dsum_{\lambda=\pm}\left[ 
\dfrac{\tilde{q}_{z}^{2}}{\tilde{q}_{\surf}^{2}}\left( 1 + \dfrac{ - \tilde{q}_{z}^{2} + 4\tilde{q}_{0}\epsilon_{\bm{k}}^{\lambda}}{ Q(\tilde{q}_{0},\tilde{k}_{\surf},\tilde{q}_{\surf}) + 2\tilde{\bm{k}}_{\surf}\cdot\tilde{\bm{q}}_{\surf} }\right) + \dfrac{\tilde{q}_{0}^{2}}{\tilde{q}_{\surf}^{2}}\dfrac{4\epsilon_{\bm{k}}^{2}}{ Q(\tilde{q}_{0},\tilde{k}_{\surf},\tilde{q}_{\surf}) + 2\tilde{\bm{k}}_{\surf}\cdot\tilde{\bm{q}}_{\surf} }\right].
\end{eqnarray}
This equality will be used to simplify other terms of the Polarization operator. \Eq{Simplify_Cos2} is easily checked for imaginary frequencies after integration over the angular variable $\varphi$ and, therefore, it is valid for all complex frequencies by analytical continuation.

For imaginary frequencies $\tilde{q}_{0} = \ii\hbar\xi = \ii\Xi$, using \Eq{Delta_T_PI_00v7} and the relation given in \Eq{Relation_L_with_00}, we have
\begin{eqnarray}%\label{Delta_T_PI_00v7}
\Pi_{L}(\qq)
= - \dfrac{g_{s}}{2\pi}\dfrac{\alpha c}{\hbar}\dfrac{\Xi^{2}}{\tilde{q}_{\surf}^{2}}\tilde{\theta}_{z}
\dint_{\delta}^{\infty}\dd u \left[ 1 - N_{\mu}\Big(\tilde{\theta}_{z}\frac{u}{2}\Big) \right]%\nonumber\\& \times
\left[ 
1 - \Real{\dfrac{ 1 - u^{2} - 2\ii\gamma u}{\sqrt{ 1 - u^{2} - 2\ii\gamma u + (1 - \gamma^{2})\delta^{2} } }}\right],
\end{eqnarray}
Once $\Pi_{L}^{(0)}(\qq)$ is regularized following \cite{Bordag2015}\cite{Bimonte2017}, we have
\begin{eqnarray}
&\Pi_{L}(\qq)
= \dfrac{g_{s}\alpha c}{8\pi\hbar}\dfrac{\Xi^{2}}{\tilde{\theta}_{z}}\Psi(\delta) + g_{s}\dfrac{\alpha c}{2\pi\hbar}
\dfrac{\Xi^{2}}{\tilde{q}_{\surf}^{2}}\tilde{\theta}_{z}
\dint_{\delta}^{\infty}\dd u N_{\mu}\Big(\tilde{\theta}_{z}\frac{u}{2}\Big)
\left[ 
1 - \Real{\dfrac{ 1 - u^{2} - 2\ii\gamma u}{\sqrt{ 1 - u^{2} - 2\ii\gamma u + (1 - \gamma^{2})\delta^{2} } }}\right],
\end{eqnarray}
Those are \Eq{Final_Pi0_L0Im} and  \Eq{Delta_T_PI_00Imv6} in \Sect{Section_QFT_Model}. The equivalent electric conductivity is ($\sigma_{L}^{\rm{NR}}(\ii\xi) = \Pi_{L}(\ii\xi)/\xi$)
\begin{eqnarray}\label{QFT_Sigma_L}
\sigma_{L}^{\rm{NR}}(\qq)
= \dfrac{g_{s}\alpha c}{8\pi}\gamma\Psi(\delta) + g_{s}\dfrac{\alpha c}{2\pi}
\dfrac{\Xi}{\tilde{q}_{\surf}^{2}}\tilde{\theta}_{z}
\dint_{\delta}^{\infty}\dd u N_{\mu}\Big(\tilde{\theta}_{z}\frac{u}{2}\Big)
\left[ 
1 - \Real{\dfrac{ 1 - u^{2} - 2\ii\gamma u}{\sqrt{ 1 - u^{2} - 2\ii\gamma u + (1 - \gamma^{2})\delta^{2} } }}\right].
\end{eqnarray}

%\newpage
\subsection{Transversal Polarization}
By using \Eq{Simplify_Cos2} in \Eq{General_Pi_P} with \Eq{Def_M_T}, the Transversal Polarization term can be simplified as
\begin{eqnarray}\label{App_Final_Pi_T}
\Pi_{T}(\qq)
= 2g_{s}e^{2}v_{F}^{2}\dfrac{\tilde{q}_{0}^{2}}{\tilde{q}_{\surf}^{2}}\dint_{\bm{k}}\dfrac{1}{2\epsilon_{\bm{k}}}\left[ 1 - N_{\mu}(\epsilon_{\bm{k}})\right]%\nonumber\\& \times
\dsum_{\lambda=\pm}\left[ 
1 + \dfrac{M_{00}(\tilde{q}_{0},\tilde{k}_{\surf},\tilde{q}_{\surf}) - 4\frac{\tilde{q}_{\surf}^{2}}{q_{0}^{2}}\left( \tilde{k}_{\surf}^{2} + \tilde{q}_{0}\epsilon_{\bm{k}}^{\lambda} \right)}{ Q(\tilde{q}_{0},\tilde{k}_{\surf},\tilde{q}_{\surf}) + 2\tilde{\bm{k}}_{\surf}\cdot\tilde{\bm{q}}_{\surf} }\right].
\end{eqnarray}
Removing the UV divergence of $\Pi_{T}^{(0)}$ \cite{Klimchitskaya2016}\cite{PRL_Mohideen}\cite{Bordag2015}\cite{Bimonte2017}, we get a finite result valid for all complex frequencies:
\begin{eqnarray}\label{UVRegularized_App_Final_Pi_T}
\Pi_{T}(\qq)
=  - g_{s}\ii\dfrac{\alpha c}{8\pi\hbar}\tilde{q}_{z}\Psi(\delta) - 2g_{s}e^{2}v_{F}^{2}\dfrac{\tilde{q}_{0}^{2}}{\tilde{q}_{\surf}^{2}}\dint_{\bm{k}}\dfrac{1}{2\epsilon_{\bm{k}}}N_{\mu}(\epsilon_{\bm{k}})
\dsum_{\lambda=\pm}\left[ 
1 + \dfrac{M_{00}(\tilde{q}_{0},\tilde{k}_{\surf},\tilde{q}_{\surf}) - 4\frac{\tilde{q}_{\surf}^{2}}{q_{0}^{2}}\left( \tilde{k}_{\surf}^{2} + \tilde{q}_{0}\epsilon_{\bm{k}}^{\lambda} \right)}{ Q(\tilde{q}_{0},\tilde{k}_{\surf},\tilde{q}_{\surf}) + 2\tilde{\bm{k}}_{\surf}\cdot\tilde{\bm{q}}_{\surf} }\right].
\end{eqnarray}
From \Eq{App_Final_Pi_T}, we study the special case of imaginary frequencies $\tilde{q}_{0} = \ii\hbar\xi = \ii\Xi$, in polar coordinates
\begin{eqnarray}
\Pi_{T}(\qq)
= g_{s}e^{2}v_{F}^{2}\dfrac{- \Xi^{2}}{\tilde{q}_{\surf}^{2}}\dint_{0}^{k_{M}}\dfrac{\dd k_{\surf}}{2\pi}\dfrac{k_{\surf}}{\epsilon_{\bm{k}}}\left[ 1 - N_{\mu}(\epsilon_{\bm{k}})\right]\dint_{0}^{2\pi}\frac{\dd\varphi}{2\pi}
\dsum_{\lambda=\pm}\left[ 
1 + \dfrac{M_{00}(\ii\Xi,\tilde{k}_{\surf},\tilde{q}_{\surf}) + 4\frac{\tilde{q}_{\surf}^{2}}{\Xi^{2}}\left( \tilde{k}_{\surf}^{2} + \ii\lambda\Xi\epsilon_{\bm{k}} \right)}{ Q(\ii\Xi,\tilde{k}_{\surf},\tilde{q}_{\surf}) + 2\tilde{k}_{\surf}\tilde{q}_{\surf}\cos(\varphi) }\right].
\end{eqnarray}
The integral in $\varphi$ can be carried our using \Eq{Angular_Integral}
\begin{eqnarray}
\int_{0}^{2\pi}\dfrac{\dd\varphi}{2\pi}\dfrac{1}{Q + a\cos(\varphi)} = \dfrac{1}{\sqrt{ Q^{2} - a^{2} }}.
\end{eqnarray}
Therefore, $\Pi_{T}(\qq)$ can be written as
\begin{eqnarray}
\Pi_{T}(\qq)
= g_{s}e^{2}v_{F}^{2}\dfrac{- \Xi^{2}}{\tilde{q}_{\surf}^{2}}\dint_{0}^{k_{M}}\dfrac{\dd k_{\surf}}{2\pi}\dfrac{k_{\surf}}{\epsilon_{\bm{k}}}\left[ 1 - N_{\mu}(\epsilon_{\bm{k}})\right]
\dsum_{\lambda=\pm}\left[ 
1 + \dfrac{M_{00}(\ii\lambda\Xi,\tilde{k}_{\surf},\tilde{q}_{\surf}) + 4\frac{\tilde{q}_{\surf}^{2}}{\Xi^{2}}\left( \tilde{k}_{\surf}^{2} + \ii\lambda\Xi\epsilon_{\bm{k}} \right)}{ \sqrt{ \Big[ Q(\ii\lambda\Xi,\tilde{k}_{\surf},\tilde{q}_{\surf}) \Big]^{2} - (2\tilde{k}_{\surf}\tilde{q}_{\surf})^{2} } }\right].
\end{eqnarray}
Similarly to what was done in \Eq{Transformation_to_real_part}, using $\lambda^{2} = 1$, the ratio can be equivalently represented as
\begin{eqnarray}%\label{Transformation_to_real_part}
\dsum_{\lambda=\pm}\dfrac{M_{T}(\ii\lambda\Xi,\tilde{k}_{\surf},\tilde{q}_{\surf})}{ N(\ii\lambda\Xi,\tilde{k}_{\surf},\tilde{q}_{\surf}) } = 2\Real{\dfrac{M_{T}(\ii\Xi,\tilde{k}_{\surf},\tilde{q}_{\surf})}{ N(\ii\Xi,\tilde{k}_{\surf},\tilde{q}_{\surf}) }},
\end{eqnarray}
therefore, we obtain
\begin{eqnarray}
\Pi_{T}(\qq)
= g_{s}e^{2}v_{F}^{2}\dfrac{- \Xi^{2}}{\tilde{q}_{\surf}^{2}}\dint_{0}^{k_{M}}\dfrac{\dd k_{\surf}}{2\pi}\dfrac{k_{\surf}}{\epsilon_{\bm{k}}}\left[ 1 - N_{\mu}(\epsilon_{\bm{k}})\right]
\left[ 
2 + 2\Real{\dfrac{M_{00}(\ii\Xi,\tilde{k}_{\surf},\tilde{q}_{\surf}) + 4\frac{\tilde{q}_{\surf}^{2}}{\Xi^{2}}\left( \tilde{k}_{\surf}^{2} + \ii\Xi\epsilon_{\bm{k}} \right)}{ \sqrt{ \Big[ Q(\ii\Xi,\tilde{k}_{\surf},\tilde{q}_{\surf}) \Big]^{2} - (2\tilde{k}_{\surf}\tilde{q}_{\surf})^{2} } } }\right].
\end{eqnarray}
Now we apply the change of coordinates $\epsilon = \sqrt{ (\hbar v_{F}k_{\surf})^{2} + m^{2} }$
\begin{eqnarray}
\Pi_{T}(\qq)
= 2g_{s}e^{2}v_{F}^{2}\dfrac{- \Xi^{2}}{\tilde{q}_{\surf}^{2}}\dint_{m}^{\epsilon_{M}}\dfrac{\dd\epsilon}{2\pi\hbar^{2}v_{F}^{2}}\left[ 1 - N_{\mu}(\epsilon)\right]
\left[ 
1 + \Real{\dfrac{ - \Xi^{2} - \tilde{q}_{\surf}^{2} + 4\ii\Xi\epsilon + 4\epsilon^{2} + 4\frac{\tilde{q}_{\surf}^{2}}{\Xi^{2}}\left( \epsilon^{2} - m^{2} + \ii\Xi\epsilon \right)}{ \sqrt{ \left( \Xi^{2} + \tilde{q}_{\surf}^{2} - 2\ii\Xi\epsilon \right)^{2} - 4( \epsilon^{2} - m^{2} )\tilde{q}_{\surf}^{2} } } }\right].
\end{eqnarray}
It is followed by the change of variable into a non-dimensional energy, using $\tilde{\theta}_{z} = \sqrt{ \Xi^{2} + \tilde{q}_{\surf}^{2} }$ (and, therefore, $\tilde{q}_{\surf}^{2} = \tilde{\theta}_{z}^{2} - \Xi^{2}$), we apply the change of variable $\epsilon = \tilde{\theta}_{z}\frac{u}{2}$, obtaining
\begin{eqnarray}
\Pi_{T}(\qq)
= 2g_{s}\dfrac{\alpha c}{\hbar}\dfrac{- \Xi^{2}}{\tilde{q}_{\surf}^{2}}\dint_{\frac{2m}{\tilde{\theta}_{z}}}^{\frac{2\epsilon_{M}}{\tilde{\theta}_{z}}}\dfrac{\dd u}{2\pi}\frac{\tilde{\theta}_{z}}{2}\left[ 1 - N_{\mu}(\tilde{\theta}_{z}\frac{u}{2})\right]
\left[ 
1 + \Real{\dfrac{ - \tilde{\theta}_{z}^{2} + 2\ii\Xi\tilde{\theta}_{z}u + \tilde{\theta}_{z}^{2}u^{2} + 4\left(\frac{\tilde{\theta}_{z}^{2}}{\Xi^{2}} - 1 \right)\left( \tilde{\theta}_{z}^{2}\frac{u^{2}}{4} - m^{2} + \ii\Xi\tilde{\theta}_{z}\frac{u}{2} \right)}{ \sqrt{ \left( \tilde{\theta}_{z}^{2} - \ii\Xi\tilde{\theta}_{z}u \right)^{2} - ( \tilde{\theta}_{z}^{2}u^{2} - 4m^{2} )( \tilde{\theta}_{z}^{2} - \Xi^{2} ) } } }\right].
\end{eqnarray}
Using the definitions $\delta = \frac{2\abs{m}}{\tilde{\theta}_{z}}$ and $\gamma = \frac{\Xi}{\tilde{\theta}_{z}}$, we can simplify this integral into
\begin{eqnarray}
\Pi_{T}(\qq)
= 2g_{s}\dfrac{\alpha c}{\hbar}\dfrac{- \Xi^{2}}{\tilde{q}_{\surf}^{2}}\dint_{\delta}^{\frac{2\epsilon_{M}}{\tilde{\theta}_{z}}}\dfrac{\dd u}{2\pi}\frac{\tilde{\theta}_{z}}{2}\left[ 1 - N_{\mu}(\tilde{\theta}_{z}\frac{u}{2})\right]
\left[ 
1 + \Real{\dfrac{ - 1 + 2\ii\gamma u + u^{2} + \left( \gamma^{-2} - 1 \right)\left( u^{2} - \delta^{2} + 2\ii\gamma u \right)}{ \sqrt{ \left( 1 - \ii\gamma u \right)^{2} - ( u^{2} - \delta^{2} )( 1 - \gamma^{2} ) } } }\right].
\end{eqnarray}
This integral can be further simplified, and the cut-off can be eliminated to obtain
\begin{eqnarray}
\Pi_{T}(\qq)
= \dfrac{g_{s}\alpha c}{2\pi\hbar}\dfrac{\Xi^{2}}{\tilde{q}_{\surf}^{2}}\tilde{\theta}_{z}\dint_{\delta}^{\infty}\dd u\left[ 1 - N_{\mu}(\tilde{\theta}_{z}\frac{u}{2})\right]
\left[ 
1 - \Real{\dfrac{ \left( 1 + \ii\gamma^{-1}u \right)^{2} + (\gamma^{-2} - 1)\delta^{2} }{ \sqrt{ 1 - 2\ii\gamma u - u^{2} + ( 1 - \gamma^{2} )\delta^{2} } } }\right].
\end{eqnarray}
Once $\Pi_{T}^{(0)}(\qq)$ is regularized following \cite{Bordag2015}\cite{Bimonte2017}, we have
\begin{eqnarray}
&\Pi_{T}(\qq)
= \dfrac{g_{s}\alpha c}{8\pi\hbar}\tilde{\theta}_{z}\Psi(\delta) - g_{s}\dfrac{\alpha c}{2\pi\hbar}
\dfrac{\Xi^{2}}{\tilde{q}_{\surf}^{2}}\tilde{\theta}_{z}
\dint_{\delta}^{\infty}\dd u N_{\mu}\Big(\tilde{\theta}_{z}\frac{u}{2}\Big)
\left[ 
1 - \Real{\dfrac{ \left( 1 + \ii\gamma^{-1}u \right)^{2} + (\gamma^{-2} - 1)\delta^{2} }{ \sqrt{ 1 - 2\ii\gamma u - u^{2} + ( 1 - \gamma^{2} )\delta^{2} } } }\right].
\end{eqnarray}
Those are \Eq{Final_Pi0_T0} and  \Eq{Delta_T_PI_T} in \Sect{Section_QFT_Model}. The equivalent transversal electric conductivity is ($\sigma_{T}^{\rm{NR}}(\ii\xi) = \Pi_{T}(\ii\xi)/\xi$)
\begin{eqnarray}\label{QFT_Sigma_T}
\sigma_{T}^{\rm{NR}}(\qq)
= \dfrac{g_{s}\alpha c}{8\pi}\gamma^{-1}\Psi(\delta) - g_{s}\dfrac{\alpha c}{2\pi}
\dfrac{\Xi}{\tilde{q}_{\surf}^{2}}\tilde{\theta}_{z}
\dint_{\delta}^{\infty}\dd u N_{\mu}\Big(\tilde{\theta}_{z}\frac{u}{2}\Big)
\left[ 
1 - \Real{\dfrac{ \left( 1 + \ii\gamma^{-1}u \right)^{2} + (\gamma^{-2} - 1)\delta^{2} }{ \sqrt{ 1 - 2\ii\gamma u - u^{2} + ( 1 - \gamma^{2} )\delta^{2} } } }\right].
\end{eqnarray}

\newpage

\end{widetext}

%%%%%%%%%%%%%%%%%%%%%%%%%%%%%%%%%%%%%%%%%%%%%%%%%%%%%%%%%%%%%%%%%%%%%%%%%%%%%%%%%%%%%%%%%%%%%%%%%%%%%%%%%%%%%%%%%%%%%%%%%%%%%%%%%%%%%%%%%%%%%%%%%%%%%%%%%%%%%%%%%%%%%%%%%%%%%%%%%%%%

%\nocite{*}
\newpage
%\bibliography{bibliography}% Produces the bibliography via BibTeX.

\providecommand{\noopsort}[1]{}\providecommand{\singleletter}[1]{#1}%

\end{document}